\documentclass{aa}  

\usepackage{txfonts}
\usepackage{graphicx}
\usepackage{txfonts}
\usepackage{natbib}
\usepackage{subfig}
\usepackage{longtable} % for 'longtable' environment
\usepackage{lscape} % for 'landscape' environment

\DeclareRobustCommand{\VAN}[3]{#2}
\let\VANthebibliography\thebibliography
\def\thebibliography{\DeclareRobustCommand{\VAN}[3]{##3}\VANthebibliography}

\newcommand{\XTone}{XRT~030511}
\newcommand{\XTtwo}{XRT~100831}
\newcommand{\XTthree}{XRT~060207}
\newcommand{\XTfour}{XRT~070618}
\newcommand{\se}{{\sc Source Extractor}}
\def\approxgt{\ifmmode \rlap{$>$}{}_{{}_{{}_{\textstyle\sim}}} \else%
$\rlap{$>$}{}_{{}_{{}_{\textstyle\sim}}}$\fi} 

\begin{document} 

   \title{Redshifts of candidate host galaxies of four fast X-ray transients using VLT/MUSE}

   % \subtitle{I. Overviewing the $\kappa$-mechanism}

   \author{Anne Inkenhaag
          \inst{1, 2}
          \and
          Peter G. Jonker\inst{1, 2}
          \and
          Andrew J. Levan\inst{1,3}
          \and
          Jonathan Quirola-V\'asquez\inst{1,4,5}
          \and 
          Franz E. Bauer\inst{5,4,6}
          \and
          Deepak Eappachen\inst{1}
          }

   \institute{Department of Astrophysics/IMAPP, Radboud University Nijmegen, P.O.~Box 9010, 6500 GL Nijmegen, The Netherlands \\
              \email{a.inkenhaag@astro.ru.nl}
         \and
             SRON, Netherlands Institute for Space Research, Niels Bohrweg 4, 2333~CA, Leiden, The Netherlands
        \and
            Department of Physics, University of Warwick, Gibbet Hill Road, Coventry, CV4 7AL, UK
        \and
            Millennium Institute of Astrophysics (MAS), Nuncio Monse$\tilde{n}$or S\'otero Sanz 100, Providencia, Santiago, Chile
        \and
            % Instituto de Astrofísica, Pontificia Universidad Católica de Chile, Casilla 306, Santiago 22, Chile
            Instituto de Astrof{\'{\i}}sica and Centro de Astroingenier{\'{\i}}a, Facultad de F{\'{i}}sica, Pontificia Universidad Cat{\'{o}}lica de Chile, Campus San Joaquín, Av. Vicuña Mackenna 4860, Macul Santiago, Chile, 7820436
        \and
            Space Science Institute, 4750 Walnut Street, Suite 205, Boulder, Colorado 80301
             }

   \date{Received April 5, 2024; accepted July 3, 2024}

% \abstract{}{}{}{}{} 
% 5 {} token are mandatory
 
  \abstract
   {Fast X-ray transients (FXTs) are X-ray flares lasting minutes to hours. Multi-wavelength counterparts to these FXTs have been proven hard to find. As a result distance measurements are through indirect methods such as host galaxy identification. Out of three of the main models proposed for FXTs;  supernova shock breakout emission (SN SBO), binary neutron star (BNS) mergers and tidal dirsuption events (TDEs) of an intermediate mass black hole (IMBH) disrupting a white dwarf (WD), the SN SBO predicts a much lower maximum peak X-ray luminosity (L$_{X, peak}$). Therefore obtaining the distance to FXTs will be a powerful probe to investigate the nature of these FXTs.
   }
   {We aim to obtain distance measurements to four FXTs by identifying candidate host galaxies. Through a redshift measurement of candidate host galaxies we derive L$_{X, peak}$ and the projected offset between the candidate host galaxy and the FXT.}
   {We have obtained Very Large Telescope (VLT)/Multi Unit Spectroscopic Explorer (MUSE) observations of a sample of FXTs. We report the redshift of between 13 and 22 galaxies per FXT. We use these redshifts to calculate the distance, L$_{X, peak}$ and the projected offsets between the FXT position and the position of the sources. Additionally, we also compute the chance alignment probabilities for these sources with the FXT postitions. }
   {We find L$_{X, peak}>10^{44}$~erg~s$^{-1}$ if we assume that any of the sources with a redshift measurement is the true host galaxy of the corresponding FXT. For XRT~100831 we find a very faint galaxy (m$_\mathrm{R, AB}$ = 26.5$\pm$0.3, $z\sim1.22$, L$_{X, peak} \approx 8\times10^{45}$~erg~s$^{-1}$ if the FXT is at this distance) within the 1$\sigma$ uncertainty region with a chance alignment probability of 0.04. For XRT~060207 we find a candidate host galaxy at $z=0.939$ with a low chance alignment probability within the 1$\sigma$ uncertainty region. However, we also report the detection of a late-type star within the 3$\sigma$ uncertainty region with a similar chance alignment probability. For the remaining FXTs (XRT~030511 and XRT~070618) we find no sources within their 3$\sigma$ uncertainty regions. The projected offsets between the galaxies and the FXT positions is >33~kpc at 1$\sigma$ uncertainty. Therefore, if one of these candidate host galaxies turns out to be the true host galaxy, it would imply that the FXT progenitor originated from a system that received a significant kick velocity at formation.}
   {We rule out a SN SBO nature for all FXTs based on L$_{X, peak}$ and the projected offsets between the FXT position and the sources, assuming any of the candidate host galaxies with a redshift determination is the true host galaxy to the FXT. For XRT~100831 we conclude the detected galaxy within the 1$\sigma$ uncertainty position is likely to be the host galaxy of this FXT based on the chance alignment probability. From the available information, we are not able to determine if XRT~060207 originated from the galaxy found within 1$\sigma$ of the FXT position or was due to a flare from the late-type star detected within the 3$\sigma$ uncertainty region. Based on L$_{X, peak}$ and the offsets within our sample, we are not able to distinguish between the BNS merger and the IMBD-WD TDE progenitor model. However, for the candidate host galaxies with an offset larger than $\sim30$~kpc, we can conclude the IMBH-WD TDE is unlikely due to the large offset.}


   \maketitle
%
%-------------------------------------------------------------------

\section{Introduction}

Fast X-ray transients (FXTs) are flashes of X-ray emission that can last from minutes to hours. Many FXTs have been observed since the 1970s \citep[e.g.,][]{Ambruster1986,Connors1986,CastroTirado1999,Arefiev2003}, but the first well localised one was serendipitously discovered by \cite{Soderberg2008} during scheduled \textit{Swift} observations of the galaxy NGC2770. 
Since then, roughly 30 more localised FXTs have been reported, mostly through archival searches of \textit{Chandra} and \textit{XMM-Newton} data \citep[e.g.,][]{Jonker2013, QV2022, QV2023, AL2020, Lin2022, Eappachen2023}. With the launch of X-ray satellite
\textit{Einstein Probe} \citep{Yuan2022}, we expect more to be reported in real time\footnote{This is already happening during the commissioning phase of \textit{EP}, EP240315a was reported shortly after its discovery \citep{Zhang2024}. }.

Since the majority of the FXT sample have been reported from archival searches, we lack contemporaneous, multiwavelength counterparts of these events in all but one case. Only for the first event, X-ray ourburst (XRO)~080109 has a multi-wavelength counterpart been observed, an unusual supernova (SN) \citep[SN~2008D;][]{Soderberg2008}. This lack of counterparts contributes to substantial uncertainties regarding the progenitors, distances and energetics of the remaining sample. There are no further redshift measurements made directly from the transient, and what measurements do exist rely on redshifts (either spectroscopic or photometric) from the host galaxy \citep[e.g.][]{QV2022, Eappachen2022}. 

Despite the uncertainties in the distance scale, there are three main models proposed regarding the nature of FXTs. These models include, but are not limited to:
a supernova shock breakout (SBO) \citep[e.g.,][]{Soderberg2008, AL2020, WK2017}, a tidal disruption event of a white dwarf (WD) by an intermediate mass black hole (IMBH) \citep[e.g.,][]{Jonker2013, Glennie2015} or the merger of a binary neutron star (BNS) system \citep[e.g.,][]{Jonker2013, Bauer2017, Dai2018, Xue2019, QV2024}. These three models are associated with different ranges in peak X-ray luminosities.
Roughly speaking: $L_\textrm{X, peak} \lesssim 10^{44}$ erg~s$^{-1}$ for SN SBOs \citealt{Soderberg2008, WK2017, Goldberg2022}, $L_\textrm{X, peak} \lesssim 10^{48}$~erg~s$^{-1}$ for IMBH-WD TDEs \citep{MacLeod2016} where the high luminosity is due to jetted emission, and $L_\textrm{X, peak} \approx 10^{44}-10^{51}$~erg~s$^{-1}$ for BNS mergers also considering beamed emission for the higher luminosities; \citep{Berger2014}. 
These different ranges in $L_\textrm{X, peak}$ means determining the distance to these event is a powerfil probe to uncover their nature. They also yield potentially very different locations for the FXTs relative to their host galaxies. BNS merger progenitor systems can be propelled to large distance via kicks imparted to both the binary and individual neutron stars \citep[e.g.,][]{Lai2006}. IMBH systems may lie in the nuclei of low-mass galaxies \citep[e.g.,][]{Reines2013}, globular clusters \citep{Jonker2012} or even in hypercompact stellar clusters \citep{Merritt2009}, while massive stellar explosions should lie within star-forming galaxies \citep[e.g.,][]{Hakobyan2012}. We note that the observed FXT properties are so diverse that they might well originate from different progenitors.

Substantial efforts have been made to associate FXTs with their host galaxies to obtain a distance through the redshift of the host galaxy \citep[e.g.,][]{Bauer2017, Xue2019, AL2020, Novara2020, Lin2022, Eappachen2022, Eappachen2023, Eappachen2024, QV2022, QV2023, QV2024}. From this distance to the host galaxy we can calculate the peak X-ray luminosity and use this to discern between different progenitor models. $L_\textrm{X, peak}$, along with the (projected) offset between the FXT position and the host galaxy, its stellar mass and the star formation rate (SFR) could help us distinguish between progenitor models \citep[e.g.,]{Xue2019, QV2022, QV2023, Eappachen2023, Eappachen2024}.

Previous work on associating FXTs with their host galaxies has for example been done by \cite{Bauer2017}, who identify a host for CDF-S XT1 with $z_\mathrm{ph} = 2.23 $(0.39–3.21 at 2$\sigma$ confidence) and \citep{Xue2019} who identify a host for CDF-S XT2 at $z=0.738$. \cite{AL2020} claim for a set of \textit{XMM-Newton} discovered FXTs, that all are consistent with the SN SBO model. This conclusion is based on redshifts of galaxies close to the FXT positions in projection, although they cannot confidently associate all sources with a host galaxy.

During their search for magnetar powered FXTs in the \textit{Chandra} archive, \cite{Lin2022} identified a host galaxy candidate for XRT~170901. This candidate host galaxy is a late-type galaxy with evidence for strong SF activity, although the FXTs projected position does not coincide with a SF region. They use this information, in combination with the peak luminosity of the FXT and the X-ray light curve shape, to argue this FXT is consistent with a magnetar created in a BNS merger. 

\cite{QV2022} and \cite{QV2023}, together investigate the detection of 22 FXTs from archival \textit{Chandra} searches. Between these works, 13 FXTs have a suggested host galaxy association. For these candidate host galaxies, both papers obtain the star formation rates and stellar masses of the galaxies based on existing photometry. \cite{QV2022} and \cite{QV2023} rule out a classical (non-relativistic) SBO nature for all of their FXTs with a host galaxy candidate based on the peak X-ray luminosity, and X-ray properties. \cite{QV2022} also state that for the sample they classify as "nearby" FXTs ($d\lesssim100$~Mpc) the luminosity is too low to be from any of the proposed models, but more likely to be from ultra luminous X-ray sources or X-ray binaries in close-by galaxies. 

\cite{Eappachen2024} perform a more detailed host galaxy search on a subset of the \textit{XMM-Newton} sample discussed in \cite{AL2020}. They find candidate host galaxies with $0.0928<z<0.645$, implying $10^{43} < L_\textrm{X, peak} < 10^{45}$~erg~s$^{-1}$ for this set of seven FXTs. They also investigate the SFR and the masses of the candidate host galaxies and conclude that only one event, XRT~100621 remains consistent with a SN SBO event as already reported by \cite{Novara2020} and \cite{AL2020}. Another FXT in this sample is 
a Galactic flare star. The remaining five sources are consistent with being either due to an IMBH-WD TDE or due to a BNS merger, but they are not able to distinguish between those two models. 

In this work we present Very Large Telescope (VLT)/Multi Unit Spectroscopic Explorer (MUSE) observations of the position and the environment of four FXTs, detected by \textit{XMM-Newton} and \textit{Chandra} and identified by \cite{Lin2019, AL2020} and \cite{QV2022}.
% where previous work did not reveal a clear host galaxy. 
We attempt to measure the redshifts of all extended objects within the MUSE data cube, and for those with a redshift measurement, we calculate $L_\textrm{X, peak}$ under the assumption that the object was the host galaxy of the FXT. From this we aim to constrain the nature of these events.

Throughout this paper we use H$_0 = 67.8$\,km\,s$^{-1}$\,Mpc$^{-1}$, $\Omega_{\rm m}=0.308$ and $\Omega_{\rm \Lambda} = 0.692$ \citep{Planck2016}. Magnitudes are presented in the AB magnitude system, uncorrected for Galactic extinction, and uncertainties are at the 1$\sigma$ confidence level unless stated otherwise.

%--------------------------------------------------------------------
\section{Sample under study in this Manuscript}

\begin{table*}
\centering
\caption{The sample of FXTs studied in this work using VLT/MUSE observations (ID 109.236W). The VLT/MUSE exposure time was 2804 seconds divided over four exposures. F$_{X, peak}$ is reported in the 0.3--10~keV energy range and given in units of $10^{-13}$~erg s$^{-1}$ cm$^{-2}$. The positional uncertainty is given at a 1$\sigma$ confidence level.}
\label{tab:sample}
 \hspace*{-.7cm}\begin{tabular}{lccccccc}
\hline
Name & RA & dec & Pos. Unc. & F$_{X, peak}$ & MUSE Obs Date & Seeing  & Ref \\
 & (degrees) & (degrees) & (arcsec) &  &  & (arcsec) &  \\
\hline 
\XTone$^\ddagger$  & 76.77817 & -31.86967 & 0.53 & $23\pm3^\dagger$ & 2022-04-03 & 0.88 & (a), (b) \\
\XTtwo & 90.00450 &	-52.71501 & 0.83 & $8.9\pm3.4$ & 2022-04-03 & 1.25 & (b) \\
\XTthree$^*$ & 	196.83225 & -40.46096 & 1.9 & $26.26\pm7.23$ & 2022-04-02 & 0.80 & (c) \\ 
\XTfour$^*$ & 24.27533 & -12.95260 & 0.9 & $105.92\pm15.31$ & 2022-09-20 & 1.17 & (c) \\
\hline 
\end{tabular}
\newline $^\dagger$ value taken from \cite{QV2022}.
\newline $^\ddagger$ This source was detected at a high off-axis angle in the CXO-ACIS instrument. \cite{Lin2022} and \cite{QV2022} reported this FXT simultaneously with slightly different RA and Dec due to a combination of the difficulty of performing astrometry on high off-axis sources and a different catalog used to perform the astrometry. We note that within their respective uncertainties the coordinates are fully consistent within 2$\sigma$. We use the coordinates and uncertainty presented in \cite{Lin2022}.
\newline $^*$ The RA, Dec and positional uncertainty in the X-ray position for these sources were obtained from the XMM-Newton serendipitous catalogue \citep{Webb2020}.
\newline The positional uncertainty is the square root of the RMS on the astrometric calibration as described in Section~\ref{sec:anal} plus the 1$\sigma$ uncertainty of the X-ray position as given in the literature added quadratically, as described in Section~\ref{sec:anal}. 
\newline (a) \cite{Lin2022}, (b) \cite{QV2022}, (c) \cite{AL2020}
\end{table*}

Here we describe the sample or FXTs for which we obtained MUSE observations, and a summary of their properties can be found in 
Table~\ref{tab:sample}. The core selection criteria was that in none of the cases an apparent host galaxy consistent with the FXT localisation in existing imaging of the field had been discovered. 

\subsection*{XRT 030511}

\XTone ~was first reported by \cite{Lin2019} and further investigated by \cite{Lin2022} (who refer to the event as either XRT030510 or XRT030511) and \cite{QV2022}. It was discovered in archival \textit{Chandra} data and no host galaxy was detected (\citealt{Lin2022}). \cite{QV2022} also did not detect a host galaxy and they rule out an undetected stellar counterpart based on the ratio log(L$_x$/L$_{bol}$)=log(F$_x$/F$_{bol}$) in which L$_x$ is the X-ray flare luminosity and L$_{bol}$ is the average (non-flare) bolometric luminosity. They use the upper limits on the source magnitude in the Dark Energy Camera (DECam) y-band at the position of \XTone\ to determine a minimum log(F$_x$/F$_{bol}$)$\approxgt$ 1.6-2.1 for M- and brown dwarf stars. This value is above the known range for stellar flares from late type stars (log(F$_x$/F$_{bol}$)$\lesssim-3.0$ and $\lesssim0.0$ for M- and L-dwarfs respectively \citep[e.g.,][]{GA2008, DeLuca2020}. Hence, no host galaxy has been identified for XRT 030510.

\subsection*{XRT 100831}

\XTtwo\ was first reported by \cite{QV2022}. They report a marginal detection of an object in DECam i$^\prime$-band observations, that lies just outside the $3\sigma$ X-ray uncertainty position of this FXT, but do not report a magnitude for this object. Following the same procedure as for \XTone, using the upper limit on a detection in the DECam g$^\prime$-band, \cite{QV2022} calculate log(F$_x$/F$_{bol}$)$\approxgt$ -0.8 to -0.3 depending on the type of late-type star. These values are just consistent with a stellar flare nature, therefore a stellar flare nature cannot completely be ruled out for \XTtwo, although it would require extrema in both the X-ray to optical flux ratio and location within the source error box. 

\subsection*{XRT 060207}

\XTthree\ was reported by \cite{AL2020} from searches of archival \textit{XMM-Newton} data. They do not report a host galaxy for this source down to 10$\sigma$ limits of 18~mag for the \textit{ugriz}-bands and down to 5$\sigma$ limits of $\sim21$ and $\sim20$~mag for the \textit{J}- and \textit{K} bands, respectively. They use a fiducial redshift of 0.3 to determine the peak X-ray luminosity for the FXT of L$_{X, peak} = (6.4^{+2.7}_{-2.0})\times10^{44}$~erg~s$^{-1}$. Under the assumed redshift, they argue this event is consistent with the SBO model, although the absence of host galaxies to relatively deep limits would be unusual given the typical range of supernova host galaxy magnitudes at $z < 0.3$ \citep[e.g.,][for the host galaxies of Type IIn SNe]{Cold2023}.

\subsection*{XRT 070618}

\XTfour\ was also reported by \cite{AL2020}. There is no clear host galaxy within the uncertainty region of this event, down to signal-to-noise of 10 limits of 24.3, 24.1, 23.4 and 22.7~mag for the \textit{g-,r-,i-} and \textit{z}-bands, respectively and 3$\sigma$ limits of 21.44, 18.01, 17.79 and 17.15~mag in the \textit{Y-, J-, H-} and \textit{K}-bands, respectively. 
However, there are two galaxies in the vicinity (offsets of 11 and 21~arcsec) of the transient. They assume these galaxies are at the same redshift and perform SED fits to these two galaxies simultaneously, finding $z = 0.37$. Under the assumption that \XTfour\ lies at this same redshift, they calculate L$_{X, peak} = (49^{+17}_{-11})\times10^{44}$~erg~s$^{-1}$, which they state is consistent with an SBO nature. The assumptions that these two galaxies lie at the same redshift and that one of them is the host of \XTfour\ are highly uncertain. We consider the validity of these assumptions in section~\ref{sec:disc_xt4}.

\section{Very Large Telescope MUSE observations and analysis} \label{sec:anal}

We observed the four FXTs with MUSE on the VLT between 2 April 2022 and 20 September 2022, with a exposure time of 2804 seconds per FXT divided over four dithered exposures. The seeing was between 0.8 and 1.3~arcsec. Details of the observations can be found in Table~\ref{tab:sample}.

We obtained the reduced (ESO MUSE pipeline v2.8.6) and calibrated MUSE data from the ESO archive\footnote{\url{https://archive.eso.org/scienceportal/home}}. These data cubes are bias subtracted, flat-field corrected, wavelength calibrated, sky subtracted and flux calibrated. We first run the {\sc Zurich Atmosphere Purge} \citep[{\sc ZAP}v2.1][]{Soto2016} on the data cubes, which is a high precision subtraction tool to improve the already performed sky subtraction.  
{\sc ZAP} uses principle component analysis combined with filtering to construct a sky residual for each spaxel which is subtracted from the original data cube.

We refine the astrometry of the cubes by aligning the sources to identified sources in the \textit{Gaia} DR3 catalog, using a fit geometry appropriate to the number of Gaia sources contained within the field of view.
The root mean square (RMS) of the new word coordinate system (WCS) solution is 0.6, 0.07 and 0.05~arcsecs for \XTtwo, \XTthree and \XTfour, respectively. For \XTone\ we use the uncertainty on the position of the only \textit{Gaia} source in the field as the RMS on the WCS solution, which is 0.0002~arcsec. The use of a single astrometric standard star implies we cannot check or correct for any rotation of the field. The RMS on the WCS solution is added to the X-ray positional error obtained from the literature to obtain the full astrometric uncertainty as follows:
\begin{equation*}
    \sigma = \sqrt{\sigma_{WCS}^2+\sigma_{X-ray}^2}
\end{equation*}
where $\sigma$ is the complete positional uncertainty, $\sigma_{WCS}$ is the RMS of the new WCS solution and $\sigma_{X-ray}$ is the 1$\sigma$ positional uncertainty on the X-ray position as reported in the literature.

\subsection{Object detection}

We use the \texttt{get\_filtered\_image()} function within the {\sc python} package {\sc PyMUSE} \citep[v0.4.8;][]{Pessa2018} to create broadband images in the Johnson R-band. {\sc PyMUSE} uses the transmission curve of the Johnson R-filter (reduced slightly so the long, red tail still fits the MUSE spectral range) to convolve the cube, adds the convolved spectra per spaxel and multiplies the sum with the central wavelength of the filter transmission curve (to obtain the correct units of flux instead of flux density) to create the broadband image. Using R-band images allows to use the procedure described in \cite{Bloom2002} to calculate chance alignment probabilities later on in this work without the need to redetermine the mean surface density of galaxies brighter than a certain magnitude, as their work also uses R-band data. Additionally, due to the long red tail of the R-band filter, it covers a large fraction of the wavelength range (4650-9300~\AA) of the cubes.

We also create white light images by loading the cubes as {\sc PyMUSE} \texttt{MuseCube} objects. To create these white light images, the cubes are collapsed by summing the flux within one spaxel. These white light images are necessary for the extraction of the spectra. 

To detect sources, we use the {\sc python} implementation of \se\ \citep{Bertin1996}, called {\sc SEP} \citep{Barbary2016}. When we refer to \se\ in this paper, we mean this {\sc python} implementation. We trim 13 pixels from the edge of edge cube to enable cleaner source detection. 
% will not detect objects at the edges of the images that are edge effects.

Next, we make a background image for the R-band image using the default settings in \se, which we subtract from the R-band image. We subsequently run the object detection function of \se\ on the image, with a detection threshold of 2.0$\sigma$ times the global background root mean square (RMS). 57, 55, 67 and 35 objects are detected for \XTone, \XTtwo, \XTthree\ and \XTfour, respectively. 

\subsection{Object selection}\label{sec:selection}

\subsubsection{Flags}

We remove sources that have a \se\ flag >3. \se\ flags are a sum of values representing different potential issues for (aperture) photometry\footnote{For an explanation of all the flag values, see \url{https://sextractor.readthedocs.io/en/latest/Flagging.html}}. We only keep sources that are marked with 1 (aperture photometry is likely to be biased by a neighbour or by more than 10~per cent of bad pixels in the aperture) and/or 2 (the object is deblended). Flag >3 typically are sources that e.g., fall too close to the border and/or have too many saturated pixels.

\subsubsection{Shape parameters}

To remove the sources that were found in the regions where the four quadrants of the four MUSE detectors meet, we also remove sources with high elongation. Elongation is defined as the ratio between the semi-major axis (or "a" in \se) and the semi-minor axis (or "b" in \se).  We remove objects with a/b>3.

\subsection{Extraction of spectra}\label{sec:extraction}

For the spectral extraction we again use {\sc PyMUSE}. We use the elliptical parameters from {\sc SEP} to define the pixels that need to be extracted and combined for the final spectrum of an object. 
We are most interested in those sources that are galaxies, therefore we use 
the Kron radius \citep{Kron1980} to define the aperture. {\sc SEP} has a built in function to calculate the Kron radii of objects from their semi-minor and semi-major axis. We then use these radii to define the extraction region of the objects using elliptical apertures in which "a" and "b" of each object is scaled by N times the Kron radius of the object to obtain the elliptical aperture. 
We adopt N=1.0 to reduce overlap between the apertures of different sources to reduce the possibility of spectra being contaminated by light from neighbouring sources.
For combining pixels within the spectral extraction aperture we use the mode "White Weighted Mean". In this mode, the spectrum from an aperture is a weighted sum of the spaxels by a brightness profile obtained from the white light image. 

We perform a visual inspection of the extracted spectra to split the sample into stars, dwarf stars, spectra with clearly detected emission/absorption lines, and those without emission/absorption lines. Additionally, we extract a spectrum of the circular 1$\sigma$ positional uncertainty of the FXTs, to check for emission line-only sources at the positions of the FXTs.

\subsection{Redshift} \label{sec:redshift}

For the spectra where emission lines were detected through visual inspection, we fit a Gaussian function to each emission line using the {\sc Python} package {\sc lmfit}. In most of these spectra we can identify the [OII] doublet or the [OIII] $\lambda$5007\AA\ emission line as the brightest line. We then use the redshift of this brightest line to search for other lines present in the spectra. 
The resolution of the MUSE instrument (2.6\AA) is comparable to the difference in wavelength between the two lines in the [OII] doublet at rest-frame wavelength (2.8\AA). However, the restframe separation between the doublet lines is known, so we fit two Gaussians to the doublet with a fixed separation dependent on the redshift. We also force the full-width at half maximum (FWHM) to be the same for both lines in the doublet, but we leave the normalisation of the two lines free, as it is known to vary based on electron density \citep{Pradhan2006}. The H$\alpha$ emission line is only detected in eight spectra. Due to redshift, it may have moved outside the wavelength range covered by the MUSE instrument (4650-9300~\AA) in the other spectra.  

The mean value ($\mu$) of the Gaussian fit on any line is taken to be the central wavelength of the emission line. We then use the vacuum rest wavelengths of the emission lines\footnote{\url{https://classic.sdss.org/dr6/algorithms/linestable.php}} to calculate the redshift of the lines. We average the redshift obtained from different lines if multiple emission lines are detected. The error on the redshift is taken to be the standard deviation of the redshift distribution for all fitted lines. These errors are of the order $10^{-4}$ to $10^{-6}$, which is most likely an underestimation of the error on the redshift, as they for example do not include any uncertainty in the wavelength calibration which is of the order of 0.01\AA\ \citep{Weilbacher2020}. 

For the sources where no emission lines are detected, we attempt to fit the shape of the continuum of the galaxy spectrum using the penalised pixel fitting method ({\sc pPXF}; \cite{Cappellari2017}) used with the MILES stellar library \citep{Vazdekis2010}. {\sc pPXF} does, however, require an estimate of the redshift.
% to mask the regions where emission lines from the galactic nucleus may be. 
We look for the Calcium H+K absorption lines to estimate the redshift to use in {\sc pPXF}.
For eight sources we are able to obtain the redshift using both the emission lines and absorption lines because the Calcium H+K lines are detected as well as at least one emission line. We calculate the unweighted average of the redshifts obtained through both methods to obtain the best-fit redshift. For four sources we are able to obtain the redshift through using only the Calcium H+K lines and {\sc pPXF}, without the detection of emission lines. 

\subsection{$L_\textrm{X, peak}$ calculation}

We convert from redshift to luminosity distance using the {\sc Distance} package from {\sc astropy.cosmology} \citep{astropy:2013, astropy:2018, astropy:2022} and the cosmological parameters mentioned in the introduction. To obtain $L_\textrm{X, peak}$ for the FXTs, we assume the FXT lies at the same distance as the galaxy and use the luminosity distance and $F_\textrm{X, peak}$ given in Table~\ref{tab:sample} to calculate $L_\textrm{X, peak}$. We also calculate the offset between the FXT and the galaxy assuming the angular distance of the galaxy, calculated using the \texttt{angular\_diameter\_distance} function from {\sc astropy.cosmology}. 

\subsection{Photometry}\label{sec:phot_Pch}

We perform aperture photometry on the R-band images using elliptical apertures. The parameters of the apertures are the same as used for the spectral extraction, as we assume all objects are galaxies and therefore we obtain the Kron magnitudes. 
We set \texttt{sub\_pix = 1}, so the pixels are not divided into sub pixels in the photometry, only whole pixels are used. 

Since in MUSE cubes a unit (or count) corresponds to 10$^{-20}$ erg s$^{-1}$ cm$^{-2}$ \AA$^{-1}$, we can convert the flux from the aperture photometry (which is the pixel values within the aperture added) to F$_\lambda$ and convert this to AB magnitude. We then apply a $-0.055$~mag correction to obtain the Johnson R band magnitude, based on the star Vega \citep{Frei1994}. This is necessary for the chance alignment probability calculation.

We calculate the chance alignment probability of the objects and the FXT position following the prescription from \cite{Bloom2002}:
\begin{equation}
    P_{i,ch} = 1-e^{-\pi (r_{i}^2 \sigma(\le m_i)}.
\end{equation}
in which 
\begin{multline*}
      \sigma(\le m_i) = \frac{1}{(3600^2\times0.334\mathrm{log}_e 10)} \\
    \times 10^{0.334(m_i-22.963)+4.320} \textrm{galaxy arcsec}^{-2}  
\end{multline*}
is defined as the mean surface density of galaxies brighter than magnitude $m_i$ in the R-band \citep{Hogg1997}. 
Here, $P_{i, ch}$ is the chance alignment probability, $r_{i}$ is the distance in arcsec between the centre of the FXT position and the x and y coordinates of the object as determined by \se\ and $m_i$ is the Kron magnitude of this same object. 

%--------------------------------------------------------------------

\section{Results}

\begin{figure*}
\centering
\includegraphics[width=\textwidth]{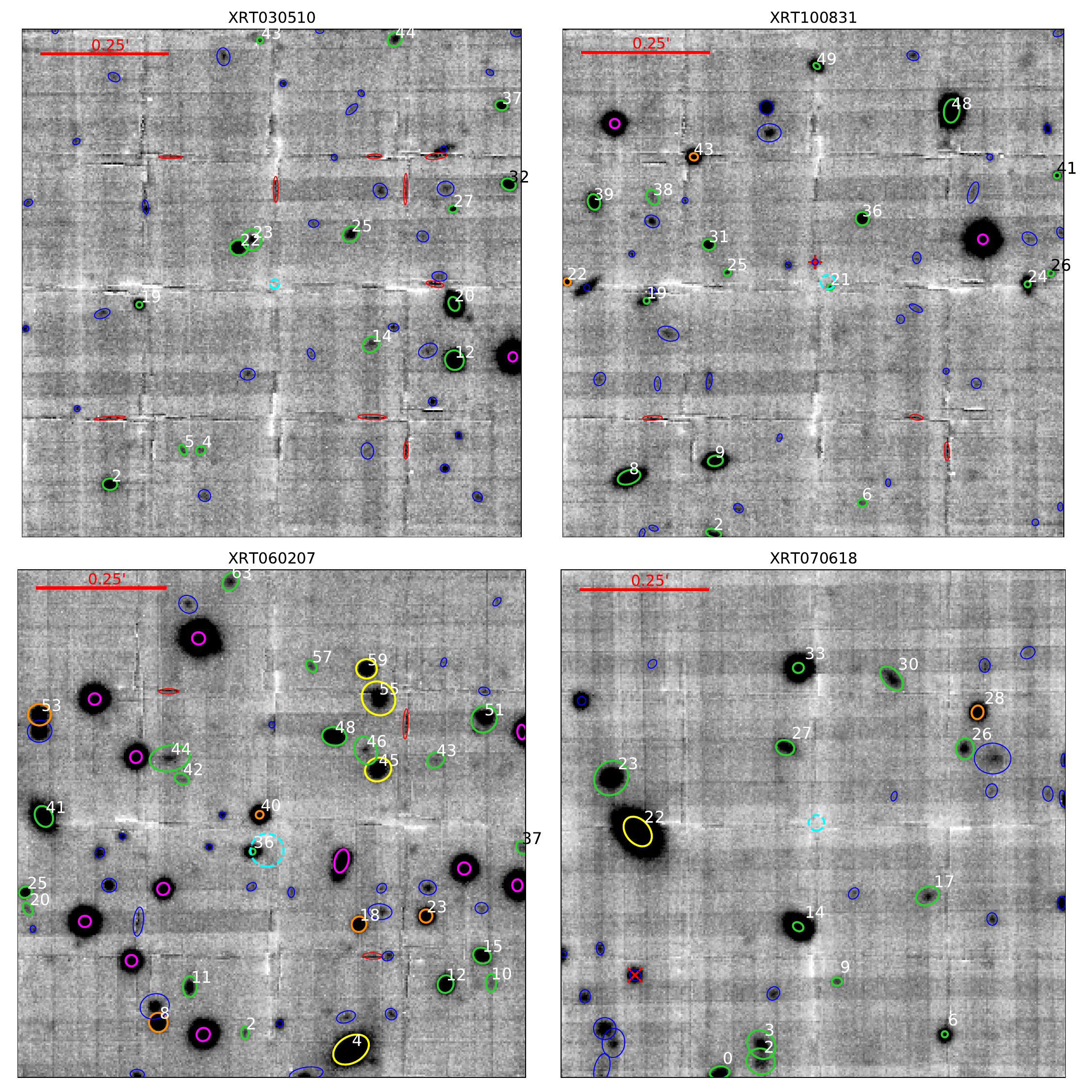}
      \caption{High contrast, grey-scale 1x1~arcmin R-band images obtained from the MUSE data cubes. The position of each FXT is indicated by a cyan, dashed circle near the centre of the image. Its size represents the 1$\sigma$ uncertainty on the X-ray position of the FXT. Every ellipse indicates the position of a source detected by {\sc SEP}, with different colours for different types of sources. Green ellipses are sources for those where we detected emission lines in the spectra to which we fit Gaussian functions to determine the redshift. Yellow ellipses indicate sources for which we used the Ca H+K absorption lines to estimate a redshift and used {\sc pPXF} to obtain a more accurate redshift measurement. Orange ellipses indicate sources identified as dwarf stars and pink ellipses as normal stars. The red ellipses were filtered out by our selection criteria based on flags and elongation. For blue ellipses we did extract a spectrum, but the signal-to-noise ratio was too low to detect emission lines or use {\sc pPXF} to obtain the redshift. We note that due to the high contrast of the image needed to show all faint sources, not all sources that appear bright have a high S/N spectrum. We have plotted the spectrum of the source with an R-band magnitude of 22.76$\pm$0.06, marked with a red `x' in the image of \XTfour\ in Figure~\ref{fig:single_spec}, to give an example of a bright source for which we are not able to determine the redshift. The sizes of the ellipses show the extraction region used to obtain the spectrum for each source, see Section~\ref{sec:extraction} for how the size and shape were determined. The sources for which we were able to obtain a redshift are labelled with numbers, these correspond to the numbers listed in Table~\ref{tab:results}. Orange ellipses are labelled with numbers corresponding to the source numbers listed in Table~\ref{tab:dwarfs}. North is up and East is left in these images.
              }
    \label{fig:images_sources}
\end{figure*}

\begin{figure*}
\centering
\includegraphics[width=\textwidth]{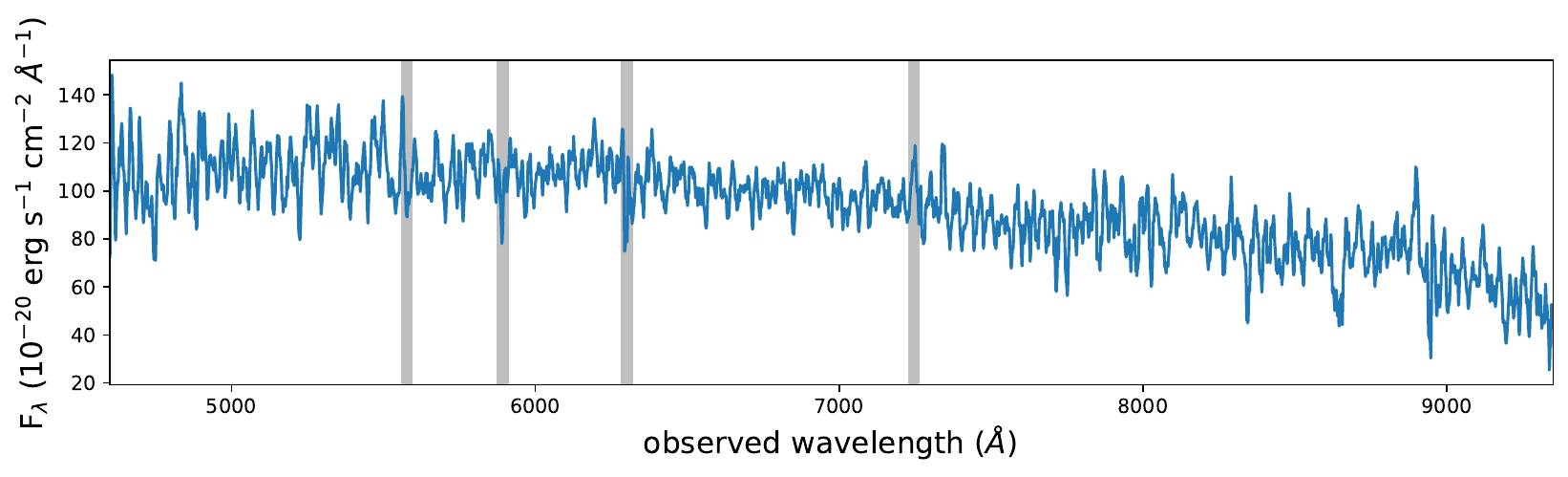}
      \caption{The spectrum of the source with an R-band magnitude of 22.76$\pm$0.06, marked with a red `x' in the image of \XTfour\ in Figure~\ref{fig:images_sources}, smoothed using \texttt{Box1DKernel} with a kernel width of 10 pixels, to give an example of a bright source for which we are not able to determine the redshift. 
              }
    \label{fig:single_spec}
\end{figure*}

\longtab{%
\onecolumn
\begin{landscape}
\begin{longtable}{lcccccccccc}
    \caption{The sources per FXT for which we extracted spectra from the MUSE cubes, according to the selection criteria explained in Section~\ref{sec:selection}. The source numbers are indicated in Figure~\ref{fig:images_sources}. RA and Dec are the source's centroid coordinates in the R-band images and the apparent magnitude is the Kron magnitude in the R-band images, uncorrected for Galactic extinction. We also indicate what emission/absorption lines were identified in the spectra to obtain the redshift. The redshift is presented with the estimated uncertainty in the last digit between brackets. The distance is the luminosity distance calculated using the {\sc Distance} module within {\sc astropy.cosmology} using the redshift listed in the table. The offset is calculated between the centroid positions of the FXT and the candidate host galaxy, using only the error on the FXT position to calculate the error on the offset. L$_{X, peak}$ is reported in the 0.3--10~keV energy range. The absolute magnitude (uncorrected for Galactic extinction) is calculated using the distance listed in the table and $P_{ch}$ is calculated as described in Section~\ref{sec:phot_Pch}.
    }\\
    \label{tab:results}\\
    \hline
    Source & RA & Dec & m$_\mathrm{R, AB}$ & Lines detected & $z$ & D$_L$ & offset & L$_{X, peak}$ & M$_\mathrm{R, AB}$ & $P_{ch}$ \\
     & (degrees) & (degrees) & & & & (Gpc) & (kpc) & (10$^{44}$ erg s$^{-1}$) &  & \\
    \hline
    \endfirsthead
    \caption{continued.}\\
    \hline
    Source & RA & Dec & m$_\mathrm{R, AB}$ & Lines detected & $z$ & D$_L$ & offset & L$_{X, peak}$ & M$_\mathrm{R, AB}$ & $P_{ch}$ \\
     & (degrees) & (degrees) & & & & (Gpc) & (kpc) & (10$^{44}$ erg s$^{-1}$) &  & \\
    \hline
    \endhead
    \hline
    \endfoot
    \hline 
    \multicolumn{11}{c}{\textbf{XRT~030510}}\\ 
    \hline 
    2 & 76.78438 & -31.8761 & 22.54$\pm$0.05&[OII], H$\beta$, [OIII] & 0.65510(1) & 4.0 & 216$\pm$4 & 45 & -20.49$\pm$0.05 & 0.9846\\
    4 & 76.78093 & -31.875 & 24.91$\pm$0.15 &[OII], [OIII] & 0.628611(6) & 3.8 & 149$\pm$4 & 40 & -18.01$\pm$0.15 & 1.0\\
    5 & 76.7816 & -31.87499 & 25.11$\pm$0.17&[OII], H$\beta$, [OIII] & 0.73252(6) & 4.6 & 165$\pm$4 & 59 & -18.22$\pm$0.17 & 1.0\\
    12 & 76.77125 & -31.87208 & 21.7$\pm$0.03 &[OII], H$\beta$, [OIII]$\lambda$5007\AA, Ca H+K & 0.48551(3) & 2.8 & 141$\pm$3 & 21 & -20.54$\pm$0.03 & 0.7111\\
    14 & 76.77442 & -31.87156 & 23.8$\pm$0.09 &[OIII] & 0.54879(2) & 3.3 & 87$\pm$3 & 29 & -18.77$\pm$0.09 & 0.8783\\
    19 & 76.78326 & -31.87029 & 24.63$\pm$0.13 &[OII], [OIII], H$\alpha$ & 0.30944(3) & 1.7 & 75$\pm$2 & 7 & -16.47$\pm$0.13 & 0.9969\\
    20 & 76.77127 & -31.87026 & 21.1$\pm$0.03&[OII], H$\beta$, [OIII] & 0.62900(4) & 3.8 & 148$\pm$4 & 40 & -21.83$\pm$0.03 & 0.4858\\
    22 & 76.77946 & -31.86843 & 22.52$\pm$0.05&[OII], H$\beta$, [OIII] & 0.46611(3) & 2.7 & 36$\pm$3 & 19 & -19.62$\pm$0.05 & 0.1466\\
    23 &  76.77897 & -31.86818 & 22.580.05 & [OII], H$\beta$, [OIII] & 0.4370(1) & 2.5 & 343 & 17 & -19.40.05 & 0.1461\\
    25 & 76.7752 & -31.86799 & 23.36$\pm$0.07 &[OII] & 1.357 & 9.9 & 92$\pm$5 & 269 & -21.62$\pm$0.07 & 0.6186\\
    27 & 76.7713 & -31.86718 & 24.82$\pm$0.15 &H$\beta$, [OIII] & 0.68966(3) & 4.3 & 165$\pm$4 & 51 & -18.35$\pm$0.15 & 1.0\\
    32 & 76.76918 & -31.86638 & 23.28$\pm$0.07 &[OII] & 0.997 & 6.8 & 244$\pm$4 & 125 & -20.87$\pm$0.07 & 0.9992\\
    37 & 76.76946 & -31.86383 & 23.50.08 & [OII], [OIII] & 0.30729(9) & 1.6 & 1572 & 7 & -17.580.08 & 1.0\\
    43 & 76.77866 & -31.86172 & 26.290.29 & [OII], H$\beta$, [OIII] & 0.5047(1) & 2.9 & 1803 & 23 & -16.060.29 & 1.0\\
    44 & 76.77353 & -31.86169 & 23.86$\pm$0.09 &[OII] & 1.1 & 7.6 & 267$\pm$4 & 160 & -20.55$\pm$0.09 & 1.0\\
    \hline 
    \multicolumn{11}{c}{\textbf{XRT~100831}}\\ 
    \hline 
    2 & 90.0105 & -52.72306 & 23.79$\pm$0.09&[OII], H$\beta$, [OIII] & 0.56071(6) & 3.3 & 214$\pm$6 & 11 & -18.83$\pm$0.09 & 1.0\\
    6 & 90.00257 & -52.72208 & 25.26$\pm$0.18&[OII], H$\beta$, [OIII] & 0.72433(3) & 4.6 & 193$\pm$6 & 22 & -18.04$\pm$0.18 & 1.0\\
    8 & 90.015 & -52.72125 & 21.17$\pm$0.03&[OII], H$\beta$, [OIII], Ca H+K & 0.55973(9) & 3.3 & 215$\pm$6 & 11 & -21.45$\pm$0.03 & 0.81\\
    9 & 90.0104 & -52.72072 & 21.7$\pm$0.03&[OII], H$\beta$, [OIII], H$\alpha$, Ca H+K & 0.27709(6) & 1.5 & 107$\pm$4 & 2 & -19.13$\pm$0.03 & 0.7624\\
    19 & 90.01407 & -52.71556 & 25.90.24 & [OII] & 1.100 & 7.6 & 1787 & 62 & -18.520.24 & 1.0\\
    21$^\dagger$ & 90.00428 & -52.71513 & 26.470.31 & [OII] & 1.22 & 8.7 & 67 & 80 & -18.220.31 & 0.0435\\
    24 & 89.99377 & -52.71502 & 24.27$\pm$0.11 &[OII] & 1.011 & 6.9 & 191$\pm$7 & 50 & -19.92$\pm$0.11 & 0.9999\\
    25 & 90.00977 & -52.71465 & 24.98$\pm$0.16 &[OII], H$\beta$, [OIII]$\lambda$5007\AA & 0.61107(2) & 3.7 & 82$\pm$6 & 14 & -17.87$\pm$0.16 & 0.9834\\
    26 & 89.99254 & -52.71467 & 26.24$\pm$0.28 &[OII] & 1.009 & 6.9 & 214$\pm$7 & 50 & -17.94$\pm$0.28 & 1.0\\
    31 & 90.01075 & -52.71374 & 23.61$\pm$0.08 &H$\beta$, [OIII], H$\alpha$ & 0.2265(8) & 1.2 & 54$\pm$3 & 1 & -16.71$\pm$0.08 & 0.8881\\
    36 & 90.00258 & -52.71291 & 23.23$\pm$0.07&[OII], H$\beta$, [OIII] & 0.6031(2) & 3.7 & 58$\pm$6 & 14 & -19.58$\pm$0.07 & 0.4209\\
    38 & 90.01374 & -52.71222 & 24.41$\pm$0.12 &[OII] & 0.610 & 3.7 & 157$\pm$6 & 14 & -18.44$\pm$0.12 & 0.9999\\
    39 & 90.01687 & -52.71237 & 22.61$\pm$0.05 &H$\beta$, [OIII]$\lambda$5007\AA, H$\alpha$ & 0.14225(2) & 0.7 & 74$\pm$2 & 0 & -16.6$\pm$0.05 & 0.981\\
    41 & 89.9922 & -52.71152 & 25.36$\pm$0.19 &[OII] & 1.385 & 10.2 & 254$\pm$7 & 109 & -19.68$\pm$0.19 & 1.0\\
    48 & 89.99782 & -52.70943 & 20.63$\pm$0.02&[OII], H$\beta$, [OIII], H$\alpha$ & 0.323714(5) & 1.7 & 119$\pm$4 & 3 & -20.58$\pm$0.02 & 0.4692\\
    49 & 90.00501 & -52.70799 & 23.31$\pm$0.07&[OII], H$\beta$, [OIII], H$\alpha$ & 0.32371(4) & 1.7 & 121$\pm$4 & 3 & -17.9$\pm$0.07 & 0.9946\\
    \hline 
    \multicolumn{11}{c}{\textbf{XRT~060207}}\\ 
    \hline 
    2 & 196.83312 & -40.46675 & 25.05$\pm$0.16 &[OII] & 0.795 & 5.1 & 163$\pm$15 & 82 & -18.5$\pm$0.16 & 1.0\\
    4 &196.82867 &-40.46728 & 20.23$\pm$0.02& Ca H+K  & 0.58811(4) & 3.5 & 170$\pm$13 & 39 & -22.52$\pm$0.02 & 0.3802\\
    10 & 196.82279 & -40.46511 & 24.29$\pm$0.11 &[OII] & 1.429 & 10.5 & 259$\pm$16 & 349 & -20.82$\pm$0.11 & 1.0\\
    11 & 196.83542 & -40.46528 & 23.24$\pm$0.07&[OII], H$\beta$, [OIII] & 0.44222(4) & 2.5 & 106$\pm$11 & 19 & -18.76$\pm$0.07 & 0.9212\\
    12 & 196.82467 & -40.46517 & 22.00$\pm$0.04&[OII], H$\beta$, [OIII], Ca H+K & 0.53148(4) & 3.1 & 167$\pm$12 & 31 & -20.48$\pm$0.04 & 0.863\\
    15 & 196.82317 & -40.46425 & 22.82$\pm$0.06&[OII], H$\beta$, [OIII] & 0.72830(7) & 4.6 & 205$\pm$14 & 66 & -20.49$\pm$0.06 & 0.9864\\
    20 & 196.84221 & -40.46281 & 24.21$\pm$0.11 &[OII] & 0.884 & 5.8 & 226$\pm$15 & 106 & -19.62$\pm$0.11 & 1.0\\
    25 & 196.84233 & -40.46228 & 23.76$\pm$0.09&[OII], H$\beta$, [OIII] & 0.59278(2) & 3.6 & 193$\pm$13 & 40 & -19.01$\pm$0.09 & 0.9999\\
    36 & 196.83279 & -40.46094 & 24.51$\pm$0.13 &[OII] & 0.939 & 6.3 & 14$\pm$15 & 123 & -19.48$\pm$0.13 & 0.0575\\
    37 & 196.8215 & -40.46081 & 24.66$\pm$0.14 &[OII] & 1.153 & 8.1 & 248$\pm$16 & 205 & -19.88$\pm$0.14 & 1.0\\
    41 & 196.84154 & -40.45983 & 21.1$\pm$0.03 &[OII], Ca H+K & 0.591164(1) & 3.6 & 177$\pm$13 & 40 & -21.67$\pm$0.03 & 0.6364\\
    42 & 196.83575 & -40.45864 & 24.31$\pm$0.12 &[OII] & 1.157 & 8.1 & 108$\pm$16 & 207 & -20.24$\pm$0.12 & 0.9454\\
    43 & 196.82512 & -40.458 & 23.83$\pm$0.09 &[OII] & 1.385330(1) & 10.2 & 190$\pm$16 & 323 & -21.2$\pm$0.09 & 0.9974\\
    44 & 196.83625 & -40.45797 & 20.24$\pm$0.02 &[OII] & 0.986 & 6.7 & 126$\pm$16 & 139 & -23.88$\pm$0.02 & 0.1682\\
    45 &196.8275 &-40.45833 &22.0$\pm$0.04 & Ca H+K  & 0.58761(7) & 3.5 & 108$\pm$13 & 39 & -20.75$\pm$0.04 & 0.5268\\
    46 & 196.82804 & -40.45772 & 21.72$\pm$0.04 &[OII] & 0.767 & 4.9 & 123$\pm$14 & 75 & -21.73$\pm$0.04 & 0.4674\\
    48 & 196.82937 & -40.45728 & 21.92$\pm$0.04&[OII], H$\beta$, [OIII] & 0.62575(8) & 3.8 & 107$\pm$13 & 45 & -21.0$\pm$0.04 & 0.4786\\
    51 & 196.82308 & -40.45669 & 22.03$\pm$0.04&[OII], H$\beta$, [OIII] & 0.58731(4) & 3.5 & 199$\pm$13 & 39 & -20.72$\pm$0.04 & 0.9268\\
    55 &196.8275 &-40.45606 &21.38$\pm$0.03 & Ca H+K  & 0.58839(6) & 3.5 & 148$\pm$13 & 39 & -21.36$\pm$0.03 & 0.5848\\
    57 & 196.83033 & -40.45503 & 24.68$\pm$0.14 &H$\beta$, [OIII]$\lambda$5007\AA & 0.624(2) & 3.8 & 152$\pm$13 & 45 & -18.23$\pm$0.14 & 1.0\\
    59 &196.828 & -40.45511 &22.25$\pm$0.04 & Ca H+K  & 0.73265(8) & 4.6 & 178$\pm$14 & 67 & -21.08$\pm$0.04 & 0.873\\
    63 & 196.83371 & -40.45233 & 23.72$\pm$0.09 &[OII] & 1.430 & 10.6 & 270$\pm$16 & 350 & -21.4$\pm$0.09 & 1.0\\
    \hline 
    \multicolumn{11}{c}{\textbf{XRT~070618}}\\ 
    \hline 
    0 & 24.27846 & -12.96053 & 22.96$\pm$0.06&[OII], H$\beta$, [OIII] & 0.431 & 2.4 & 179$\pm$5 & 75 & -18.99$\pm$0.06 & 0.9975\\
    2 & 24.27712 & -12.96017 & 22.34$\pm$0.05 &[OII] & 1.291 & 9.3 & 243$\pm$8 & 1087 & -22.51$\pm$0.05 & 0.9554\\
    3 & 24.27712 & -12.95964 & 22.32$\pm$0.05&[OII], H$\beta$, [OIII], H$\alpha$ & 0.36744(2) & 2.0 & 138$\pm$5 & 51 & -19.22$\pm$0.05 & 0.9304\\
    6 & 24.27108 & -12.95933 & 24.87$\pm$0.15&[OII], H$\beta$, [OIII], H$\alpha$ & 0.36758(1) & 2.0 & 150$\pm$5 & 51 & -16.67$\pm$0.15 & 1.0\\
    9 & 24.27462 & -12.95764 & 24.73$\pm$0.14 &[OII] & 1.223 & 8.7 & 158$\pm$8 & 951 & -19.97$\pm$0.14 & 0.9998\\
    14 & 24.27587 & -12.95586 & 21.16$\pm$0.03&[OII], H$\beta$, [OIII], Ca H+K & 0.43200(1) & 2.5 & 71$\pm$5 & 75 & -20.78$\pm$0.03 & 0.209\\
    17 & 24.27162 & -12.95489 & 23.25$\pm$0.07&[OII], H$\beta$, [OIII] & 0.44380(3) & 2.5 & 90$\pm$5 & 80 & -18.76$\pm$0.07 & 0.8431\\
    22 &24.28117 &-12.95281 & 18.86$\pm$0.01 & Ca H+K  & 0.2080(1) & 1.1 & 72$\pm$3 & 14 & -21.26$\pm$0.01 & 0.108\\
    23 & 24.282 & -12.95108 & 21.01$\pm$0.03 &[OII], Ca H+K & 0.660367(1) & 4.1 & 174$\pm$6 & 209 & -22.04$\pm$0.03 & 0.5629\\
    26 & 24.27038 & -12.95019 & 23.19$\pm$0.07 &[OII] & 0.921 & 6.1 & 154$\pm$7 & 472 & -20.75$\pm$0.07 & 0.9369\\
    27 & 24.27629 & -12.95014 & 22.86$\pm$0.06 &[OII], [OIII] & 0.60702(3) & 3.7 & 65$\pm$6 & 170 & -19.97$\pm$0.06 & 0.4039\\
    30 & 24.27279 & -12.94792 & 22.99$\pm$0.06 &[OII], [OIII]$\lambda$5007\AA, H$\alpha$ & 0.36762(2) & 2.0 & 99$\pm$5 & 51 & -18.54$\pm$0.06 & 0.898\\
    33 & 24.27587 & -12.94758 & 21.29$\pm$0.03&[OII], H$\beta$, [OIII], Ca H+K & 0.48739(2) & 2.8 & 112$\pm$6 & 100 & -20.97$\pm$0.03 & 0.4302\\
    % \hline 
\end{longtable}
$^\dagger$See Section~\ref{sec:disc_xt2} for how this redshift was obtained.
\end{landscape}
}

Figure~\ref{fig:images_sources} shows the R-band images of the four MUSE cubes of the FXTs with the 1$\sigma$ positions of the FTXs marked with a dashed cyan circle. In these images the objects detected by \se\ are indicated by ellipses, with their colour meaning the following. Red: this source was removed because of flags or elongation. Blue: we were unable to obtain a redshift for the source.  Green: we obtained a redshift through fitting a Gaussian to the emission line(s). Yellow: we obtained a redshift using the calcium H+K lines and {\sc pPXF}. Pink or orange: the source is a star or dwarf star, respectively. The green and yellow sources are labeled with a number, as they are the sources for which we have calculated the distance and $L_\textrm{X, peak}$. In the images of \XTtwo, \XTthree~ and \XTfour, we also indicate the dwarf stars we identify with a number. We note that not all bright sources in Figure~\ref{fig:images_sources} have spectra with  emission lines, while some faint sources do, allowing for the redshift to be measured for the faint source but sometimes not for the brighter ones. As an example we show in Figure~\ref{fig:single_spec} the spectrum of the bright source (R-band mag 22.76$\pm$0.06, marked with a red `x' in the bottom right panel of Figure~\ref{fig:images_sources}). The reproduction package of this paper includes code to show all spectra so the reader can check the absence/presence of emission lines in each spectrum. 
% We have also added an example of a spectrum of a bright source without emission lines to the appendix in Figure~\ref{apfig:nolines}.}

The labeled sources are listed in Table~\ref{tab:results}, including the derived redshift (and the line identifications), absolute and apparent magnitudes, the offset between the FXT and the candidate host galaxy, the corresponding (0.3--10~keV) $L_\textrm{X, peak}$ and the chance alignment probability. For sources where only one emission line is detected we opted to identify it as originating from the [OII] doublet, and accordingly fit this line with two Gaussian functions to account for it being a doublet. An alternative often used assumption is that a single line is due to H$\alpha$, however, for the spectra where this one line is at wavelengths shorter than the restframe wavelength of H$\alpha$, this is not a valid assumption. For the remaining cases we base this assumption of this one line being [O~II] on the absence of other emission lines. For example, most of these single lines have a high signal-to-noise, which means, assuming a typical Balmer decrement of $\sim3$, we would, for example, expect to be able to detect the H$\beta$ emission line if this detected line was H$\alpha$. %This is just one example of why we assume it is not H$\alpha$ as we cannot list all possibilities.
For the sources with a larger redshift, we see the two lines in the doublet as two clearly separate lines, lending support to the doublet identification. 
We have plotted the fits to the [O~II] doublets if it was the only line used to determine the redshift in Figure~\ref{apfig:OII_lines}. However, due to the potential of a large systematic error because of mis-identification in these cases, we report these redshift determinations without an error bar. We plot the spectra of the sources for which we determined the redshift using emission lines in Figures~\ref{apfig:lines_xt1}, \ref{apfig:lines_xt2}, \ref{apfig:lines_xt3} and \ref{apfig:lines_xt4}, where we indicate the position of the emission lines we used with vertical, dashed, blue lines. 
In Figure~\ref{apfig:ppxf_galaxies} we show the best fitting galaxy found by {\sc pPXF} in red, plotted on top of the extracted spectrum in black if the redshift was determined using the calcium H+K lines and {\sc pPXF}. There are eight sources for which the redshift was determined using both methods. 

\begin{table}
\caption{The sources per FXT for which we fitted K- and M-dwarf spectra to the extracted MUSE spectra. The sources are indicated in Figure~\ref{fig:images_sources} by orange circles and their corresponding number. We list RA and Dec for each source as well as the angular offset in arcsec. The best fitting K- or M-dwarf spectral type is listed as well as the corresponding reduced $\chi^2$ of the fit. The number of degrees of freedom was 3596 for all sources, except sources 22 an 42 of \XTtwo, for which there were 3595 degrees of freedom.}
\label{tab:dwarfs}
\centering
\hspace*{-.5cm}
\begin{tabular}{lccccc}
\hline
Source & RA & Dec & offset & Type & $\chi^2_\nu$\\
Source & (degrees) & (degrees) & (arcsec) & & \\
\hline 
\multicolumn{6}{c}{\textbf{XRT~100831}}\\ 
\hline
22 & 90.01830 & -52.71495 & 30.3 & M4V & 1.4 \\
43 & 90.01155 & -52.71092 & 21.3 & M3V & 2.5 \\
\hline 
\multicolumn{6}{c}{\textbf{XRT~060207}}\\ 
\hline
8  & 196.83675 & -40.46642 & 23.4 & M3V & 1.7  \\
18 & 196.82833 & -40.46325 & 13.6 & M5V & 1.9  \\
23 & 196.82550 & -40.46300 & 19.8 & M1V & 1.7  \\
40 & 196.83250 & -40.45978 & 4.2 & M1V & 2.4  \\
53 & 196.84175 & -40.45661 & 30.5 & M5V & 2.2 \\
\hline
\multicolumn{6}{c}{\textbf{XRT~070618}}\\ 
\hline 
28 & 24.26996 & -12.94903 & 22.6 & M5V & 2.0 \\
\hline 
\end{tabular}
\end{table}

For the dwarf stars (orange sources), we use the spectra of M- and K-dwarfs from the Pickles Stellar Spectral Flux Library \cite{Pickles1998} in combination with {\sc pPXF} to find the best fitting stellar type. The best fitting dwarf star types are listed in Table~\ref{tab:dwarfs}, with the reduced $\chi^2$ for the best fit also included. The best fits to the spectra are shown in Fig~\ref{apfig:ppxf_dwarfs} in the appendix.
For the remaining sources that were not categorised as stars, we are unable to obtain a redshift through either emission line fitting or using {\sc pPXF } because the signal to noise is too low. For completeness, the extracted spectra of the 1$\sigma$ uncertainty regions are plotted in Figure~\ref{apfig:spectra_fxts} in the appendix. 

%--------------------------------------------------------------------

\section{Discussion}

If we assume that one of the galaxies for which we were able to determine a redshift is the host galaxy of the corresponding FXT L$_{X, peak} \approxgt 10^{44}$~erg~s$^{-1}$ for all cases (Table~\ref{tab:results}).
Comparing this to the different luminosities expected for different progenitor models, we derive that L$_{X, peak}$ is too high to be produced by the SBO progenitor model for each FXT. 
Additionally, from Figure~\ref{fig:images_sources}, we find that there are no clear host galaxies within the 3$\sigma$ X-ray uncertainty position detected in the MUSE data for \XTone\ and \XTfour. For \XTtwo, we detect a faint source (m$_\mathrm{R, AB}$ = 26.5$\pm$0.3) within the 1$\sigma$ uncertainty region, which we will discuss below.
For \XTthree\ there is a galaxy within the $1\sigma$ uncertainty region, which we will also discuss below. Additionally, within the 3$\sigma$ positional uncertainty region of this FXT, there are a dwarf star and an object without redshift determination. 

For \XTone\ and \XTfour\ we do not detect emission lines in the spectra extracted from the 1$\sigma$ uncertainty positions (see Figure~\ref{apfig:spectra_fxts}). For \XTtwo\ there is suggestive evidence for the detection of an emission line near 8280\AA. We will discuss this in more detail below. For \XTthree\ we detect an emission line that we associate with the source found within the $1\sigma$ uncertainty region.

For an SBO nature L$_{X, peak} \lesssim 10^{44}$~erg~s$^{-1}$, so the maximum redshift for these FXTs to be consistent with the SBO model would be $z\sim$0.12, $\sim$0.19, $\sim$0.12 and $\sim$0.06, for \XTone, \XTtwo, \XTthree\ and \XTfour, respectively. For an SBO a young massive star is needed, therefore we check if a star forming (SF) galaxy, and even an SF dwarf galaxy can be detected within these redshift upper limits. The typical absolute magnitude of a starforming (SF) dwarf galaxy is $\approxgt15$ in the B-band \citep[e.g.,][]{Annibali2022}. This translates to apparent magnitudes between <22.3 and <24.9 for our sample, which we would have (just) been able to detect in the bluer part of the cubes. Unless the SF host galaxy of any of the FXTs is uncharacteristically faint for a SF galaxy, it is unlikely that any of our FXTs stem from SF host galaxies. 

If we compare the offsets within our sample of >33~kpc (1$\sigma$ limit) to the offset distribution of other types of transients, we see that the offsets we find effectively rule out long gamma-ray bursts (GRBs) \citep{Lyman2017} and super luminous SNe \citep{Schulze2021} natures of the events. The offset distribution of CCSNe \citep{KellyKirshner2012, Schulze2021} also are inconsistent with such a large offset. 
The types of transients for which a sizeable fraction occurs at >33 kpc are Type Ia SNe \citep{Uddin2020} and short GRBs \citep{Fong2022}, which would be consistent with an IMBH-WD TDE (which could lead to thermonuclear explosions similar to Type Ia SNe) and BNS merger nature of FXTs. In the distribution function of offsets for short GRBs there is a fraction of objects that have even larger offsets (up until an offset of roughly 100~kpc for the 1\% fraction). The 1\% limit in the offset distribution function of Type Ia's is roughly at 20 to 30~kpc. 
We stress that it is uncertain if the FXTs under study are actually related to any of the candidate host galaxies.

It is uncertain if Globular clusters (GCs) host IMBHs \citep[][]{Bahcall1975}, but they are dense enough to allow for a significant TDE rate \citep[e.g.,][]{Fragione2018}. GCs have an absolute magnitude distribution that peaks between $M_{\rm V}=-7$ and $M_{\rm V}=-8$, depending among other things on the galaxy type (see \cite{Rejkuba2012} for a review), which means that for our derived redshifts we would not be able to detect GCs around those galaxies. The distribution of galactocentric offsets between GCs and their host galaxies is also not universal. If we take the distributions for the five galaxies discussed in \cite{GCdist2022}, we see there are rarely any GCs reported with an offset of >15~kpc. Therefore the offsets we report of up to 270~kpc are much larger than expected for GCs (assuming the distributions of \cite{GCdist2022} are representative for all galaxies). This difference between GC offsets and our offsets could be an argument against a IMBH-WD TDE nature for the FXTs.

\subsubsection{\XTone} \label{sec:disc_xt1}

There are no reliable host galaxy candidates for \XTone\ based on the offsets and the chance alignment probabilities. We therefore do not discuss sources listed in Table~\ref{tab:results} in detail for this FXT. If we assume the faintest source in the R-band image with a significant detection (error <0.3 mag) is the limiting magnitude of the image, we would obtain a limiting magnitude of 26.3. For a typical galaxy of absolute magnitude $\sim-20$ \citep[e.g.,][]{Chilingarian2012} this implies a redshift of $\approxgt2.2$.

\subsection{\XTtwo} \label{sec:disc_xt2}

For \XTtwo\ we consider three candidate host spectra in detail. First, \citet{QV2022} reported a possible host at 3.2$\sigma$ of the centre of the uncertainty on the position of \XTtwo. Here, this source is also detected. It lies close to the North-East of the FXT position (marked with a red '+'-sign in Figure~\ref{fig:images_sources}) at 3.3$\sigma$. However, we are not able to determine a redshift of this object.

Second, we consider the emission line detected in the spectrum extracted at the location of the FXT and we fit two Gaussian functions to this feature assuming it is due to the [O II] doublet.
% Second, we consider the emission line detected in the spectrum extracted from the [OII] doublet and we fit two Gaussian functions to this feature using {\sc lmfit} as described in Section~\ref{sec:redshift}.
Using this identification, we obtain a redshift $z\sim1.22$. For this redshift, L$_{X, peak} \approx 8\times10^{45}$~erg~s$^{-1}$, which is too high for an SBO nature. 
Third, we checked in the spectrum of source 21 that falls within the 1$\sigma$ positional uncertainty of \XTtwo\ reported in this work. During our first visual inspection, we discarded this spectrum as we were not certain the detected line was real (see Figure~\ref{apfig:OII_lines}). However, upon further inspection due to the detection of a line in the spectrum of the position of \XTtwo\ we were able to fit lines to the spectrum of source 21, because we had a guess for the redshift from the line of the FXT region spectrum. We find the same redshift of $z\sim1.22$ for source 21 and therefore we 
% with the same profile at the same redshift of $z\sim1.22$, we decided to consider this a real line.} 
% and are able to fit two Gaussian functions at the same redshift of $z\sim1.22$. We therefore 
identify the emission line detected in the spectrum extracted from the 1$\sigma$ uncertainty to have come from this source 21. The projected offset for this source is 6$\pm$7~kpc, which is consistent with all three possible progenitor models. 
This source has a chance alignment probability of 0.04 with the FXT position, so we consider it likely this is the host galaxy of \XTtwo.

% \textbf{We are able to fit lines to the spectrum of this source, marked with a red X in Figure~\ref{fig:images_sources}, because we had a guess for the redshift from the line of the FXT region spectrum. }

Sources 48 and 49 of \XTtwo\ are both at a redshift of $z\approx0.324$. The average projected offset to the position of \XTtwo\ is $120\pm4$~kpc for this duo of galaxies and there are no other sources with a similar redshift found within the sample of sources with a derived redshift. There is another duo of sources, 24 and 26, that are at redshift $z\approx1.01$ with an average projected offset to the position of \XTtwo\ of 203$\pm$7~kpc. For both sets of galaxies, the offset is too large for it to be likely that \XTtwo\ originated from either duo.

\subsection{\XTthree}
 
Source 36 is located within the 1$\sigma$ uncertainty region of \XTthree. This galaxy has a redshift of 0.939 determined through the detection of one emission line identified as the [OII] doublet. While the emission line falls within 16~\AA\ of the skyline at 7246~\AA, we consider it a solid detection.
If this galaxy is the host of \XTthree, L$_{X, peak} = 1.2\times10^{46}$~erg~s$^{-1}$, which falls within the peak luminosity range of both the BNS and the IMBH-WD TDE progenitor models. The offset at the distance of this galaxy is $14\pm15$~kpc, which is consistent with the offset distribution of all three progenitor models. The chance alignment probability between this galaxy and the FXT is 0.06. Therefore, we consider source 36 a viable host galaxy candidate for \XTthree. 

There is, however, another source within the $3\sigma$ uncertainty region of \XTthree, source 40, which is identified by us as a type M1V late-type star (see Table~\ref{tab:dwarfs}). 
The R-band magnitude of this source is 21.27$\pm$0.03~mag$_\mathrm{AB}$. This magnitude is obtained through aperture photometry with \se\ using a circular aperture of 2.5 times the FWHM, where the FWHM was determined with \texttt{imexam} in {\sc IRAF}. 
We use this magnitude to calculate log(L$_x$/L$_{bol}$) = log(F$_x$/F$_{bol}$) following the same procedure as \cite{QV2022}\footnote{In short: we normalize stellar synthetic models of dwarf stars taken from \cite{Phillips2020} ($1000 \lesssim T_{eff} \lesssim 3000$~K and $2.5 \lesssim \mathrm{log}g \lesssim 5.5$) to the R band magnitude obtained for this star, and integrate the normalised models at optical/NIR wavelengths to calculate the bolometric flux.} and find $-2.9 \lesssim$ log(F$_x$/F$_{bol}$)$ \lesssim 0.1$ depending on the spectral type of the synthetic model used. The saturation limit of M-dwarfs is reported by \cite{DeLuca2020} to be as high as log(F$_x$/F$_{bol}$)$ \lesssim 0.0$ , which is consistent with the values we find. 

We calculate the chance alignment probability of the FXT position with this late-type star by using the procedure described in Section~\ref{sec:phot_Pch}, but we determine the surface density of stars brighter than source 40 locally on the image. There are 10 stars in total brighter than source 40 in the image of 1~arcmin$^2$, leading to a chance alignment probability of 0.16 for the distance of 4.2~arcsec between the FXT position and the dwarf star. If we limit our surface density calculation to only include late-type stars, source 40 is the brightest and the chance alignment probability is 0.02. Compared to the chance alignment probability of the galaxy (source 36) with the FXT position of 0.06, there is no clear best association of the FXT with either of these sources. 

For \XTthree\ there is group of galaxies (sources 4, 45, 51 and 55) at $z\approx 0.588$. For this group of sources, the projected distance is $156\pm13$~kpc. The positions of these four sources compared to the position of the FXT (three to the North-West and one to the South-West), could indicate this FXT happened due to interactions between these galaxies. If the FXT were to belong to this group of galaxies (assuming these galaxies form a group), L$_{X, peak} = 3.9\times10^{45}$~erg~s$^{-1}$. 

There is also a duo of sources (25 and 41) at $z\approx 0.592$. This duo is  
to the East of the FXT position, with an average position of \XTthree\ $\sim40$~kpc away from this FXT position. The average projected offset is consistent with a BNS merger origin.

\subsection{\XTfour} \label{sec:disc_xt4}

\cite{AL2020} report two potential host galaxies for \XTfour, which they assume to be at the same redshift. The two sources are numbered as source 14 (the galaxy reported at 12~arcsec\ from the FXT position with m$_R$ = 21.16$\pm$0.03 mag$_\mathrm{AB}$) and source 22 (the galaxy at 21~arcsec\ from the FXT position with  m$_R$ = 18.86$\pm$0.01 mag$_\mathrm{AB}$) in the bottom right panel of Figure~\ref{fig:images_sources} in this work. They fit the SEDs of both galaxies simultaneously to obtain the redshift of the pair as $z=0.37$ which they assume as the redshift of the FXT. We, however, find these sources are not located at the same redshift, with source 14 located at $z=0.43217(1)$ and source 22 located at $z=0.20803(9)$. This results in an L$_{X, peak}$ of $7.5\times10^{45}$ and $1.4\times10^{45}$~erg~s$^{-1}$ assuming either source 14 or source 22 is the host galaxy of \XTfour, respectively. For either of these galaxies we therefore rule out an SBO nature based on L$_{X, peak} > 10^{44}$ and the offset of the FXT with respect to these host galaxies.

For \XTfour\ galaxy 3 and 6 lie at approximately the same redshift of $z\approx0.367$. 
However, the location of \XTfour\ with respect to this duo of galaxies at an average projected offset of $\sim45$~kpc still points at a progenitor model that is not an SBO due to the offset and L$_{X, peak}$.  
 
\bigskip
It is possible that none of the candidate host galaxies discussed above is the real host galaxy. For instance, the real host galaxy might be too faint to be detected in our MUSE data.
It can be intrinsically faint or appear faint due to a large distance. Assuming a limiting magnitude in the R-band images of $\sim25$~mag (this limiting magnitude is derived from the faintest detected objects), dwarf galaxies like the Large Magellanic cloud with an absolute magnitude of $M_V\approx-18$ could be detected out to a distance of $\sim4$~Gpc or $z\sim0.65$. So, for the host galaxy to be undetected at the positions of the FXTs, it either has to be at a larger distance or it has to be fainter than $M_V\approx-18$ for $z<0.65$. If the galaxy is at $z>0.65$ L$_{X, peak} \approxgt 2\times10^{45}$~erg~s$^{-1}$ for all sources.

\section{Conclusions}

We present a study of the environment of four FXTs reported by \cite{Lin2019}, \citep{AL2020} and \cite{QV2022}. We report redshifts for between 12 and 22 galaxies in the images of these FXTs and use these redshifts to constrain $L_\textrm{X, peak}$ and the nature of the FXTs. We detect candidate host galaxies for two FXTs. For \XTtwo\ we detect an emission line in the spectrum extracted from the 1$\sigma$ positional uncertainty region. We identify this line as originating from a very faint source within the 1$\sigma$ positional uncertainty. 
This source lies at $z\sim1.22$ and has a chance alignment probability of 0.04. The BNS merger model or the IMBH-WD TDE model are both viable progenitors for this FXT based on the peak X-ray luminosity if we assume this source is the host galaxy of \XTtwo. We consider this source likely to be the host galaxy of \XTtwo.

For \XTthree, we detect a galaxy at redshift $z=0.939$ within the 1$\sigma$ uncertainty region with a low chance alignment probability. If this is indeed the host galaxy of \XTthree, the most likely progenitor would be either the BNS merger model or the IMBH-WD TDE model based on the offset and the peak X-ray luminosity. There is, however, another source within the 3$\sigma$ uncertainty position, an M-dwarf star, with a similar chance alignment probability as the galaxy with the FXT position. With the available data, we cannot determine which of these two sources \XTfour\ is associated with.

All FXTs have L$_{X, peak}$ that are too large and have a large offset with respect to all other candidate host galaxies, to be produced by a SN SBO. We cannot distinguish between the other models, IMBH-WD TDE or BNS merger, based on L$_{X, peak}$ and the offsets. We do however note that for the galaxies with offsets >30~kpc, an IMBH-WD TDE is unlikely due to the large offset. The question still remains if all FXTs have the same progenitor or if there are subgroups stemming from different progenitors.

%--------------------------------------------------------------------

\begin{acknowledgements}
Based on observations made with ESO Telescopes at the La Silla Paranal Observatory under programme ID 109.236W with PI D.~Eappachen.

This work is part of the research programme Athena with project number 184.034.002, which is financed by the Dutch Research Council (NWO). This publication is part of the project Dutch Black Hole Consortium (with project number 1292.19.202) of the research programme NWA which is (partly) financed by the Dutch Research Council (NWO).

P.G.J.~has received funding from the European Research Council (ERC) under the European Union’s Horizon 2020 research and innovation programme (Grant agreement No.~101095973).

We acknowledge support from 
ANID - Millennium Science Initiative Program - ICN12\_009 (FEB, JQV), CATA-BASAL - FB210003 (FEB), and FONDECYT Regular - 1200495 (FEB) and 1241005 (FEB).

This work makes use of Python packages {\sc numpy} \citep{2020arXiv200610256H}, {\sc scipy} \citep{2020NatMe..17..261V}; {\sc matplotlib} \citep{2007CSE.....9...90H}, 
%{\sc extinction} \citep{barbary_kyle_2016_804967}, 
{\sc PyMUSE} \cite{Pessa2018}, {\sc lmfit} \cite{lmfit}
This work made use of Astropy:\footnote{http://www.astropy.org} a community-developed core Python package and an ecosystem of tools and resources for astronomy \citep{astropy:2013, astropy:2018, astropy:2022}.

\end{acknowledgements}
%-------------------------------------------------------------------

\bibliographystyle{aa}
\bibliography{references}

\newpage

%-------------------------------------------------------------------
\begin{appendix} %First appendix
\onecolumn
\section{Redshifts obtained from emission line fitting}

\begin{figure}[ht!]
\centering
\includegraphics[width=.43\textwidth]{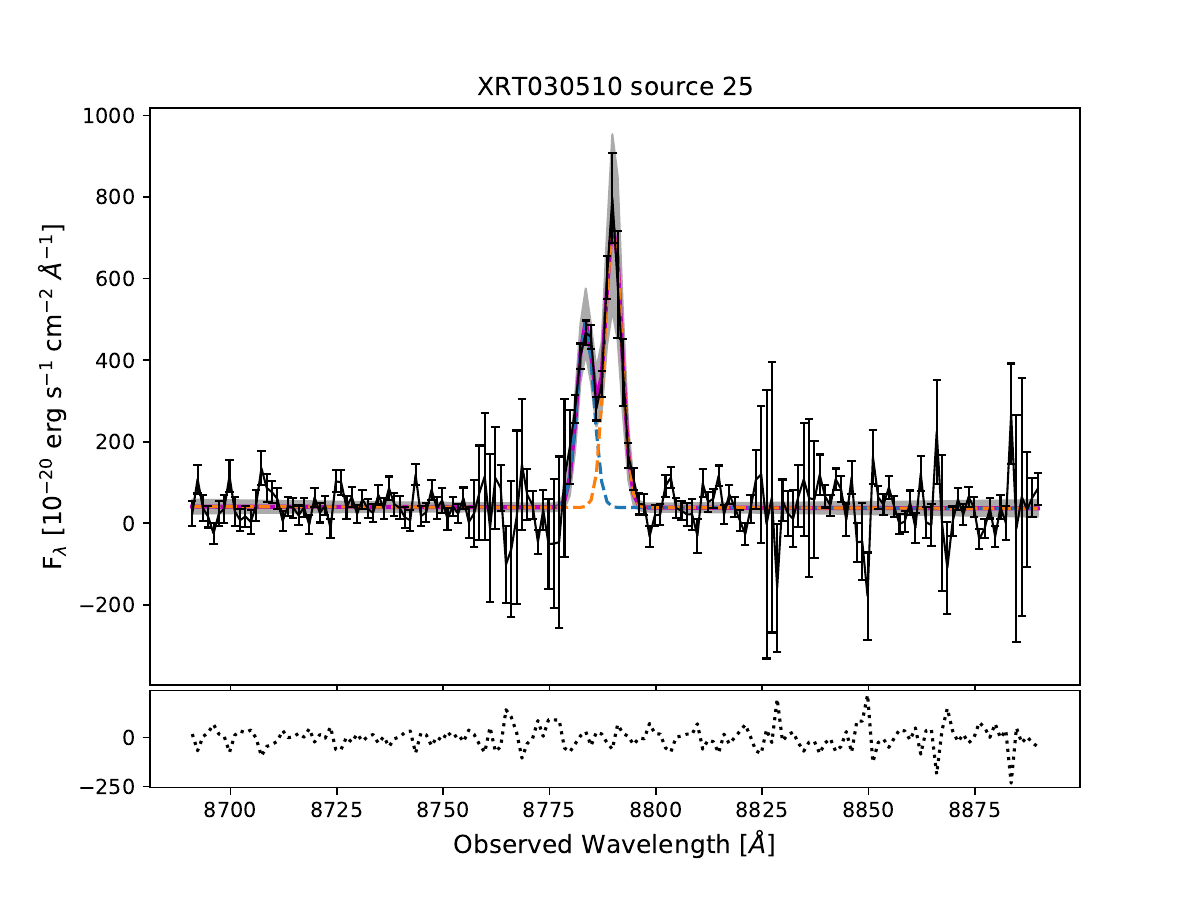}
\includegraphics[width=.43\textwidth]{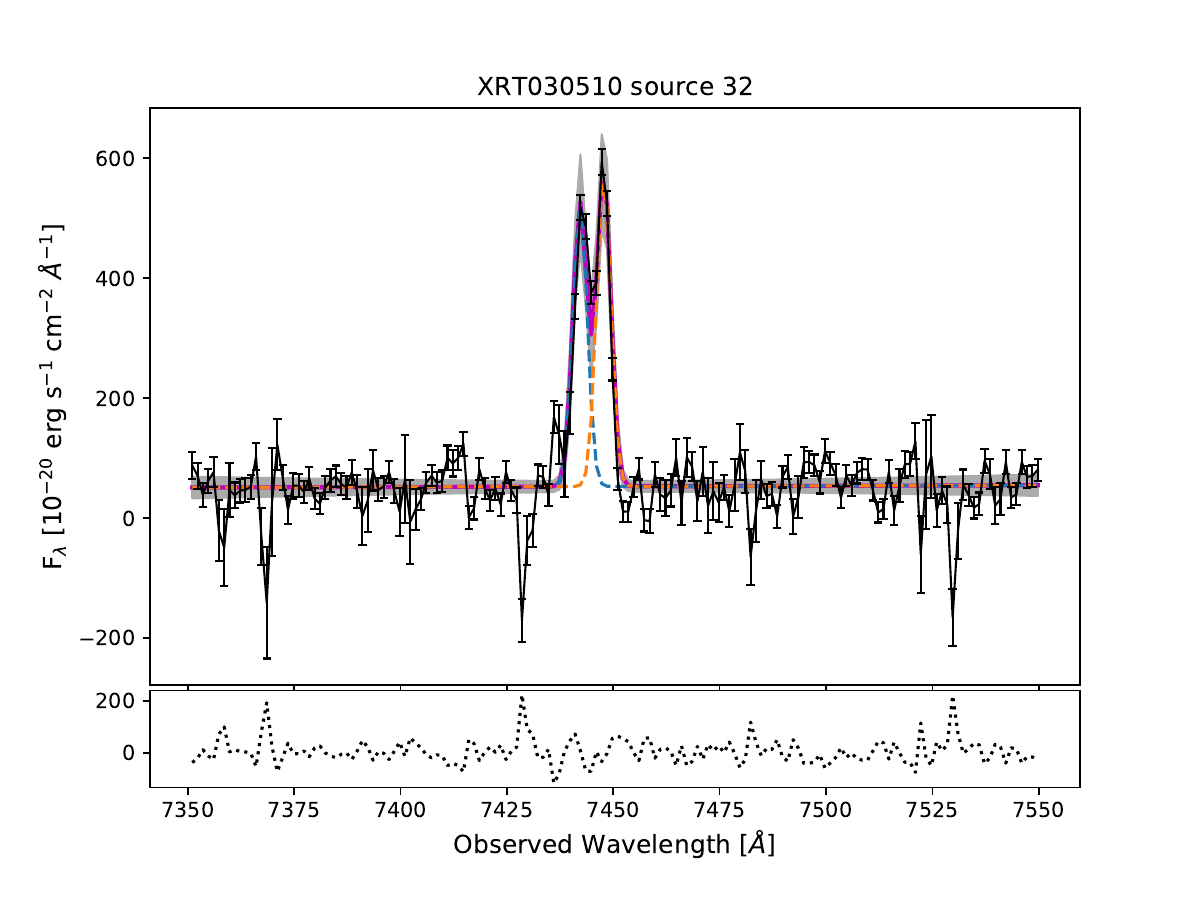}
\includegraphics[width=.43\textwidth]{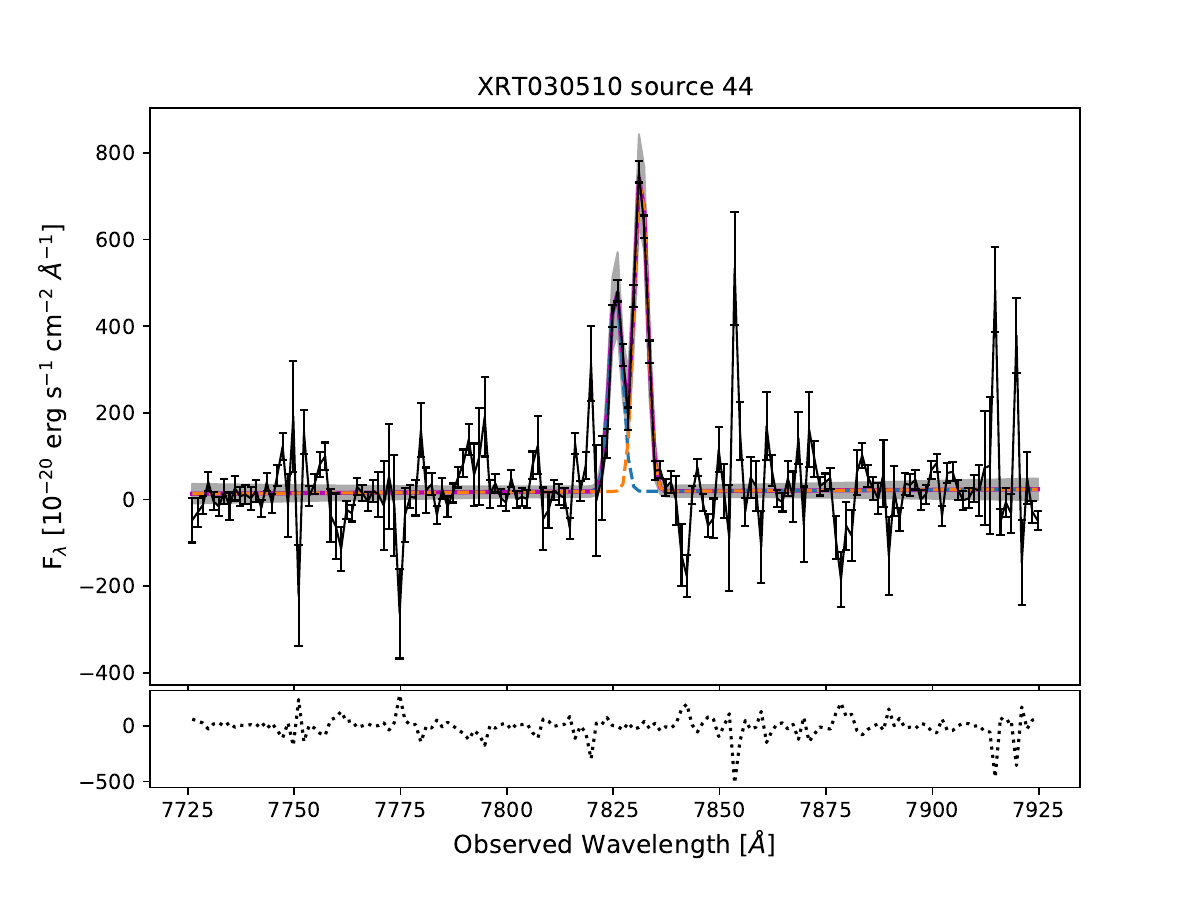}
\includegraphics[width=.43\textwidth]{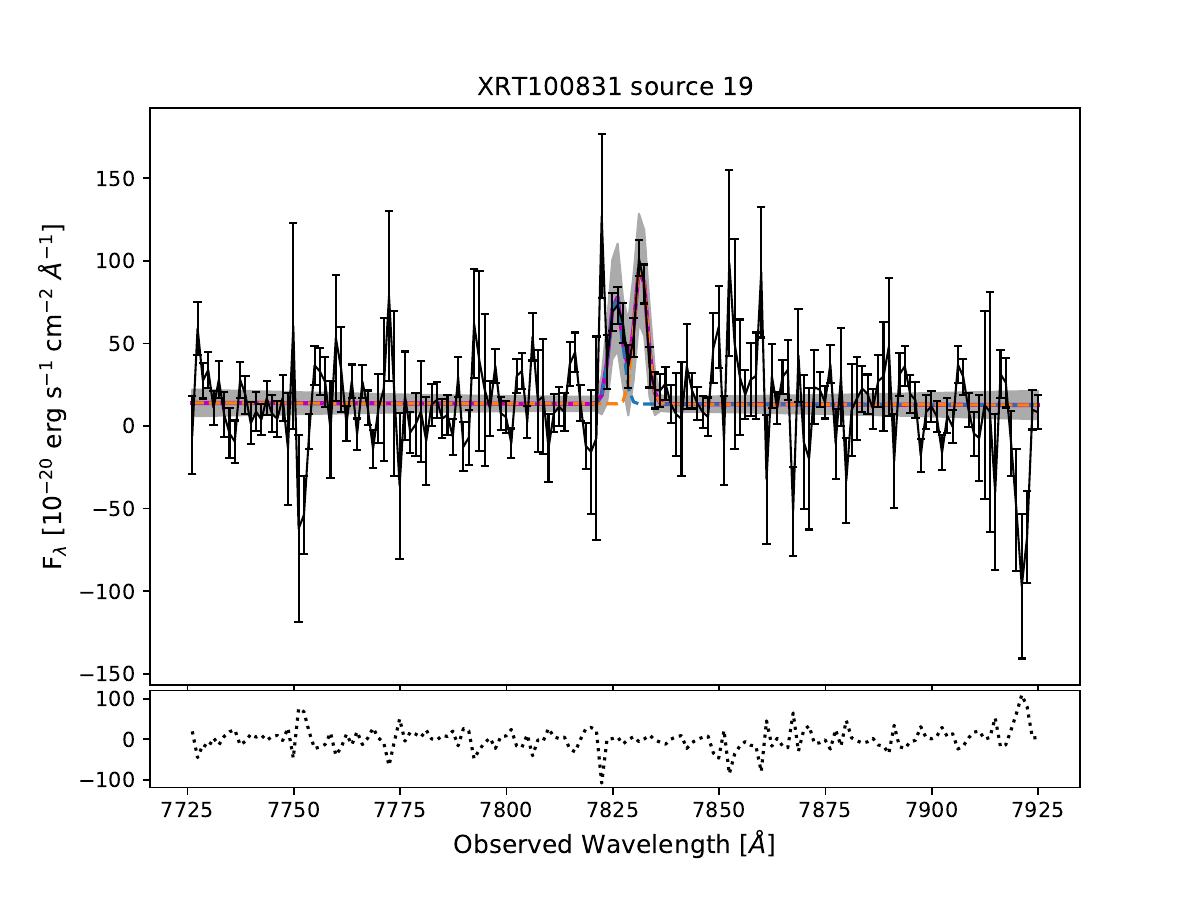}
\includegraphics[width=.43\textwidth]{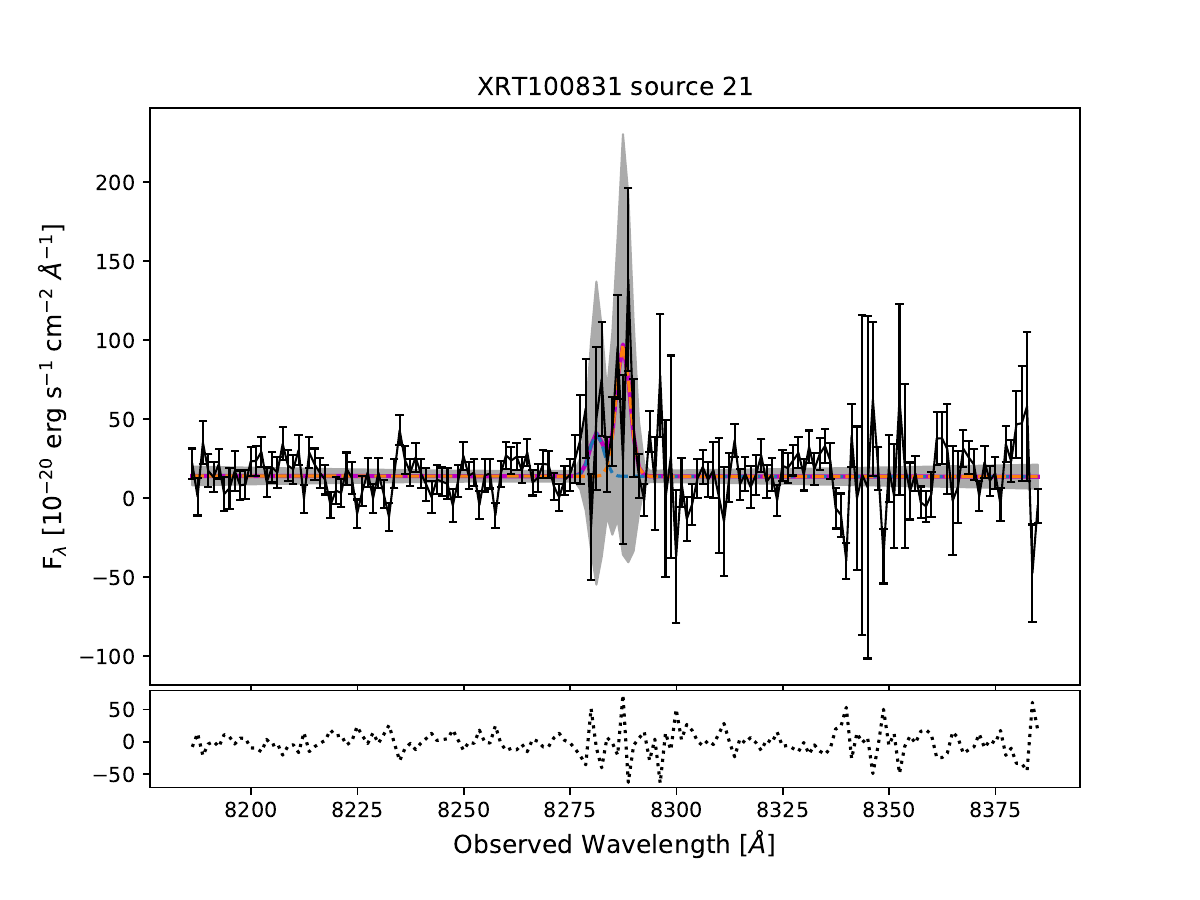}
\includegraphics[width=.43\textwidth]{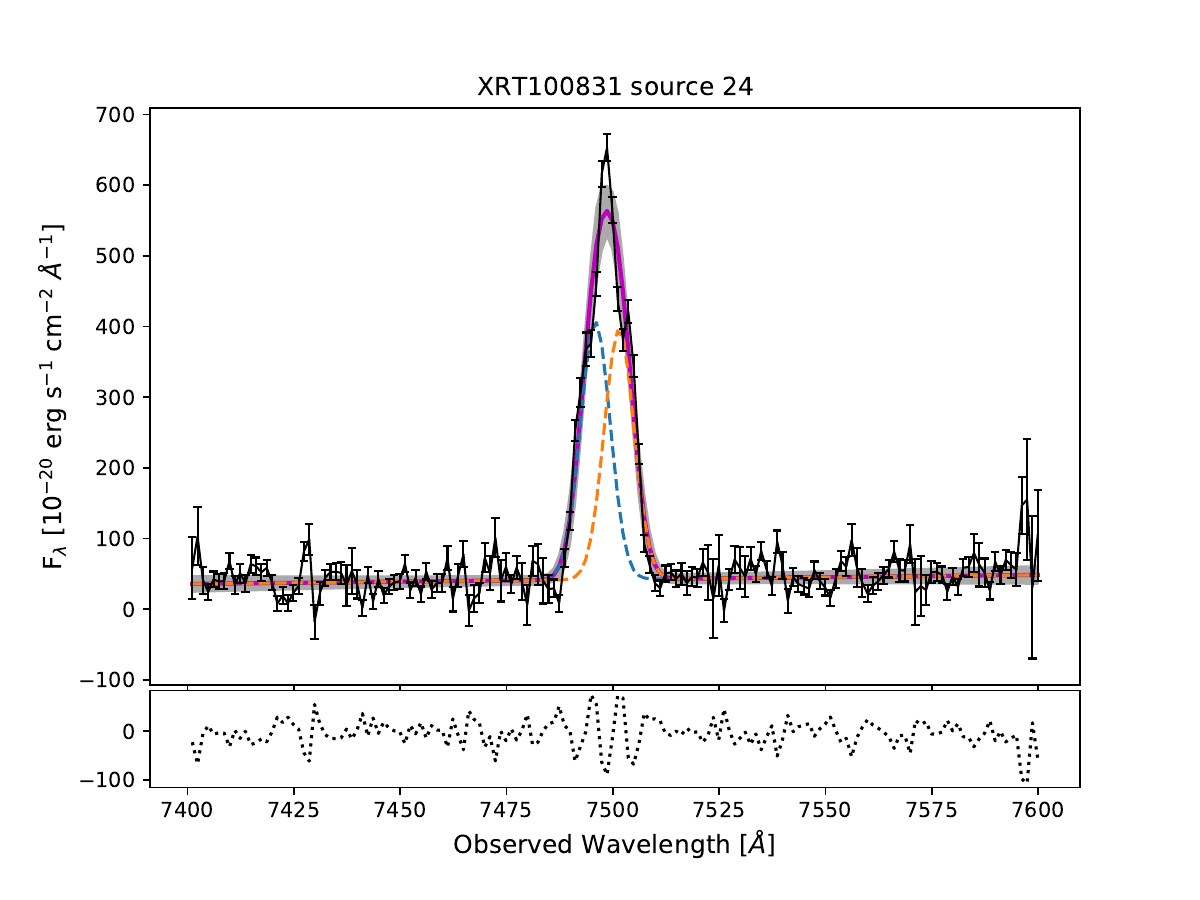}
\caption{ {Zoom in of the spectra of the emission line assumed to be due to the [O~II] doublet. For these spectra the redshift was determined only through the fit of the [O~II] doublet. The data are shown in black, the total best fit is shown in magenta and the individual lines in the doublet are shown in blue and orange. The FXT and source numbers are stated on top of each panel and correspond to the labelling in Table~\ref{tab:results} and Figure~\ref{fig:images_sources}. The error bars on the data represent 1$\sigma$ and the shaded area corresponds to the 3$\sigma$ region of the best fit}.}
\label{apfig:OII_lines}
\end{figure}

\begin{figure}[ht!]
\ContinuedFloat
\centering
\includegraphics[width=.43\textwidth]{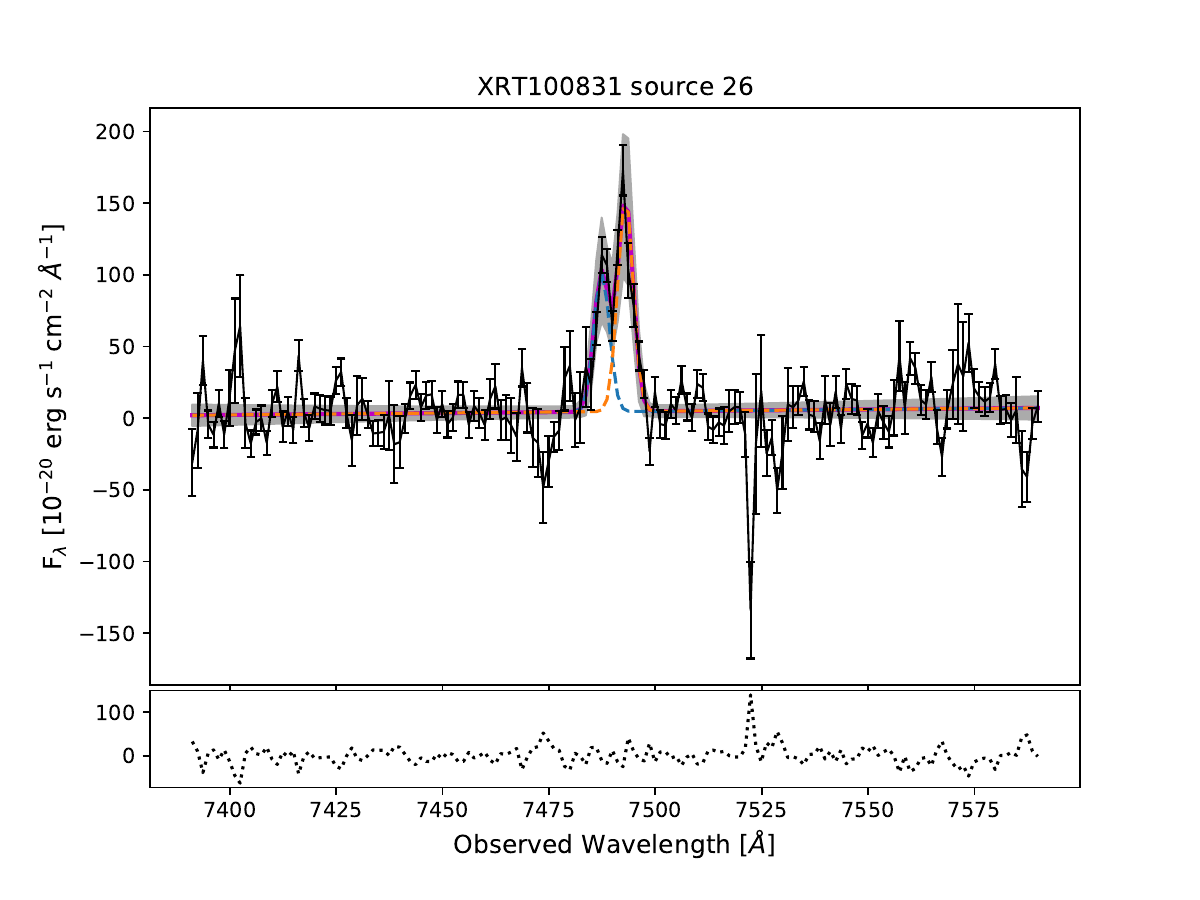}
\includegraphics[width=.43\textwidth]{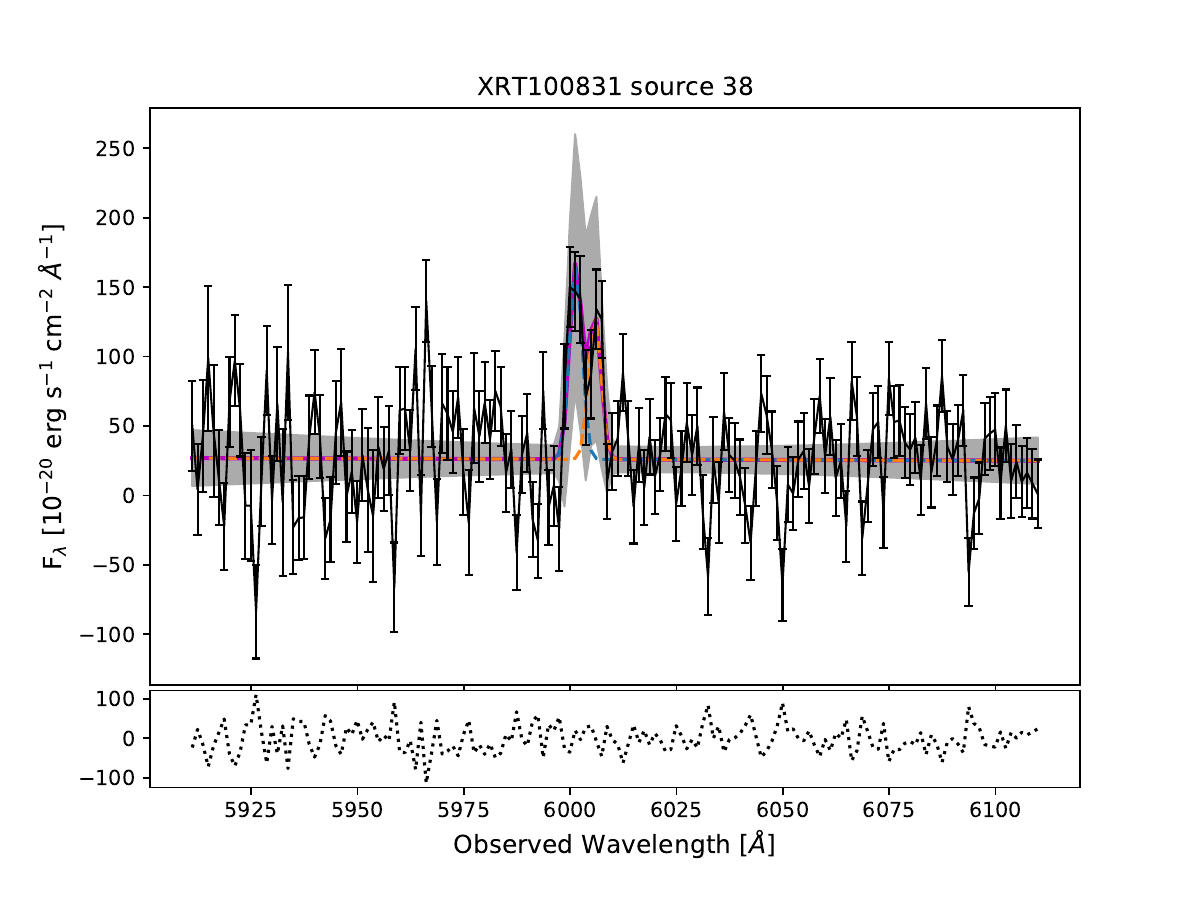}
\includegraphics[width=.43\textwidth]{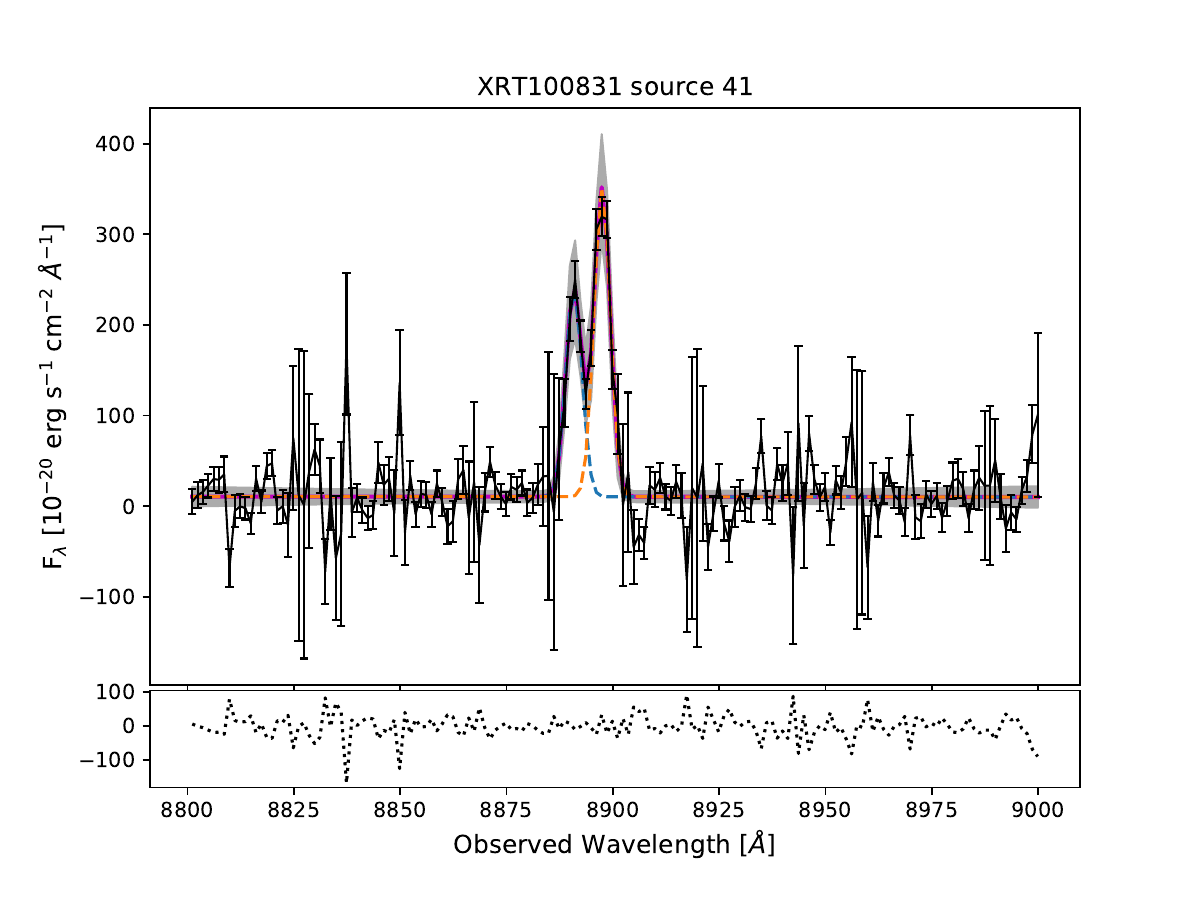}
\includegraphics[width=.43\textwidth]{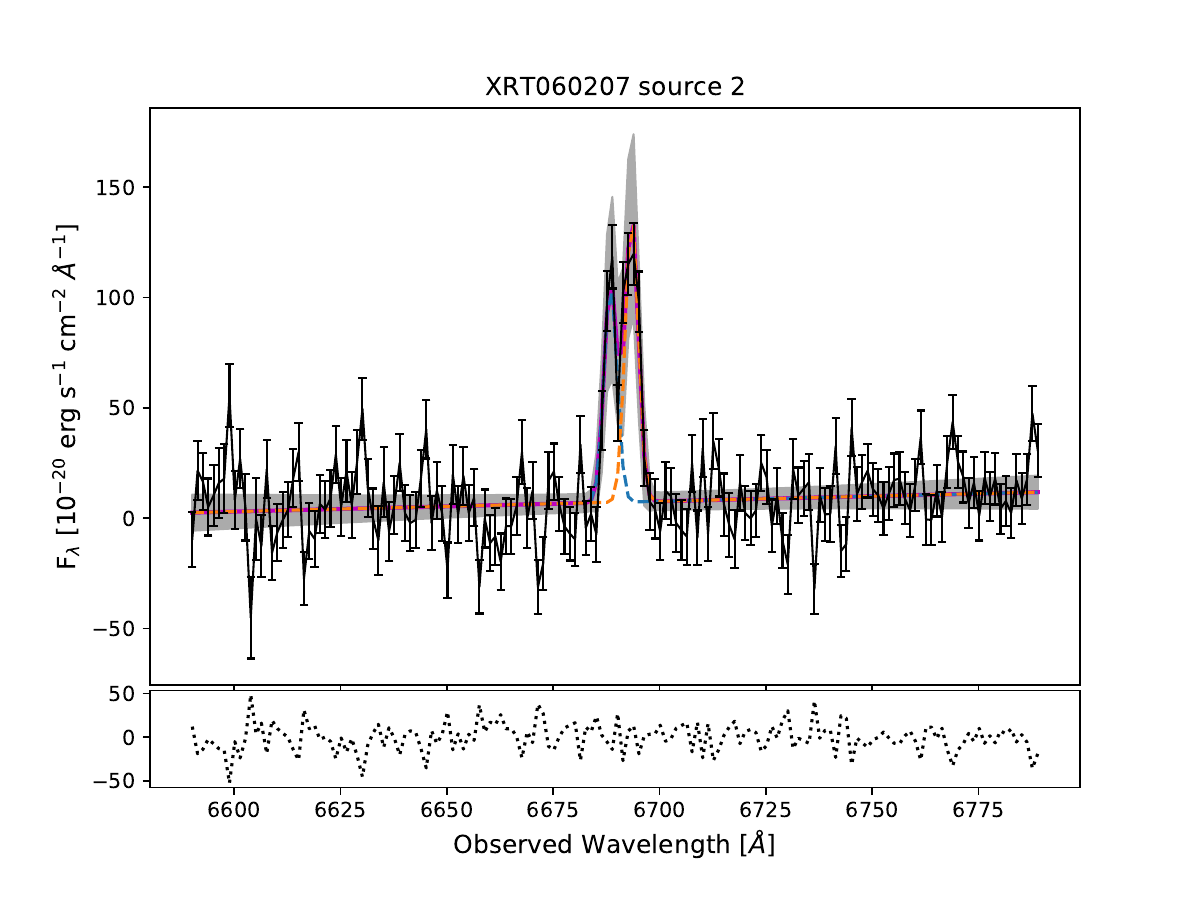}
\includegraphics[width=.43\textwidth]{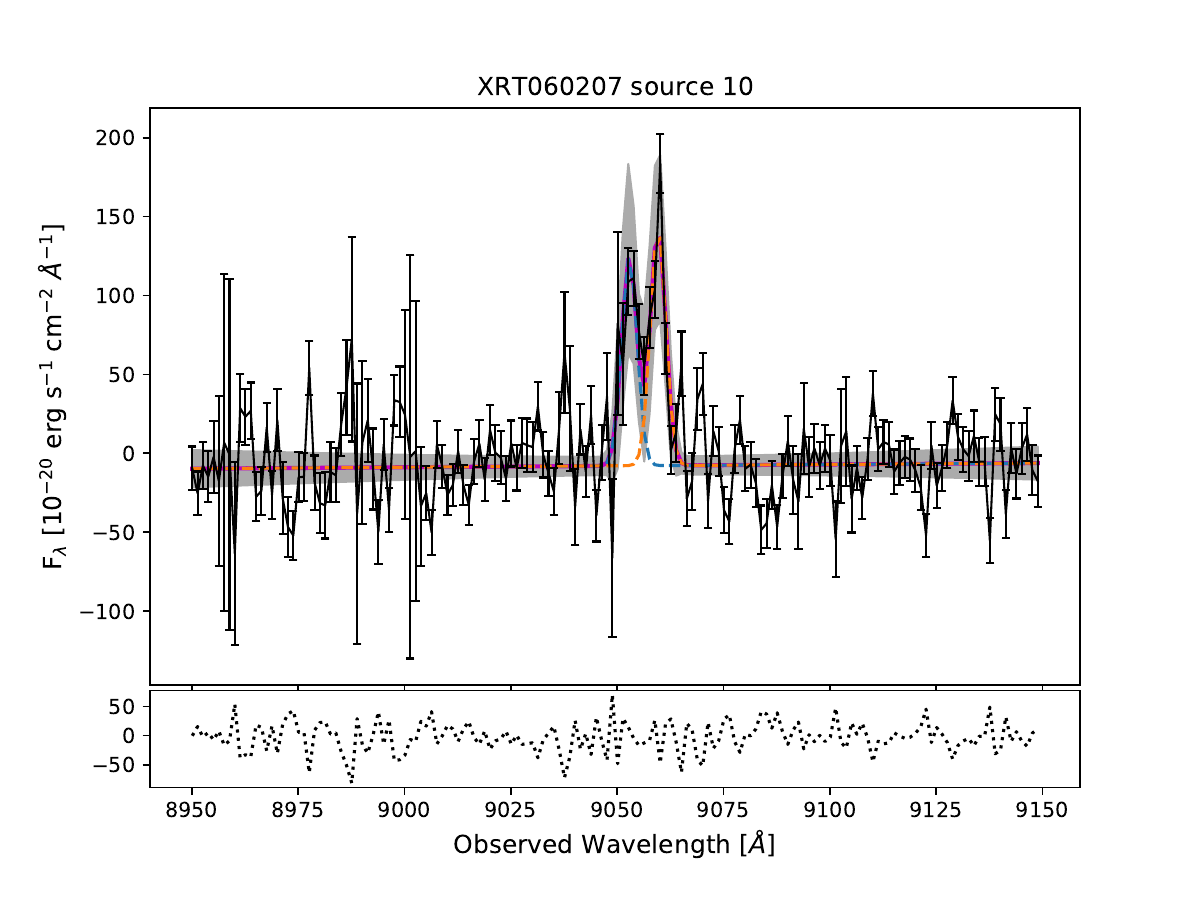}
\includegraphics[width=.43\textwidth]{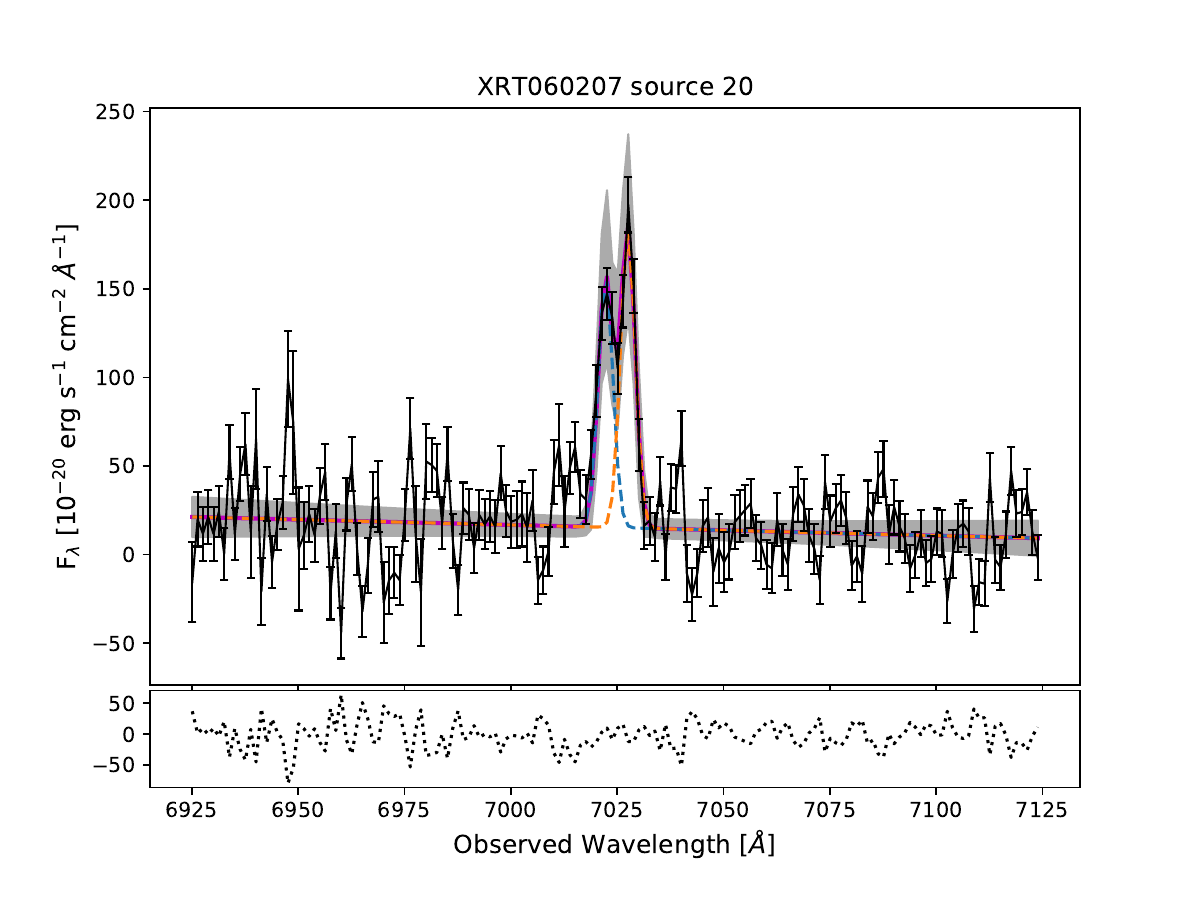}
\includegraphics[width=.43\textwidth]{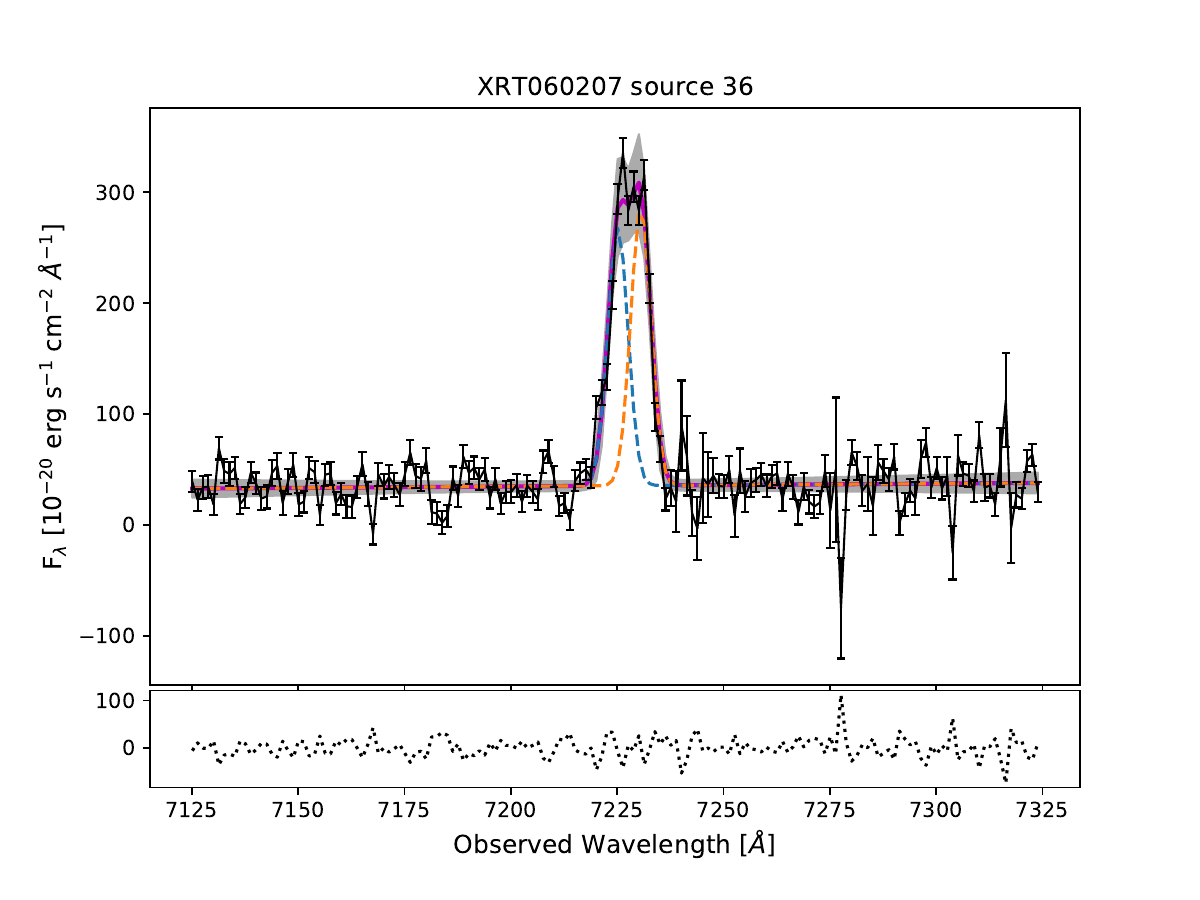}
\includegraphics[width=.43\textwidth]{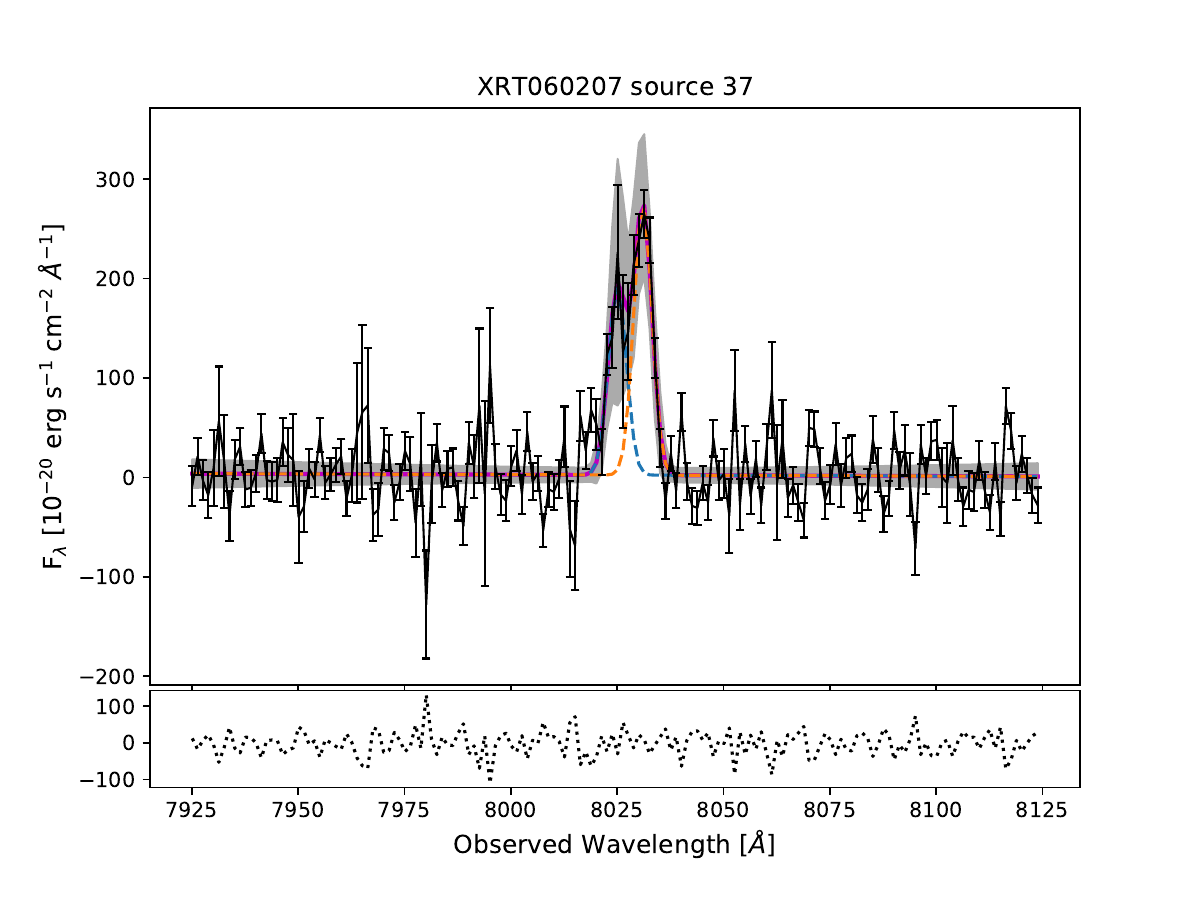}
\caption{Continued}
\end{figure}

\begin{figure}[ht!]
\ContinuedFloat
\centering
\includegraphics[width=.43\textwidth]{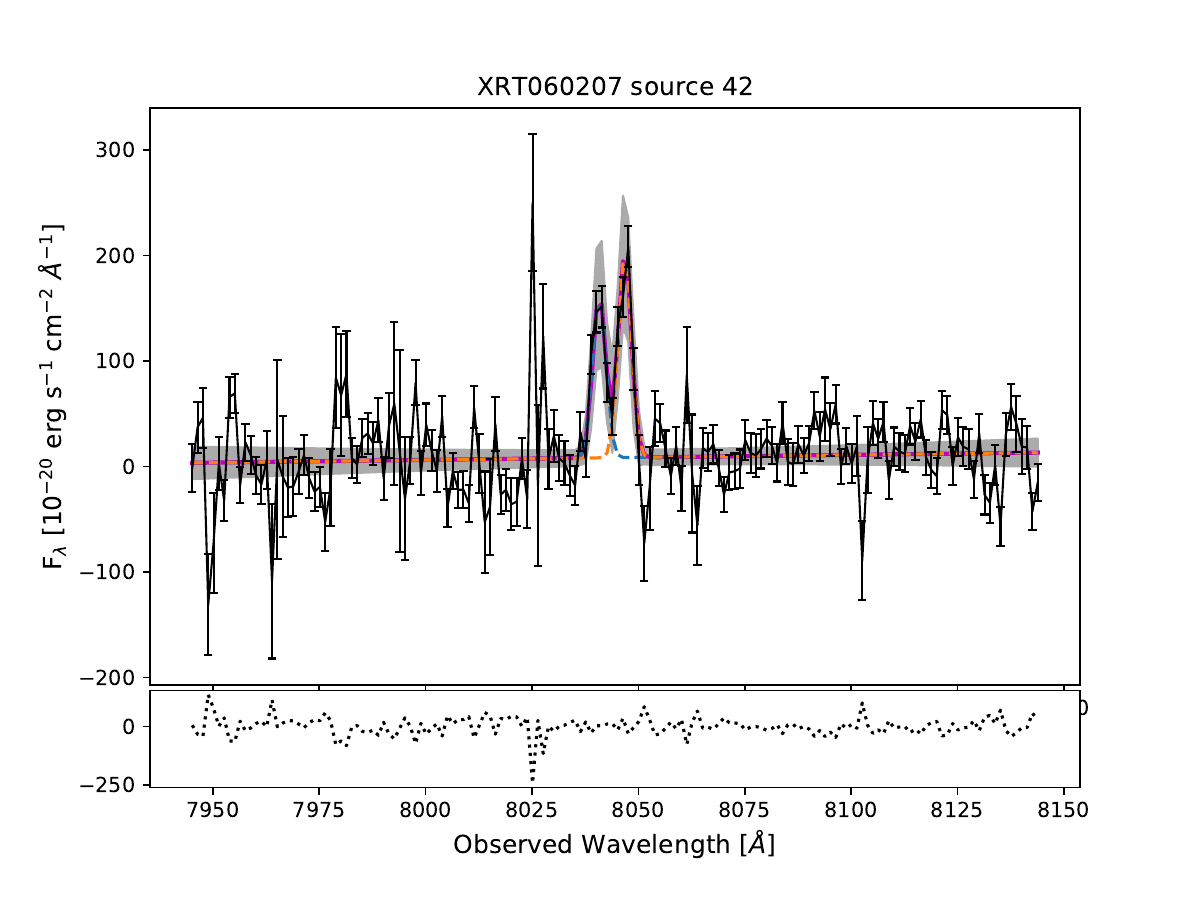}
\includegraphics[width=.43\textwidth]{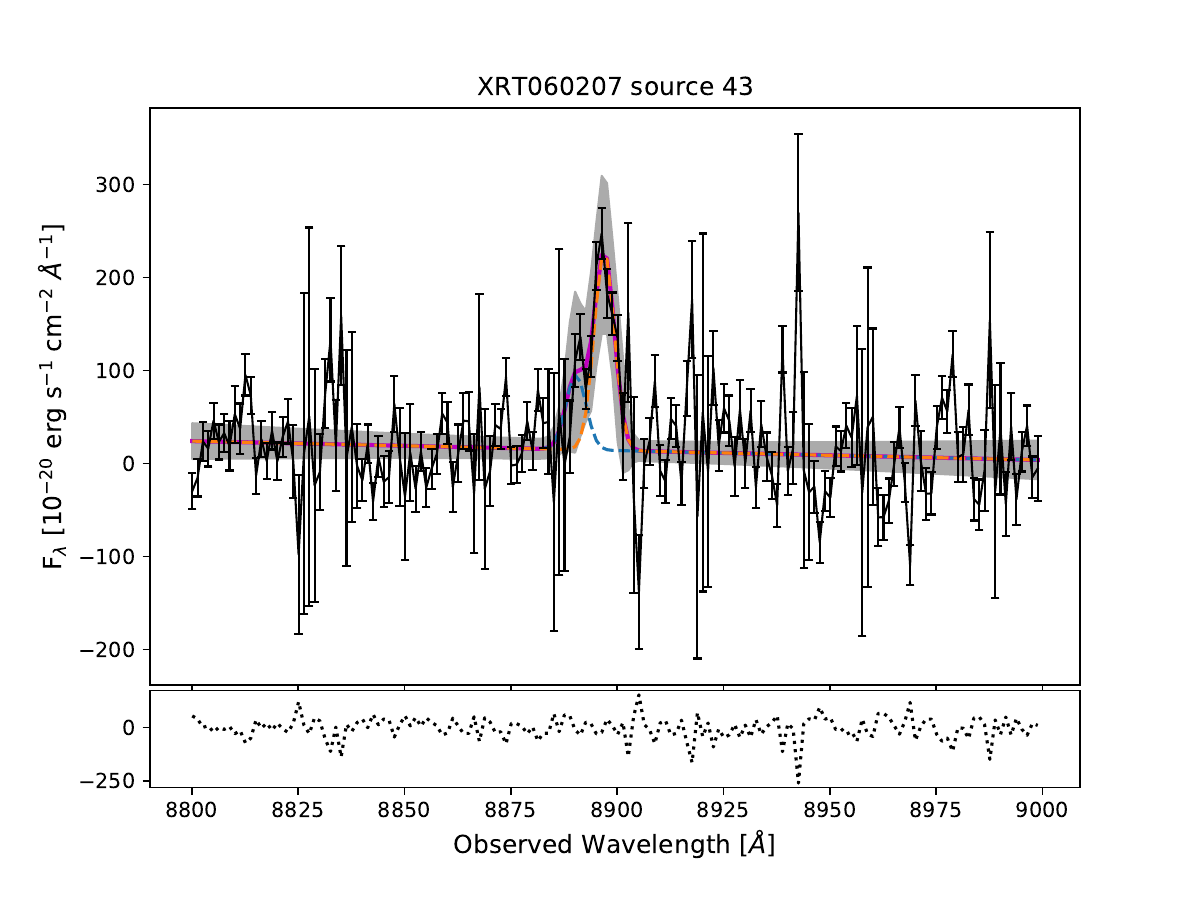}
\includegraphics[width=.43\textwidth]{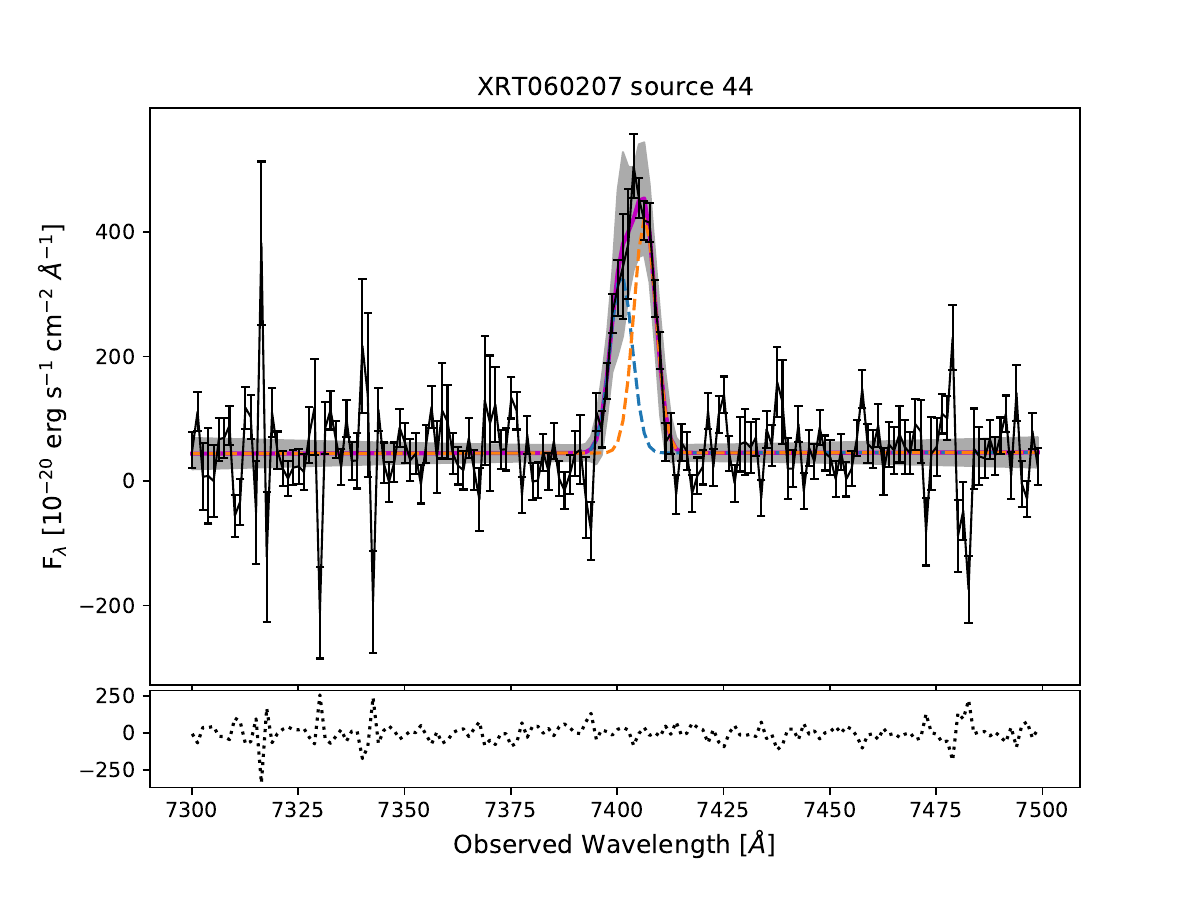}
\includegraphics[width=.43\textwidth]{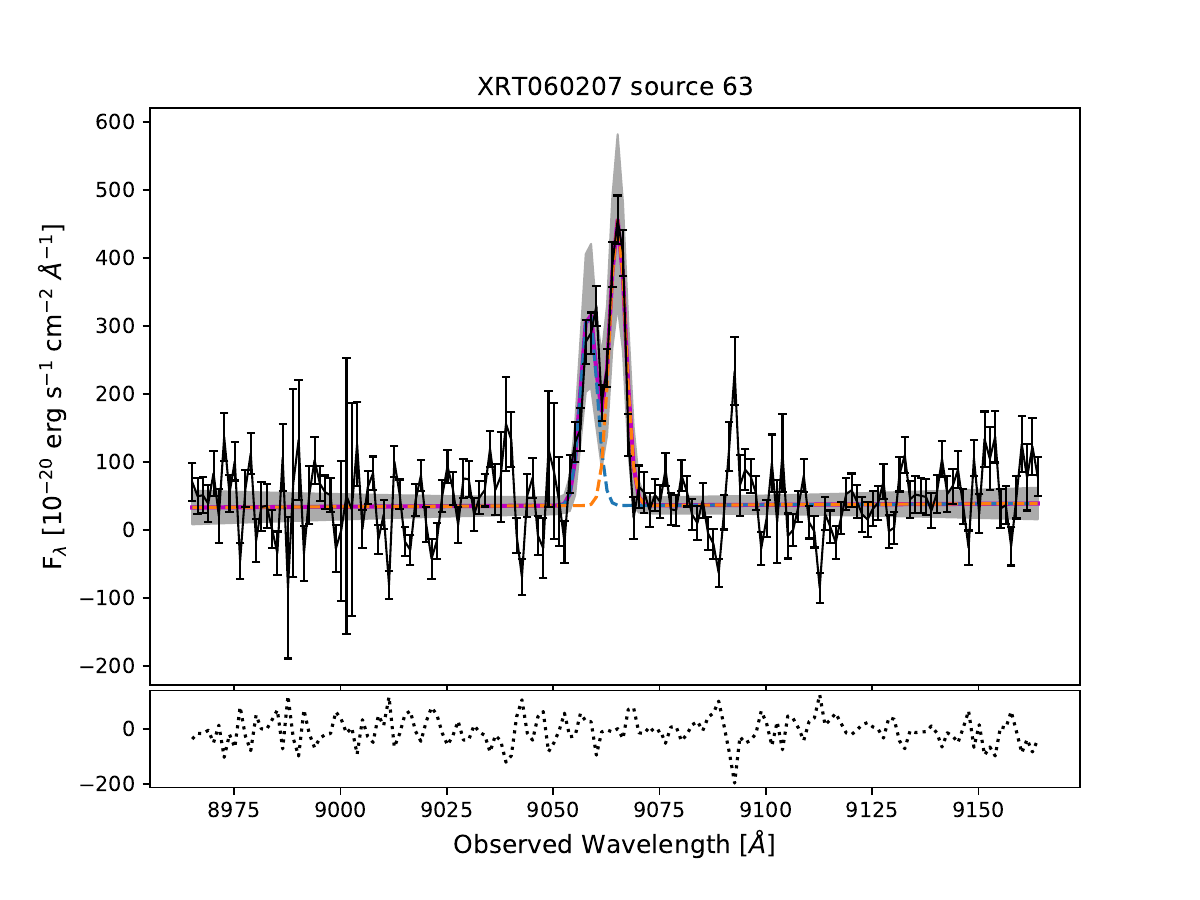}
\includegraphics[width=.43\textwidth]{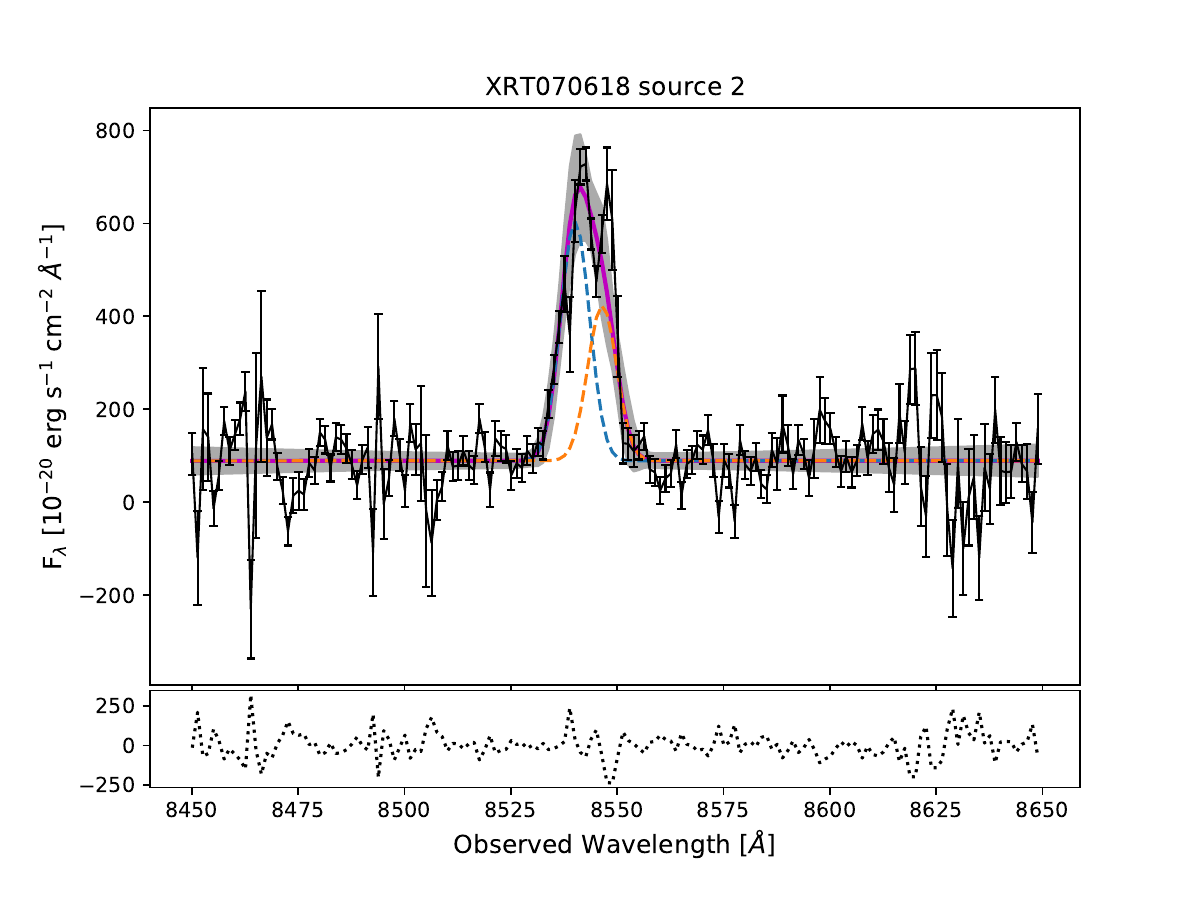}
\includegraphics[width=.43\textwidth]{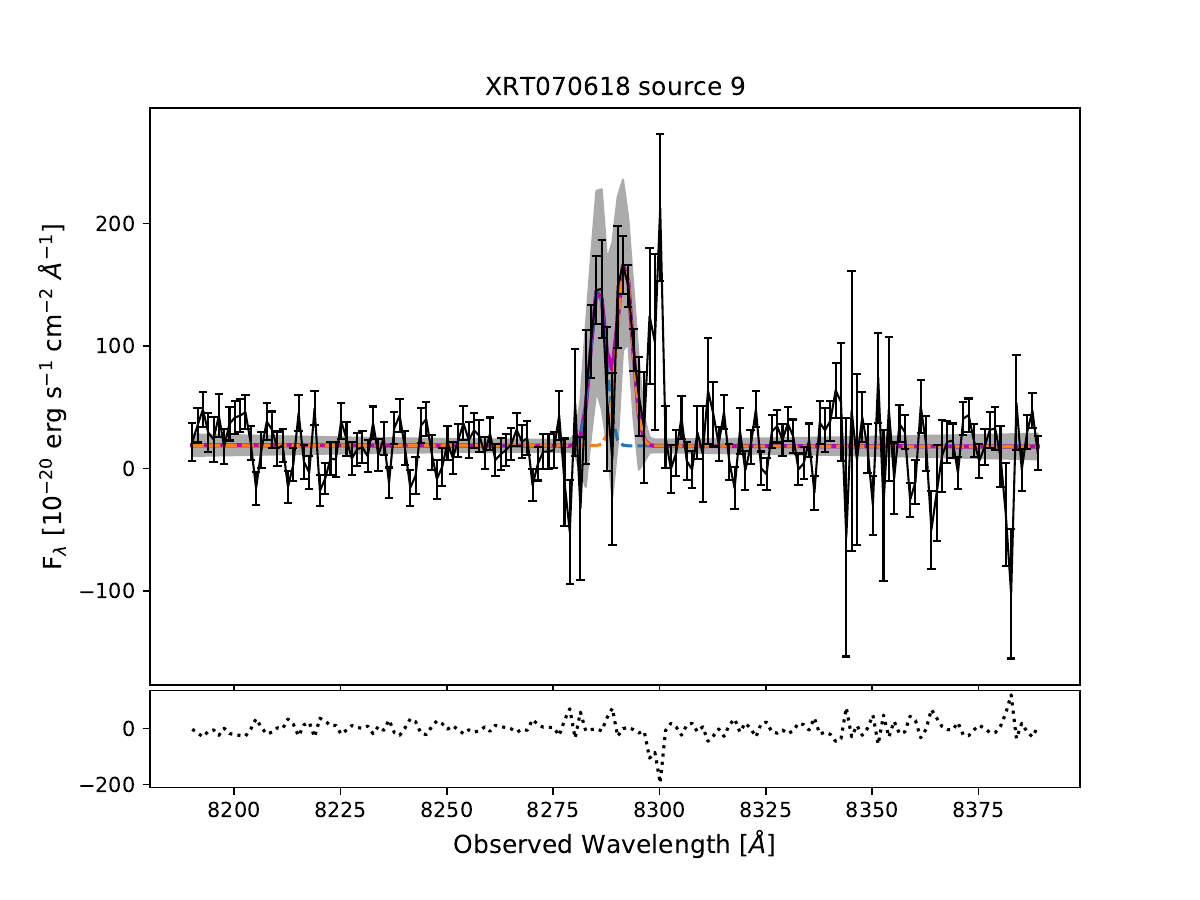}
\includegraphics[width=.43\textwidth]{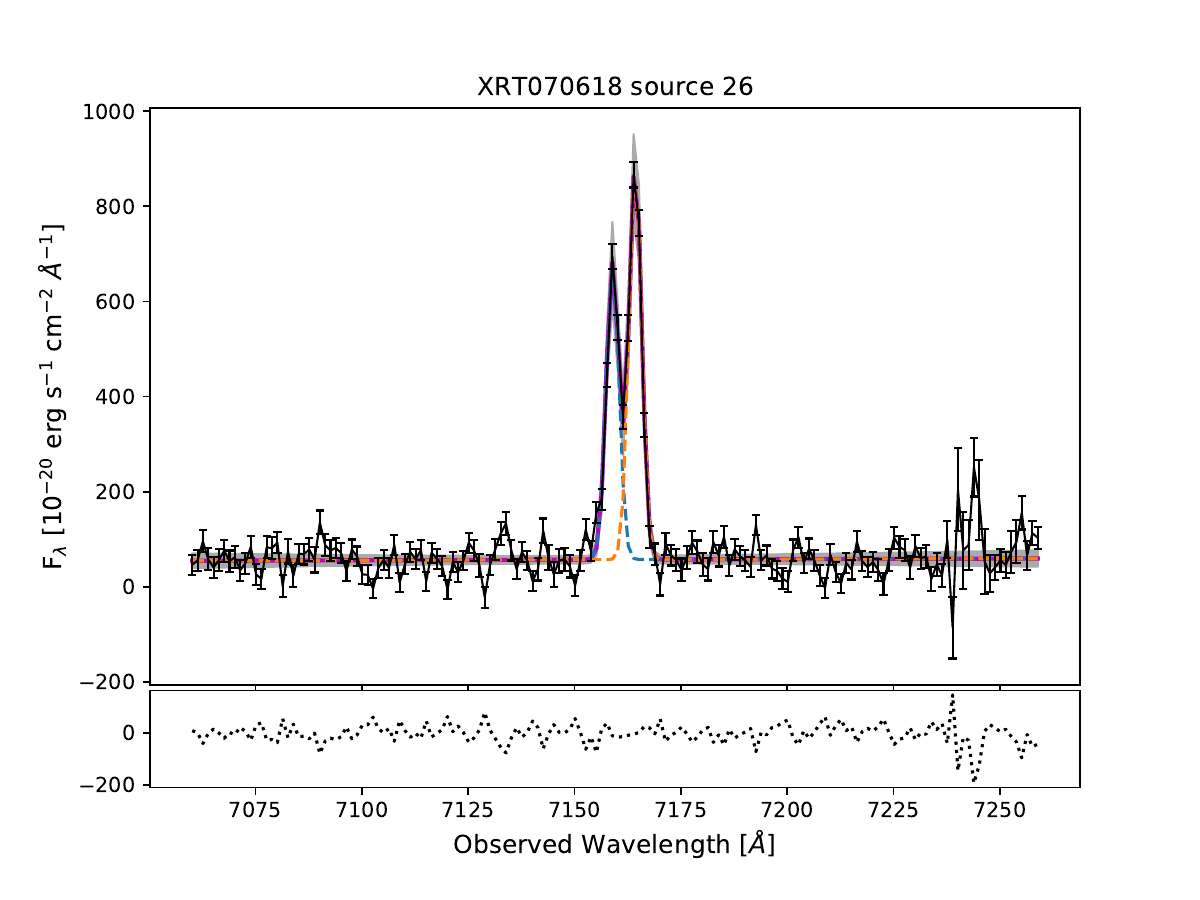}
\caption{Continued}
\end{figure}

\clearpage

\section{Redshifts obtained from emission line fitting}

\begin{figure}[ht!]
\centering
\includegraphics[width=.8\textwidth]{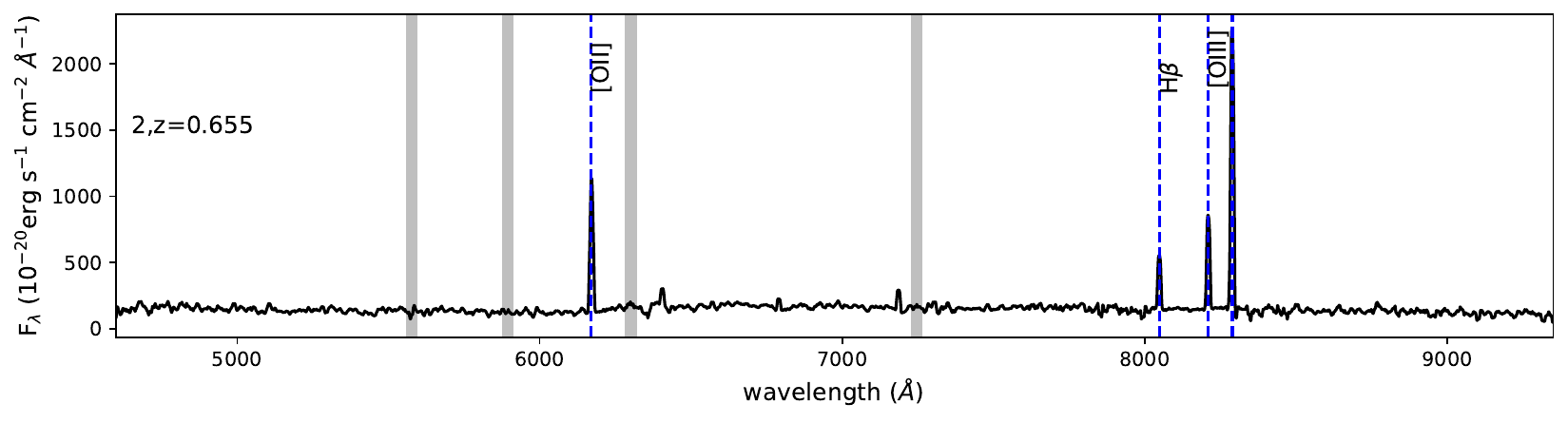}
\includegraphics[width=.8\textwidth]{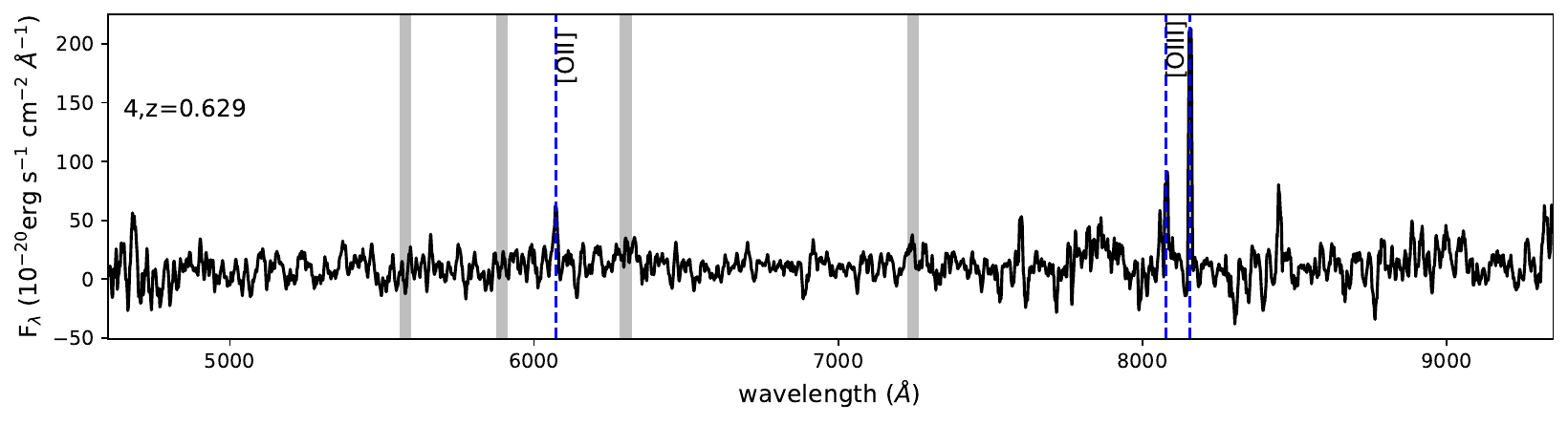}
\includegraphics[width=.8\textwidth]{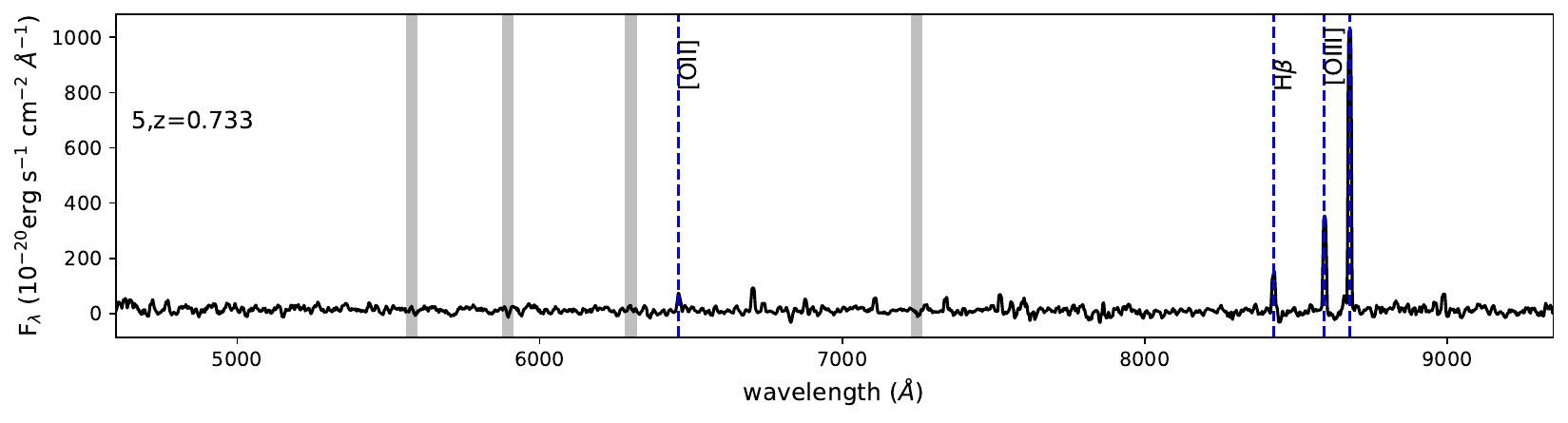}
\includegraphics[width=.8\textwidth]{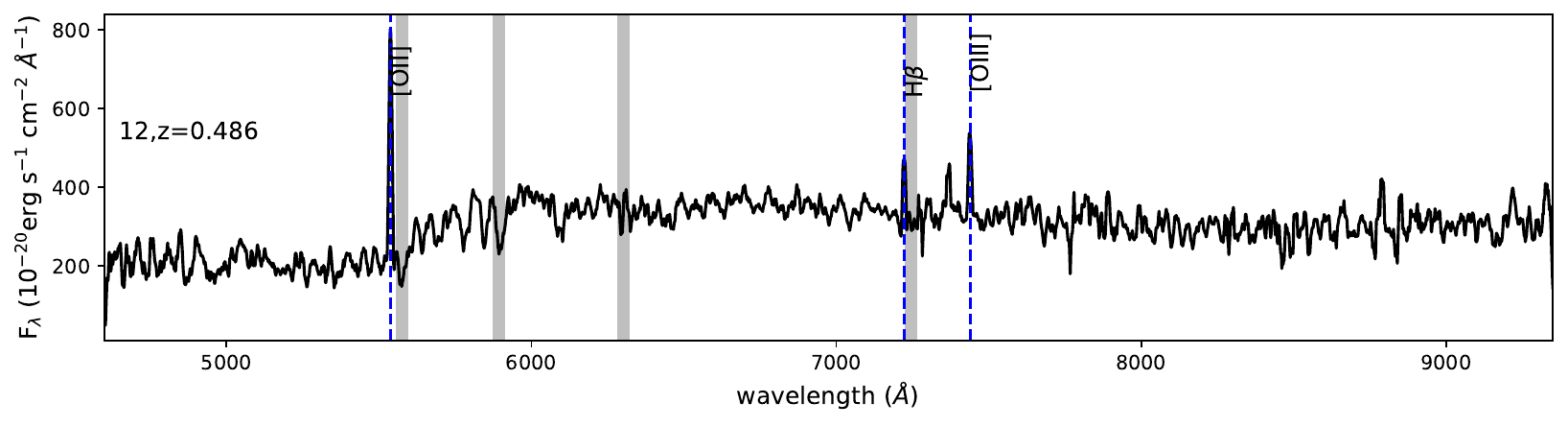}
\includegraphics[width=.8\textwidth]{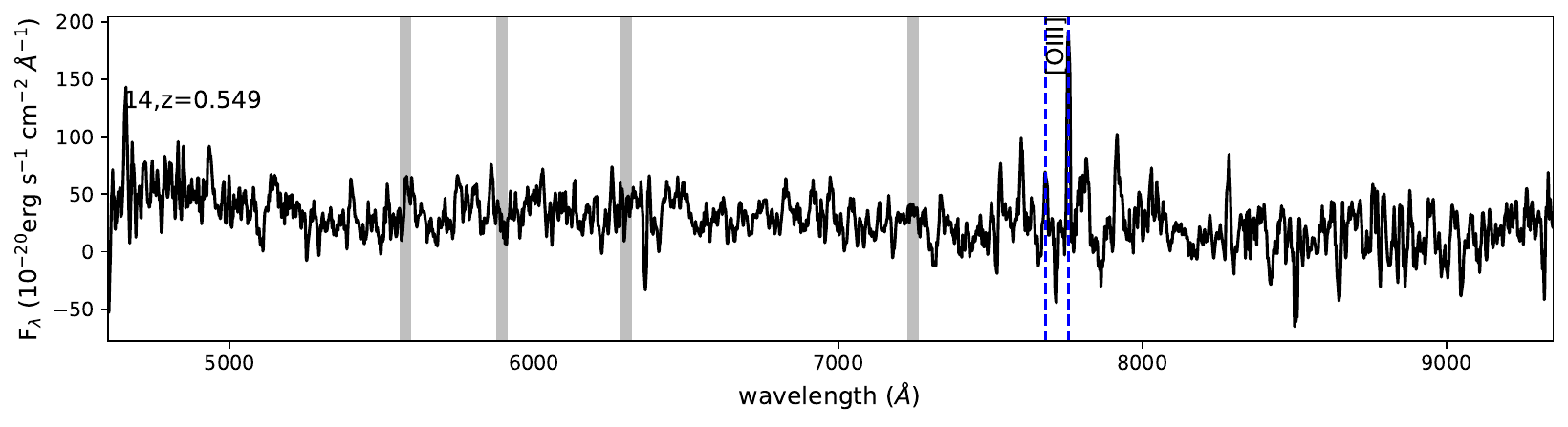}
\caption{The spectra of sources in the MUSE field of \XTone, smoothed using \texttt{Box1DKernel} with a kernel width of 10~pixels, for which we detected emission lines. The emission lines used to calculate the redshift, via the fits of Gaussians, are marked by blue vertical lines. For the spectra plotted in grey only the [OII] doublet was detected and used for the redshift determination, in the black spectra we detected multiple emission lines. The doublet is indicated by a single line here for clarity, but we did fit two separate Gaussian functions to the doublet if detected.
% Although the [OII] doublet is marked by one line as the wavelength separation is too small to plot two separate lines, we did fit two separate lines to this doublet when it was detected.} 
Each figure panel also contains the source number, corresponding to the source numbers in Table~\ref{tab:results} and Fig.~\ref{fig:images_sources} and the redshift derived (rounded off to not take too much space, for the full redshift, see Table~\ref{tab:results}). The grey vertical bands indicate the wavelengths of four prominent sky emission lines.}
\label{apfig:lines_xt1}
\end{figure}

\begin{figure}[ht!]
\ContinuedFloat
\centering
\includegraphics[width=.85\textwidth]{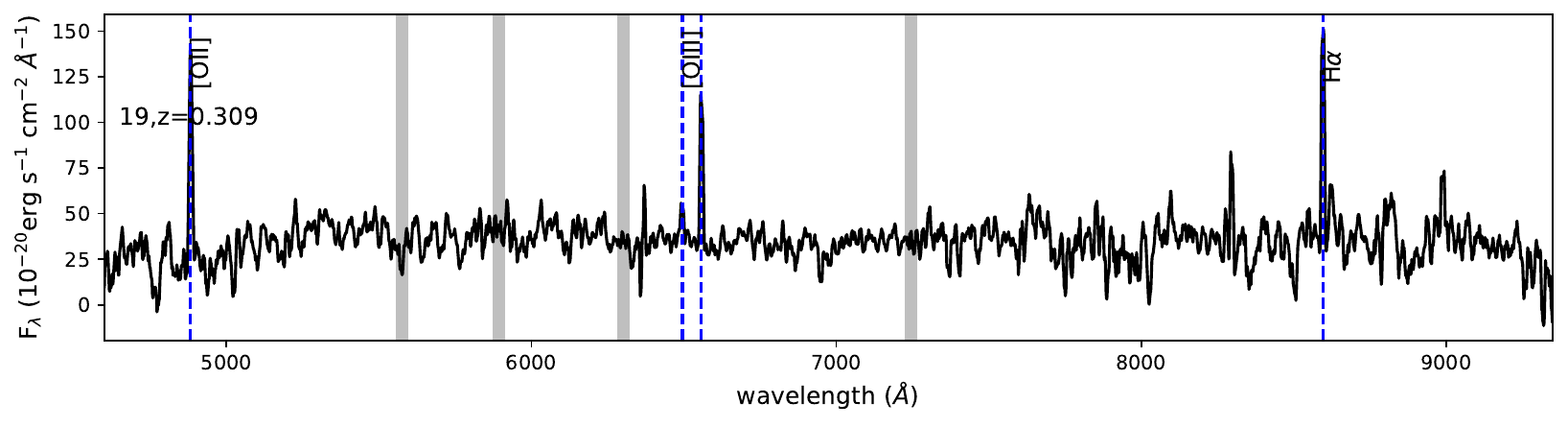}
\includegraphics[width=.85\textwidth]{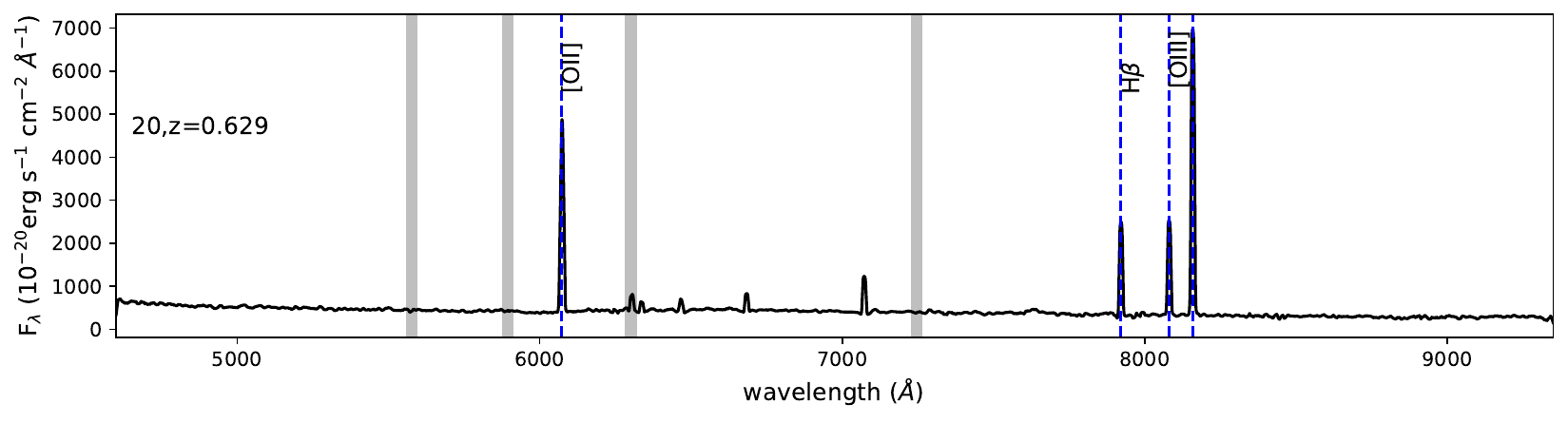}
\includegraphics[width=.85\textwidth]{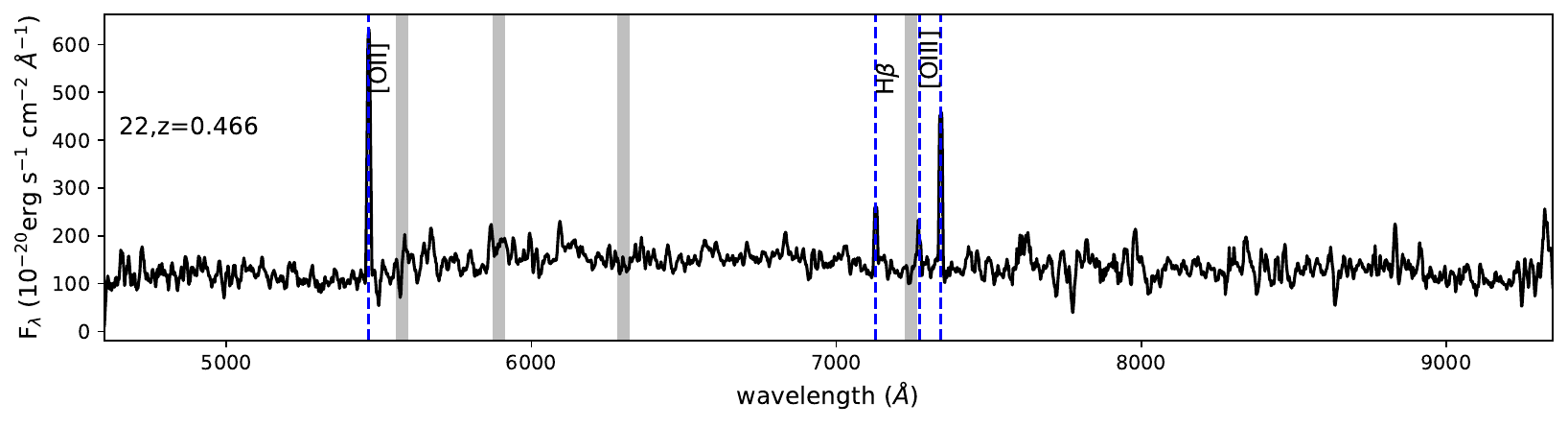}
\includegraphics[width=.85\textwidth]{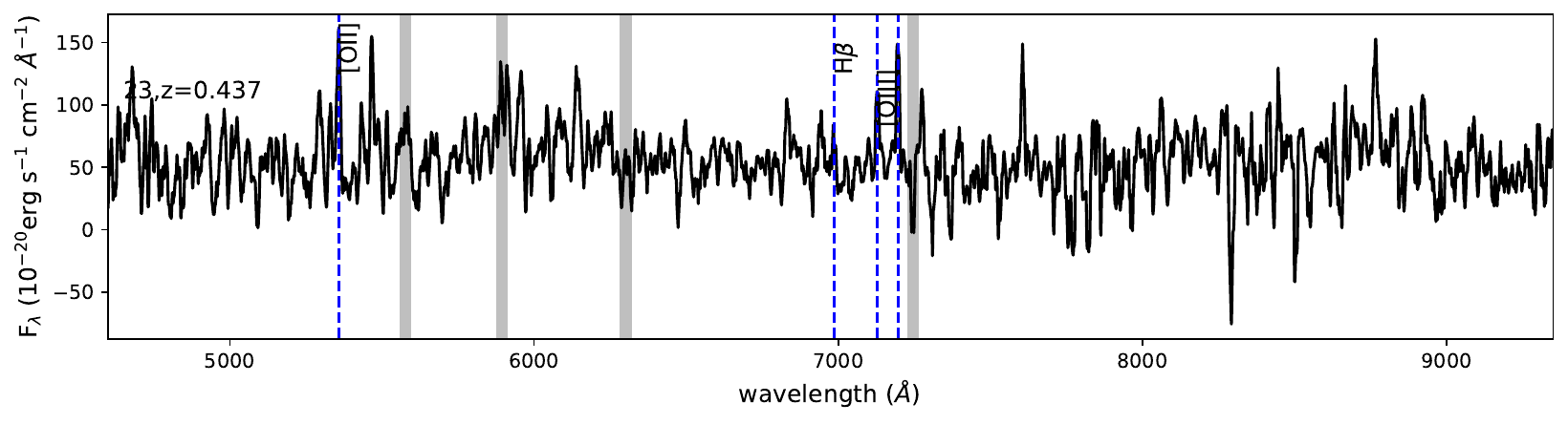}
\includegraphics[width=.85\textwidth]{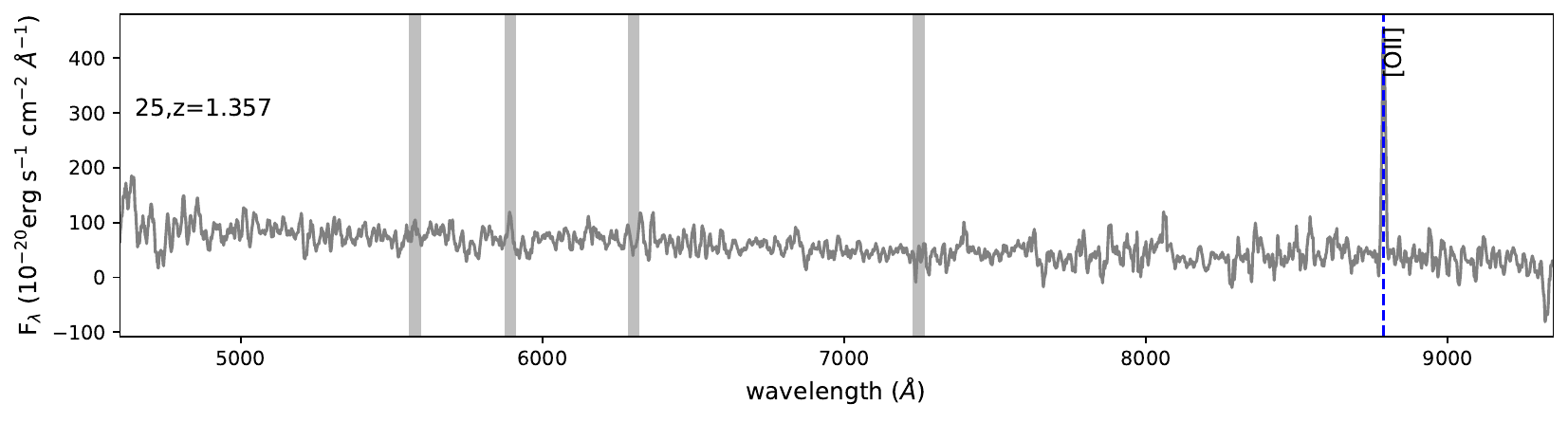}
\caption{Continued}
\end{figure}

\begin{figure}[ht!]
\ContinuedFloat
\centering
\includegraphics[width=.85\textwidth]{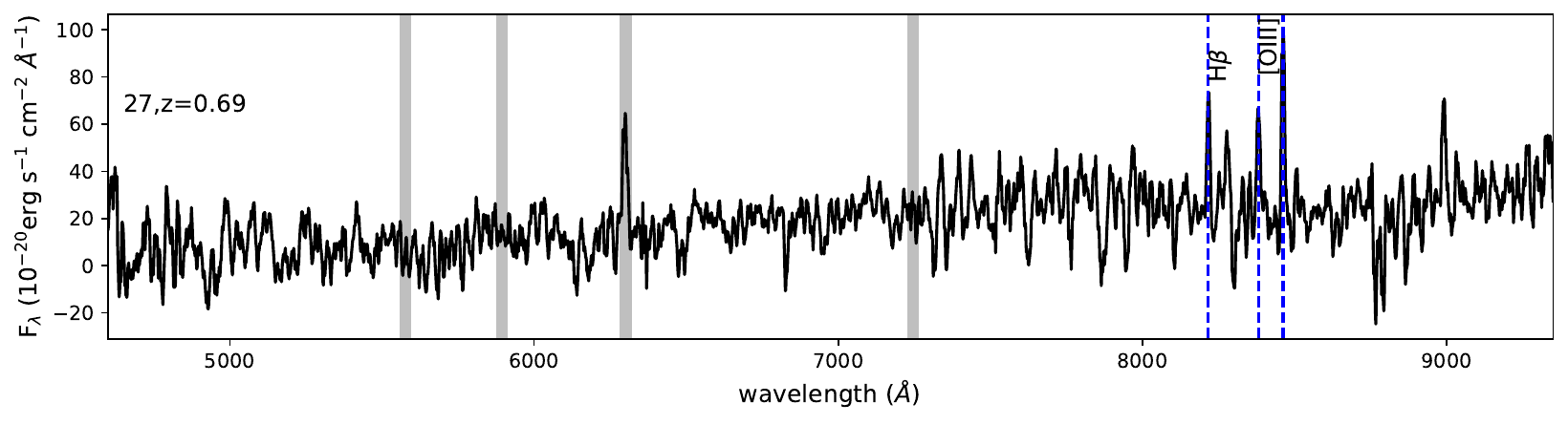}
\includegraphics[width=.85\textwidth]{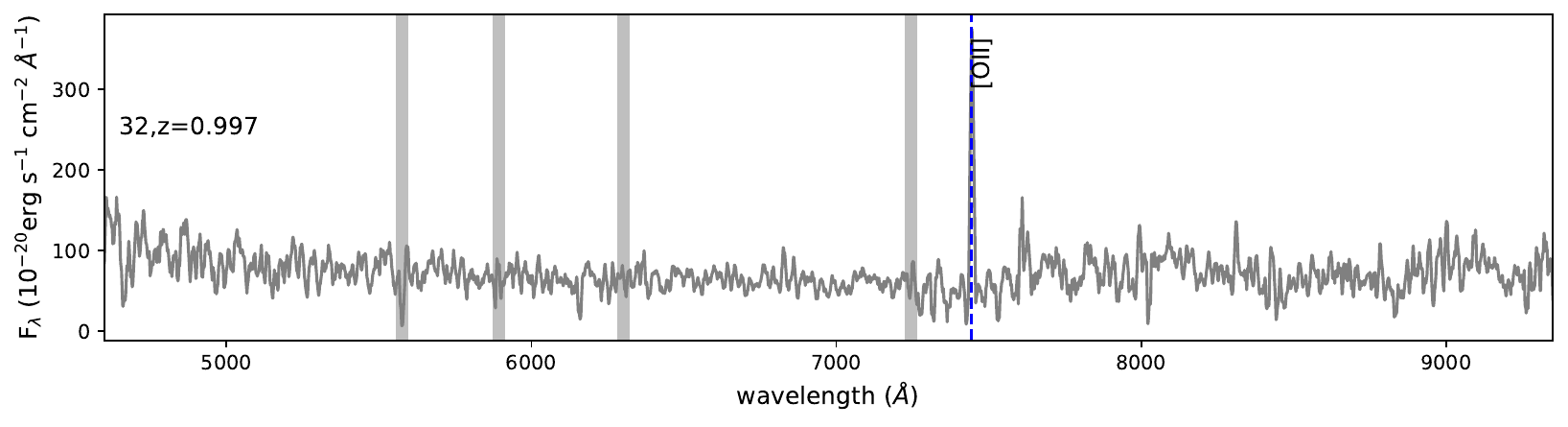}
\includegraphics[width=.85\textwidth]{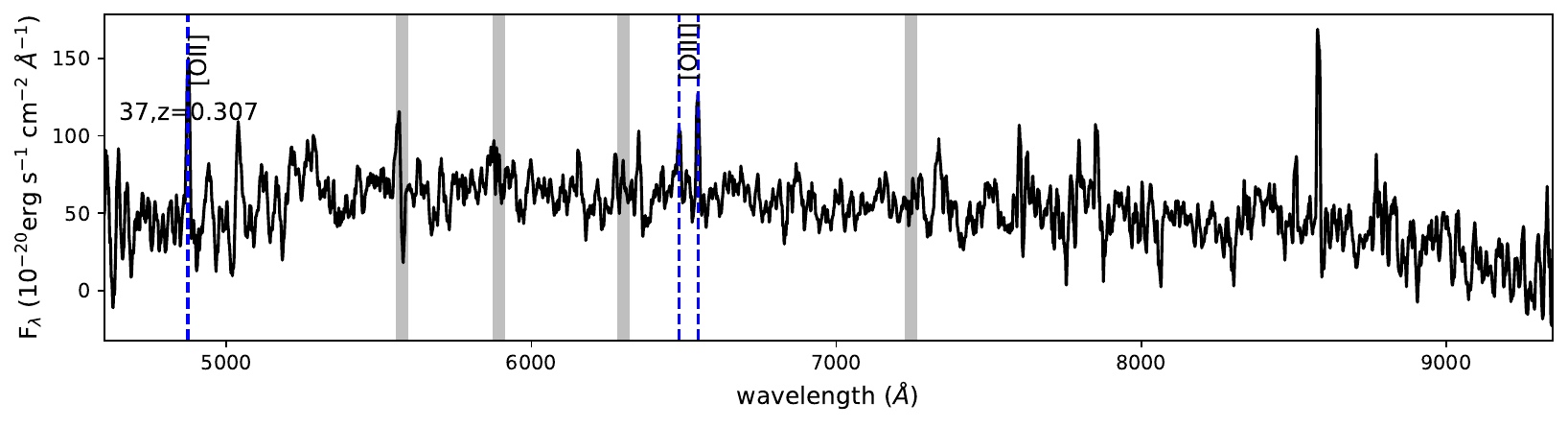}
\includegraphics[width=.85\textwidth]{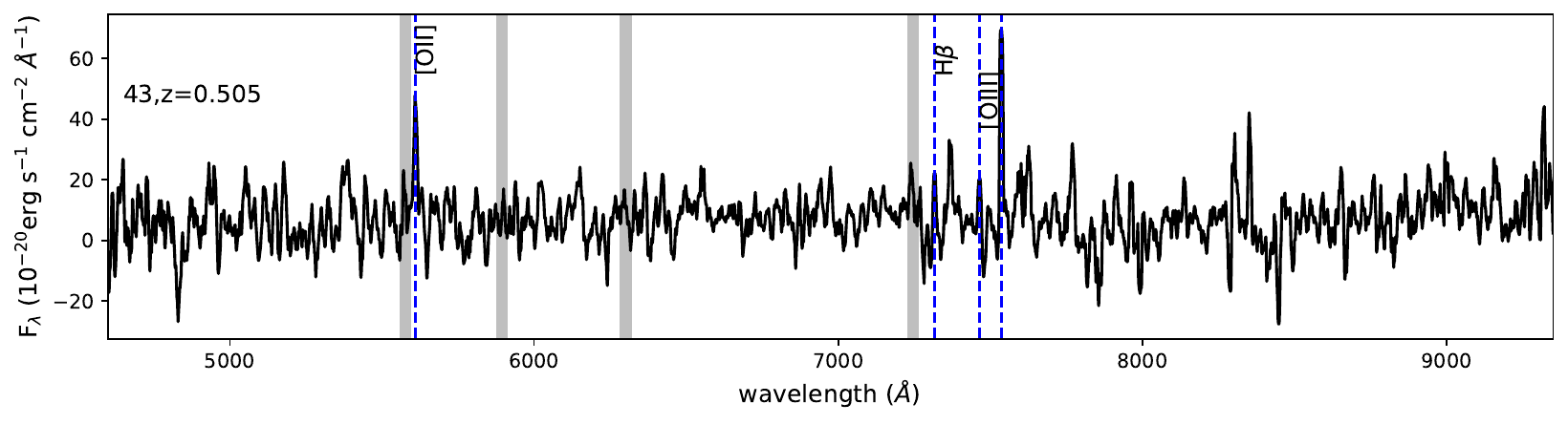}
\includegraphics[width=.85\textwidth]{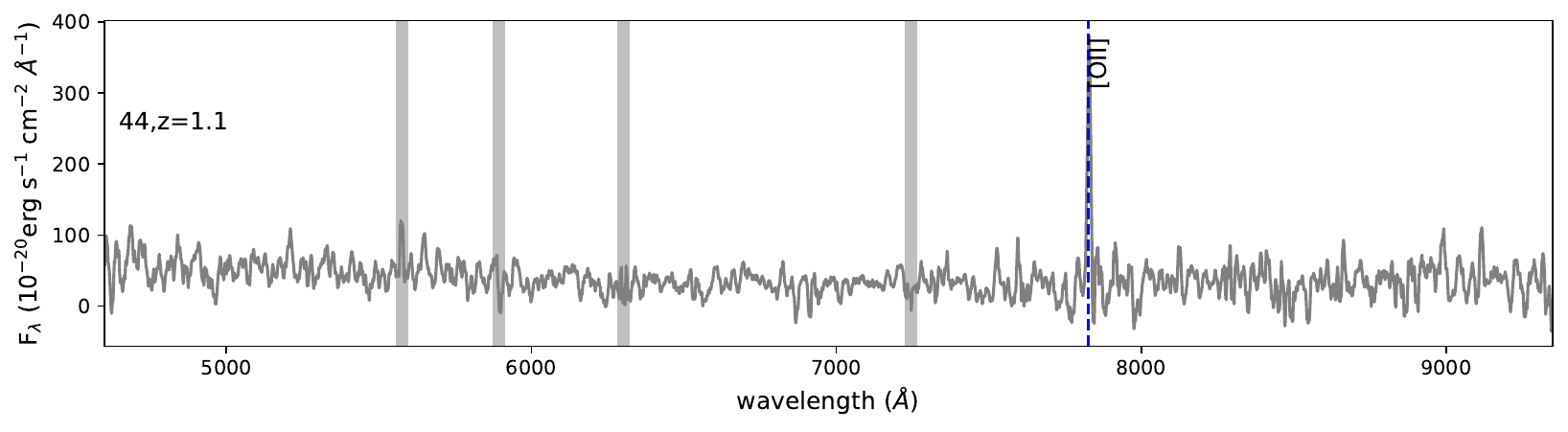}
\caption{Continued}
\end{figure}

\begin{figure}[ht!]
\centering
\includegraphics[width=.85\textwidth]{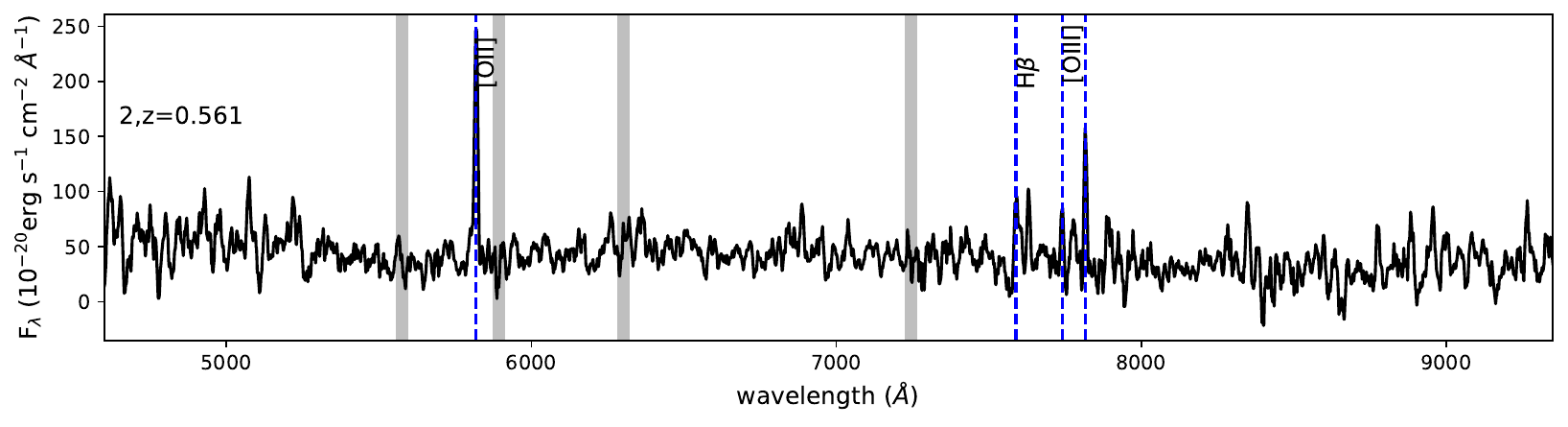}
\includegraphics[width=.85\textwidth]{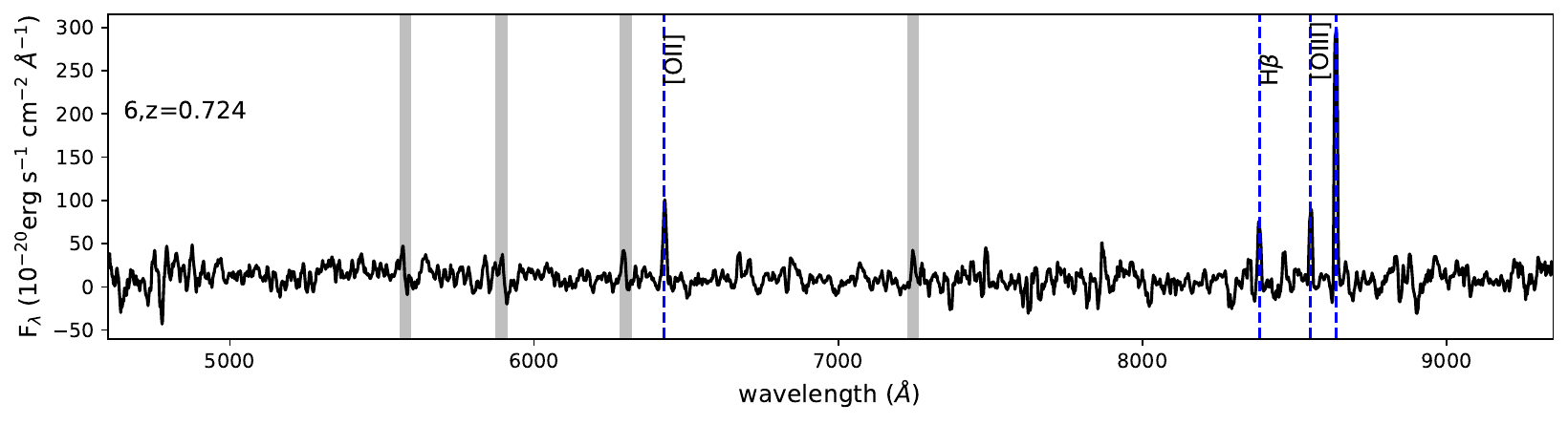}
\includegraphics[width=.85\textwidth]{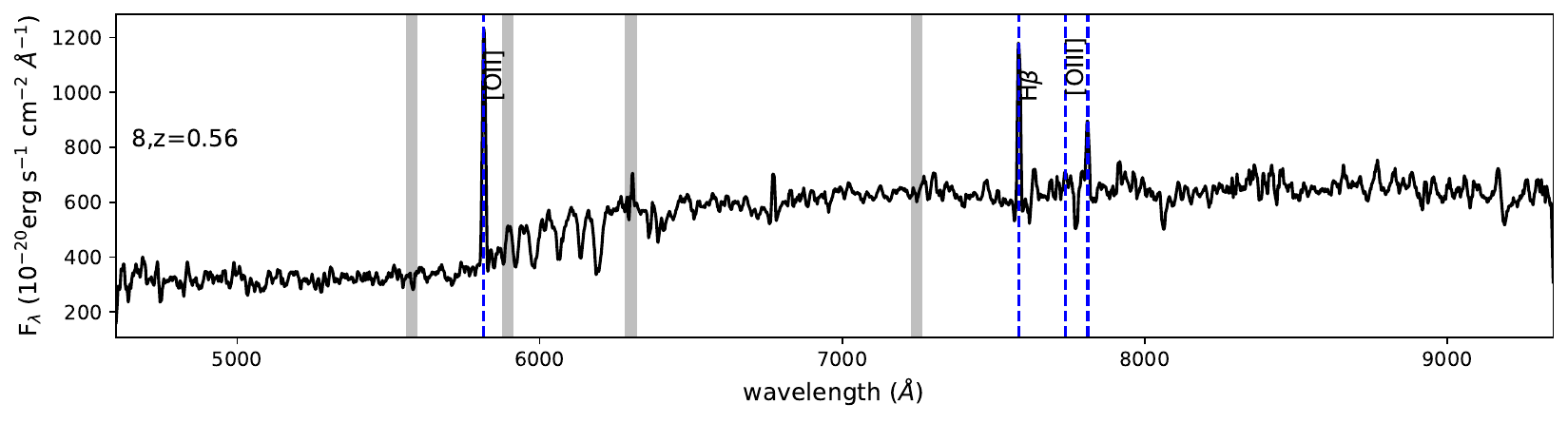}
\includegraphics[width=.85\textwidth]{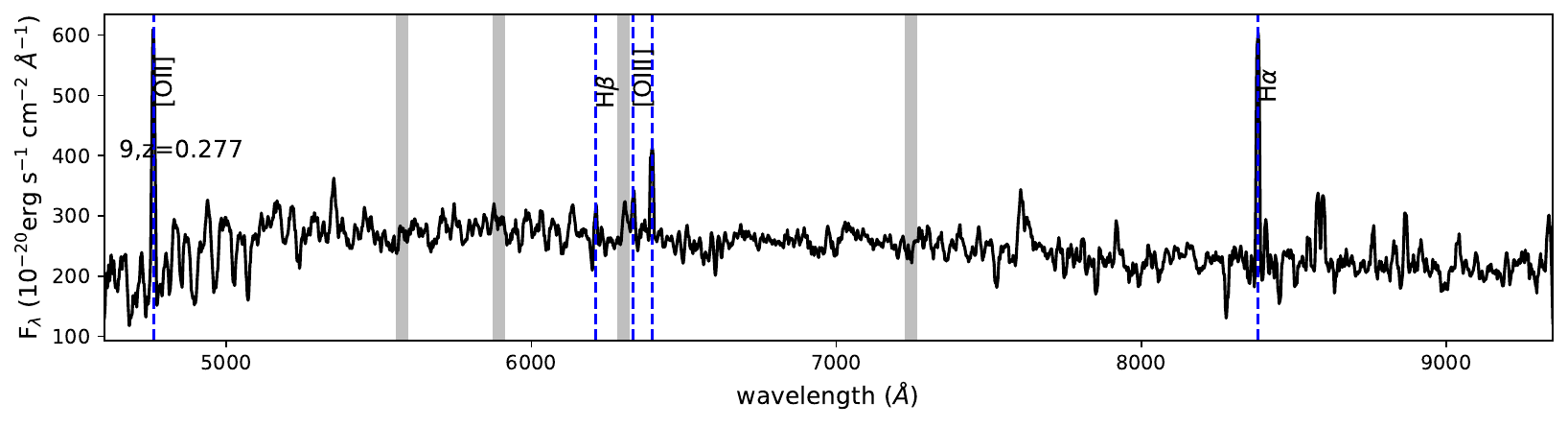}
\includegraphics[width=.85\textwidth]{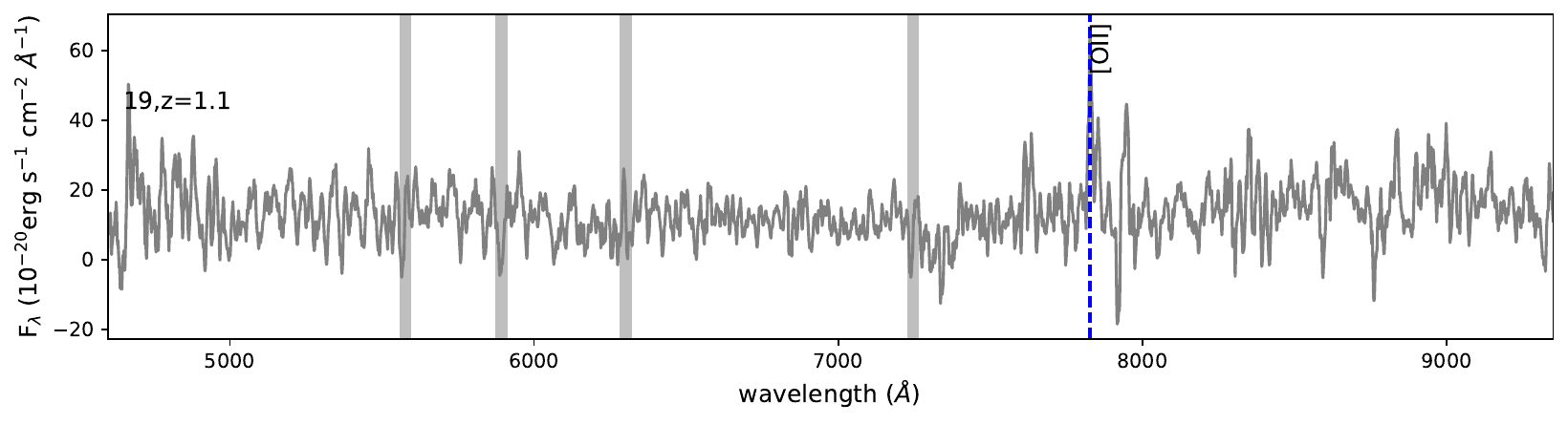}
\caption{Same as Fig.~\ref{apfig:lines_xt1}, but for \XTtwo.}
\label{apfig:lines_xt2}
\end{figure}

\begin{figure}[ht!]
\ContinuedFloat
\centering
\includegraphics[width=.85\textwidth]{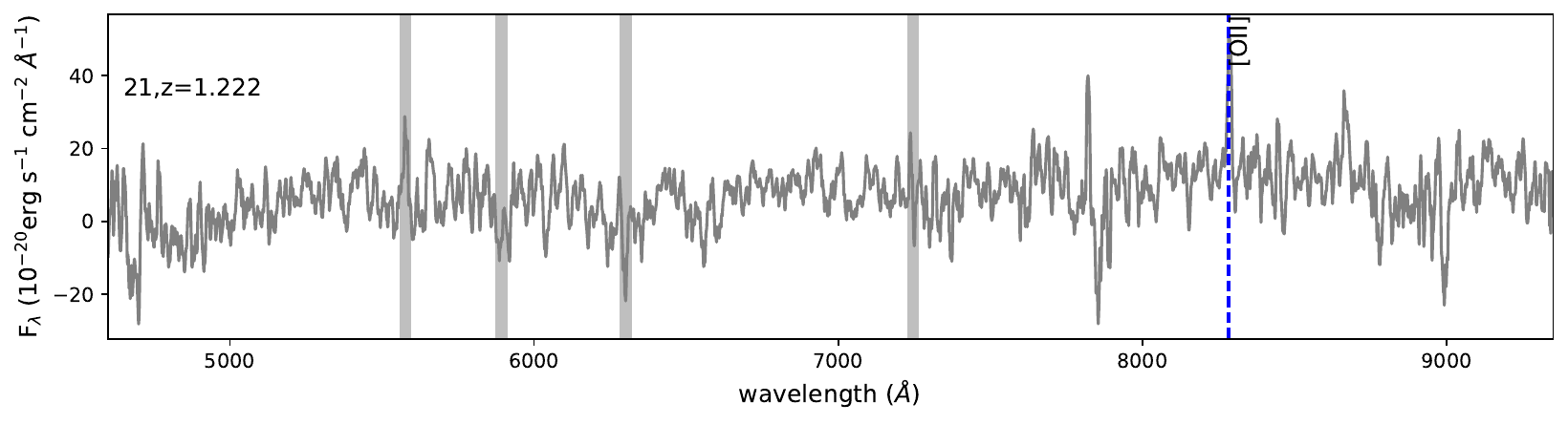}
\includegraphics[width=.85\textwidth]{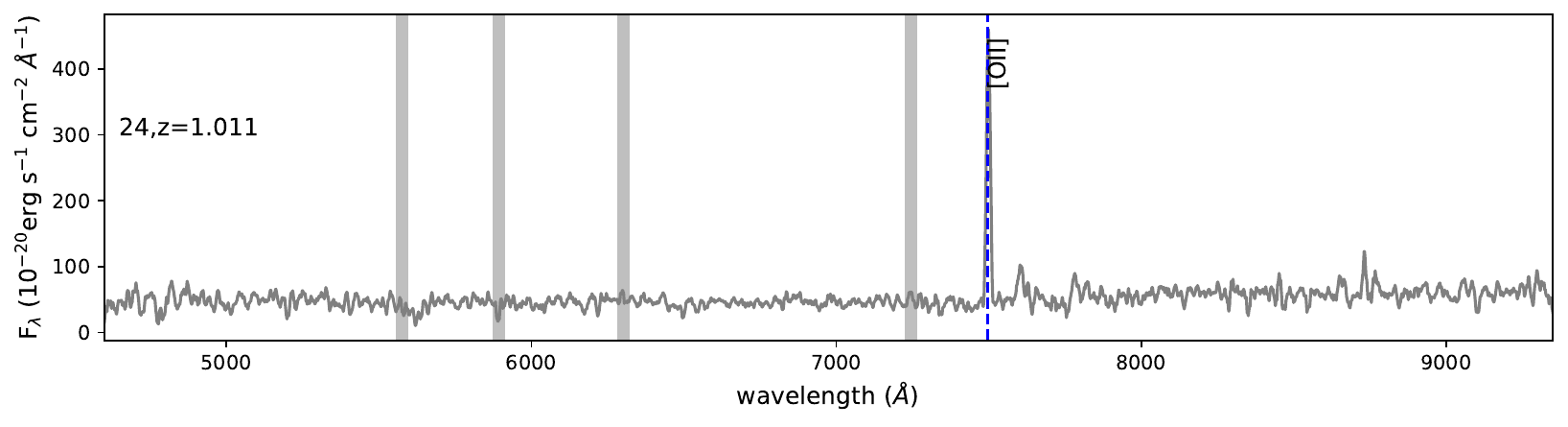}
\includegraphics[width=.85\textwidth]{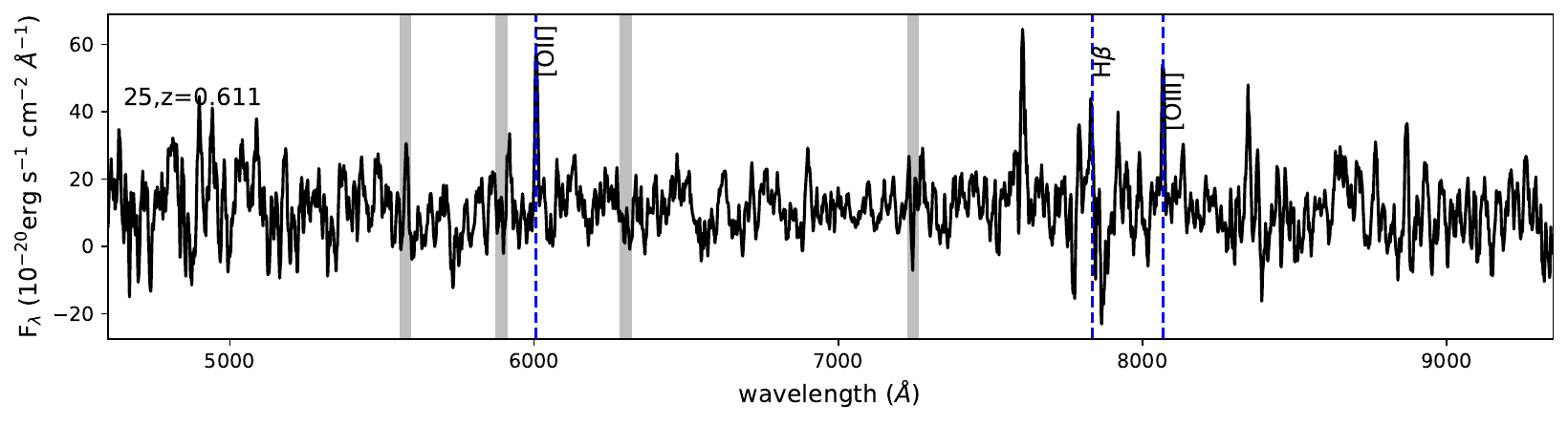}
\includegraphics[width=.85\textwidth]{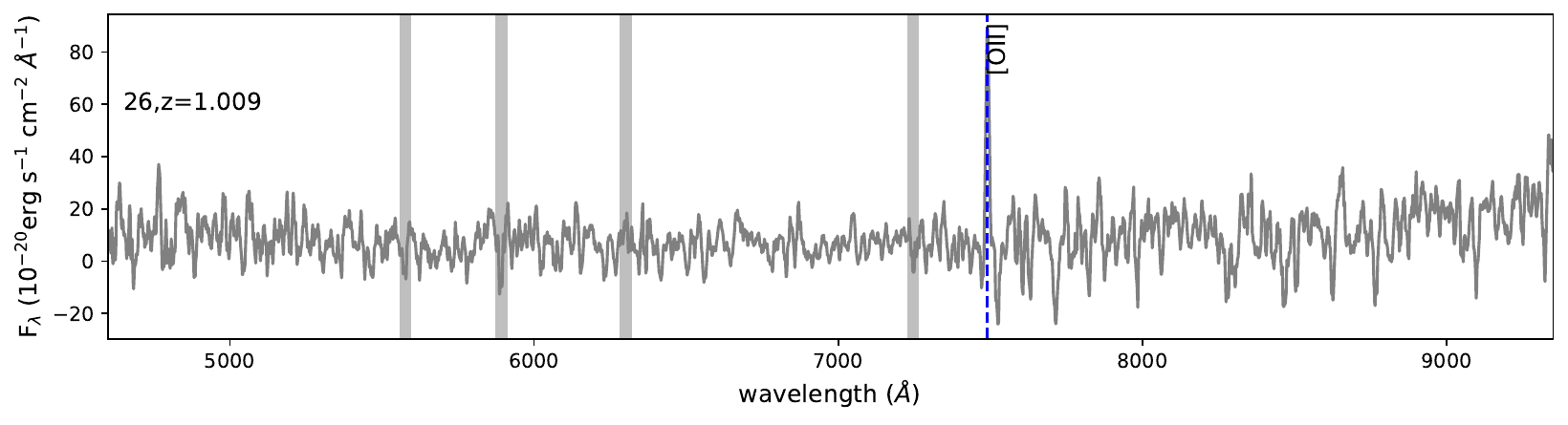}
\includegraphics[width=.85\textwidth]{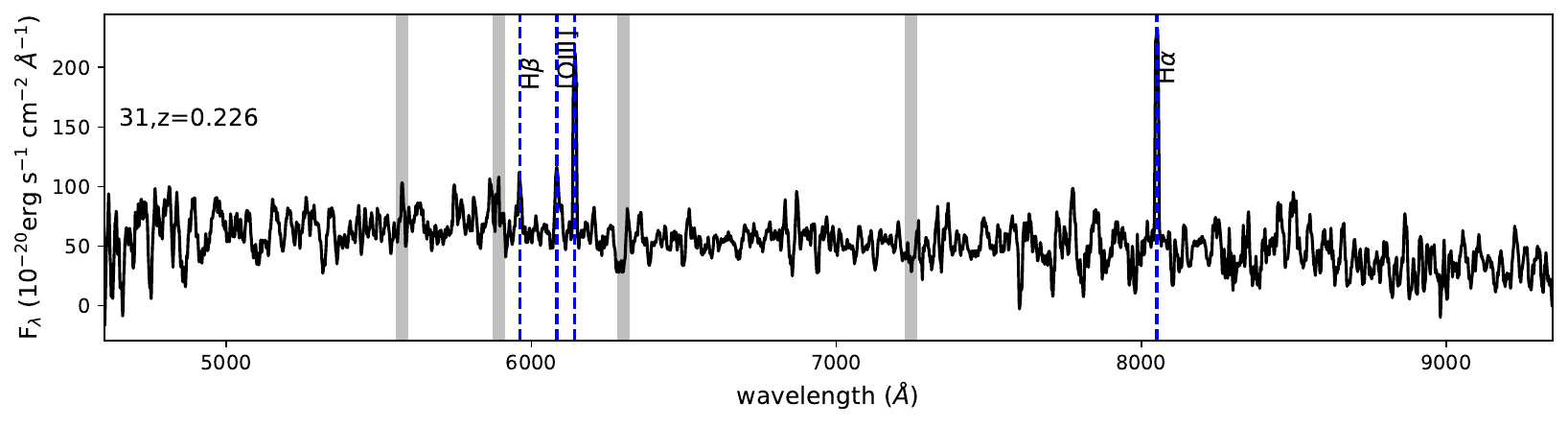}
\includegraphics[width=.85\textwidth]{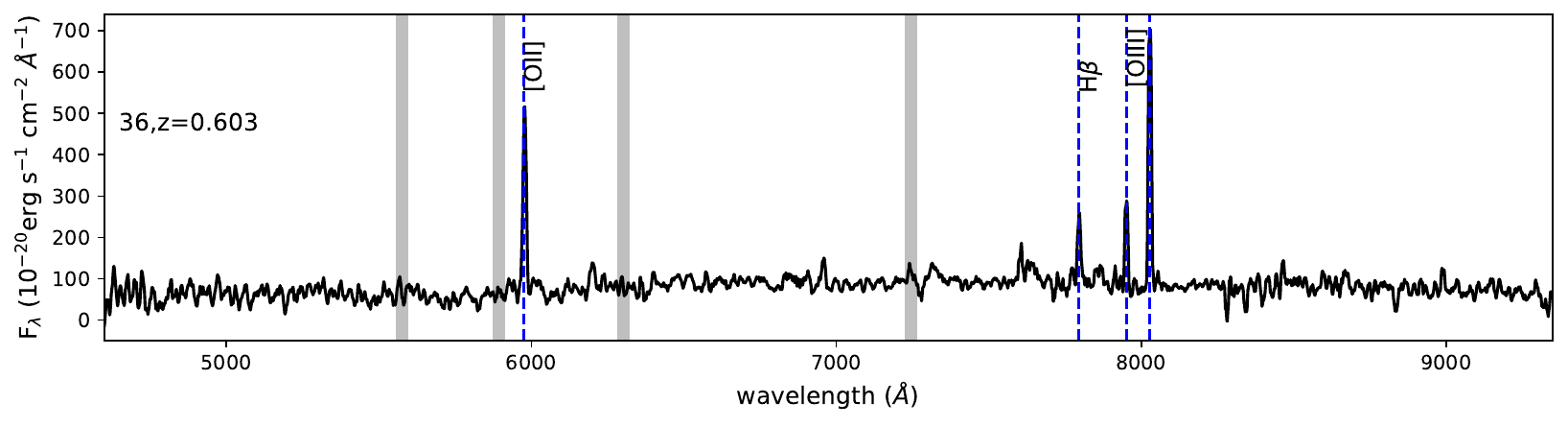}
\caption{Continued}
\end{figure}

\begin{figure}[ht!]
\ContinuedFloat
\centering
\includegraphics[width=.85\textwidth]{Figures/EM_lines/XRT100831_source36_lines.pdf}
\includegraphics[width=.85\textwidth]{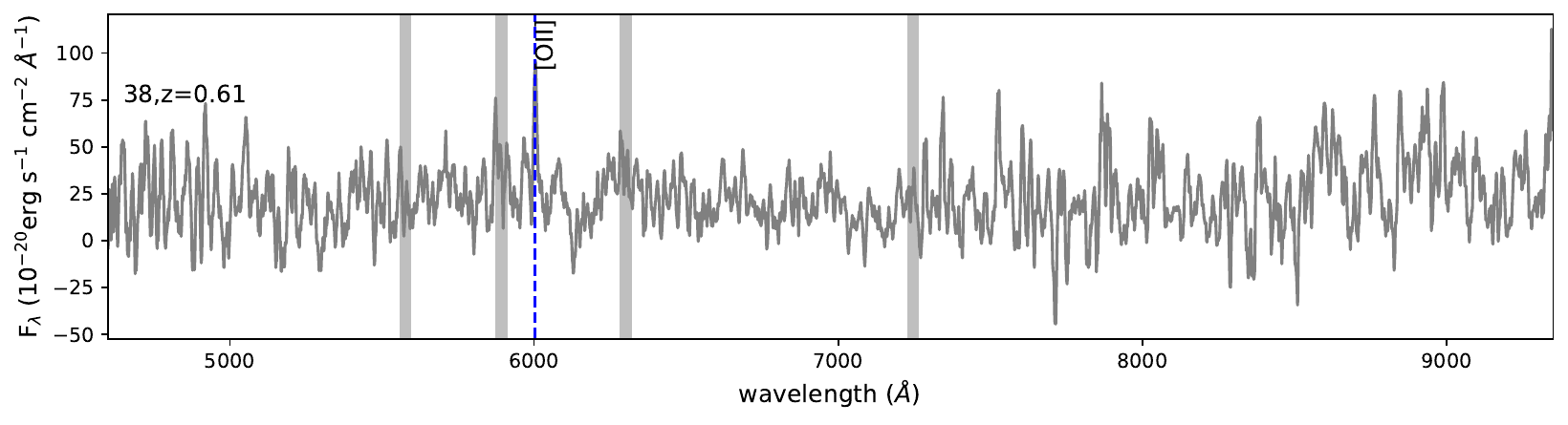}
\includegraphics[width=.85\textwidth]{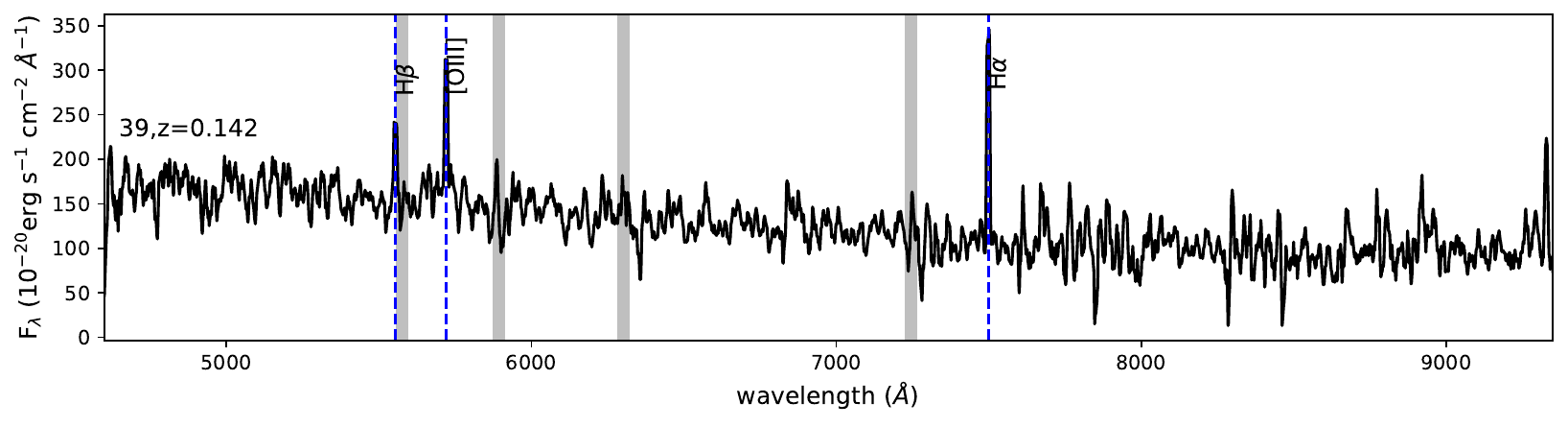}
\includegraphics[width=.85\textwidth]{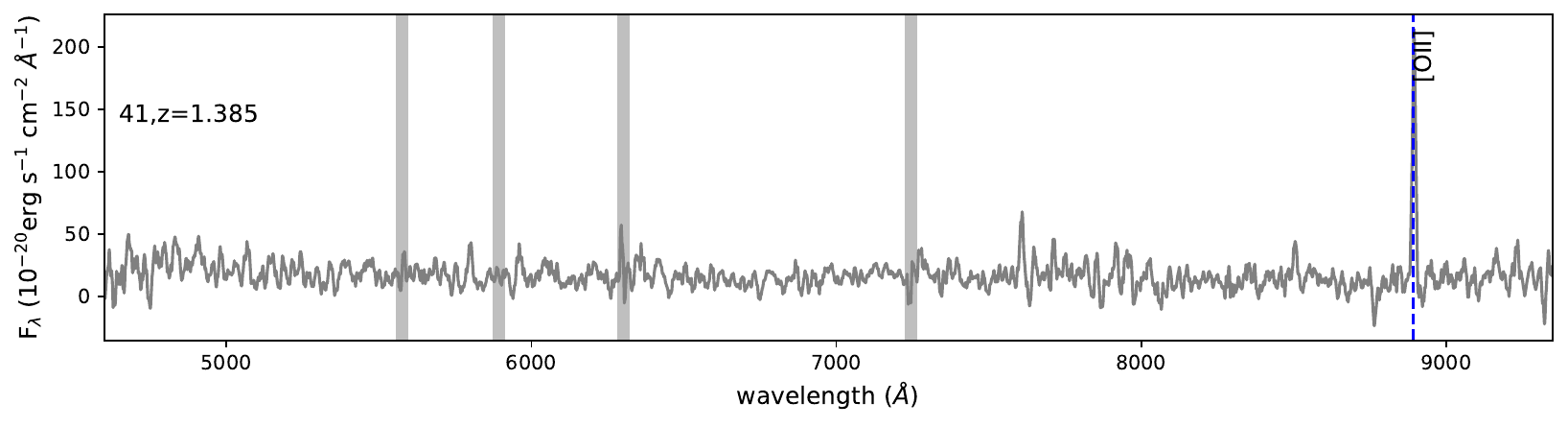}
\includegraphics[width=.85\textwidth]{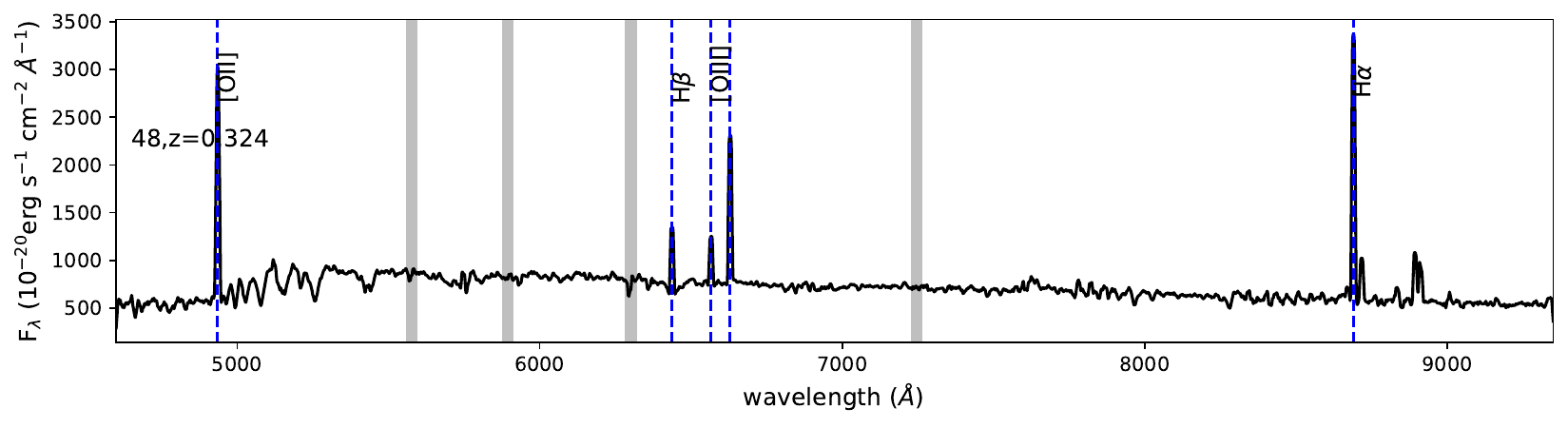}
\includegraphics[width=.85\textwidth]{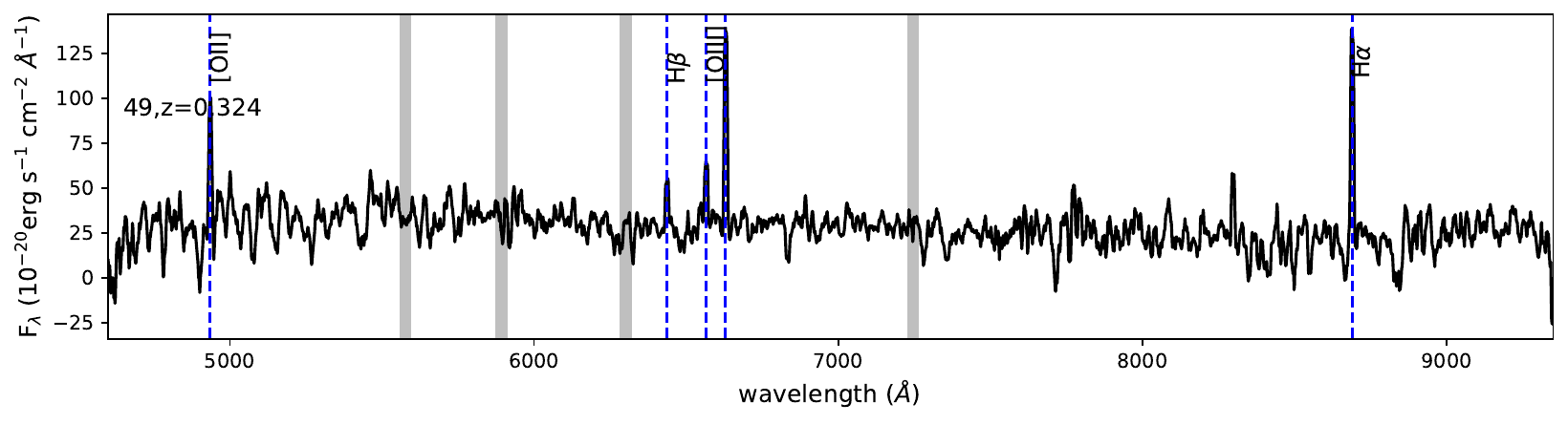}
\caption{Continued}
\end{figure}

\begin{figure}[ht!]
\centering
\includegraphics[width=.85\textwidth]{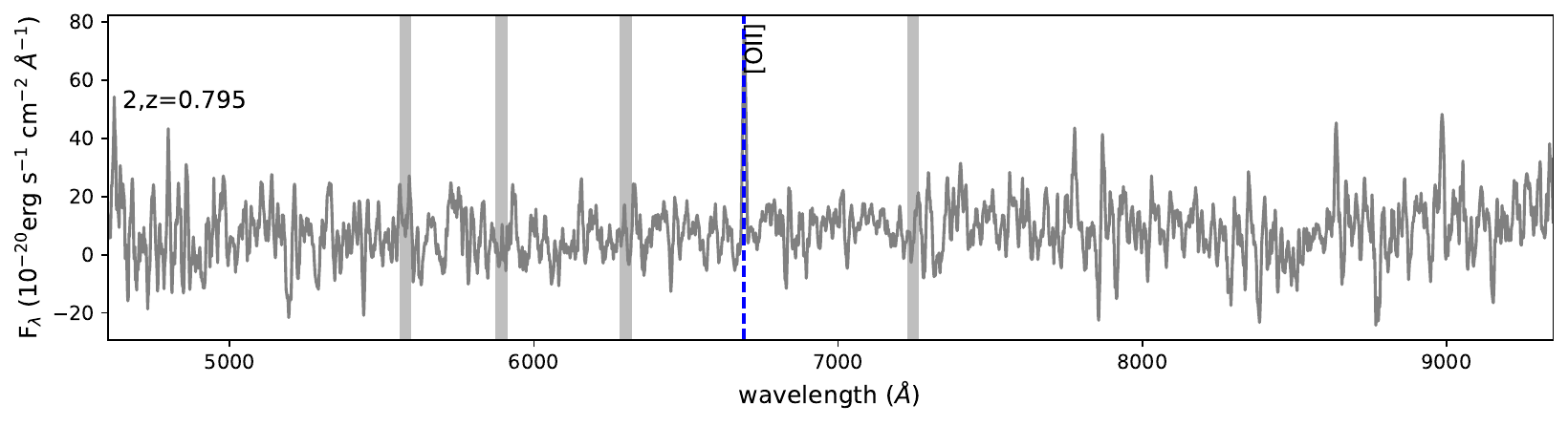}
\includegraphics[width=.85\textwidth]{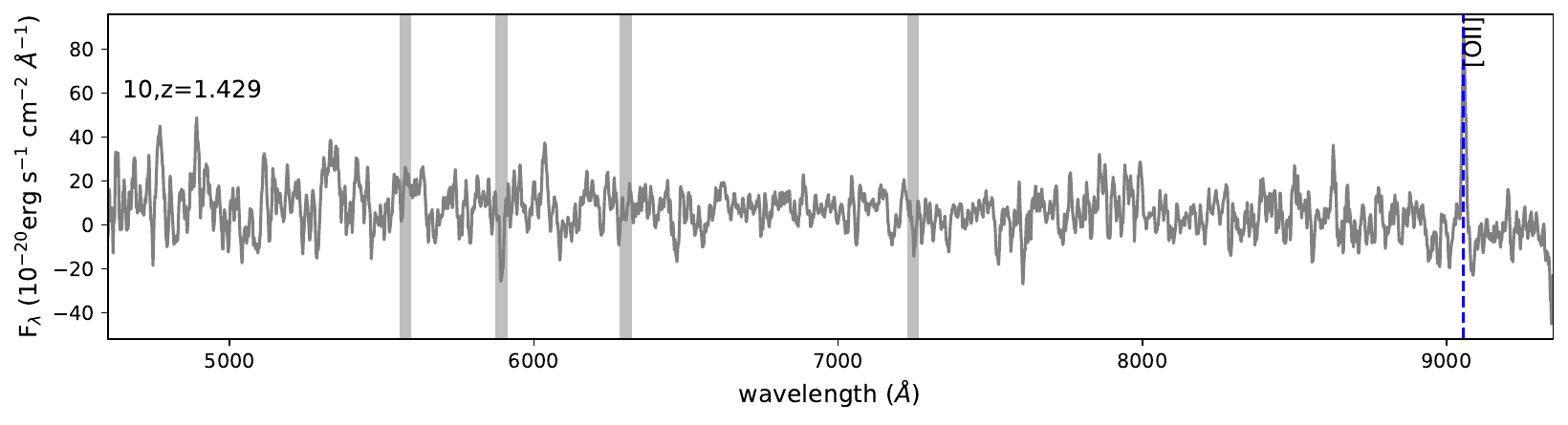}
\includegraphics[width=.85\textwidth]{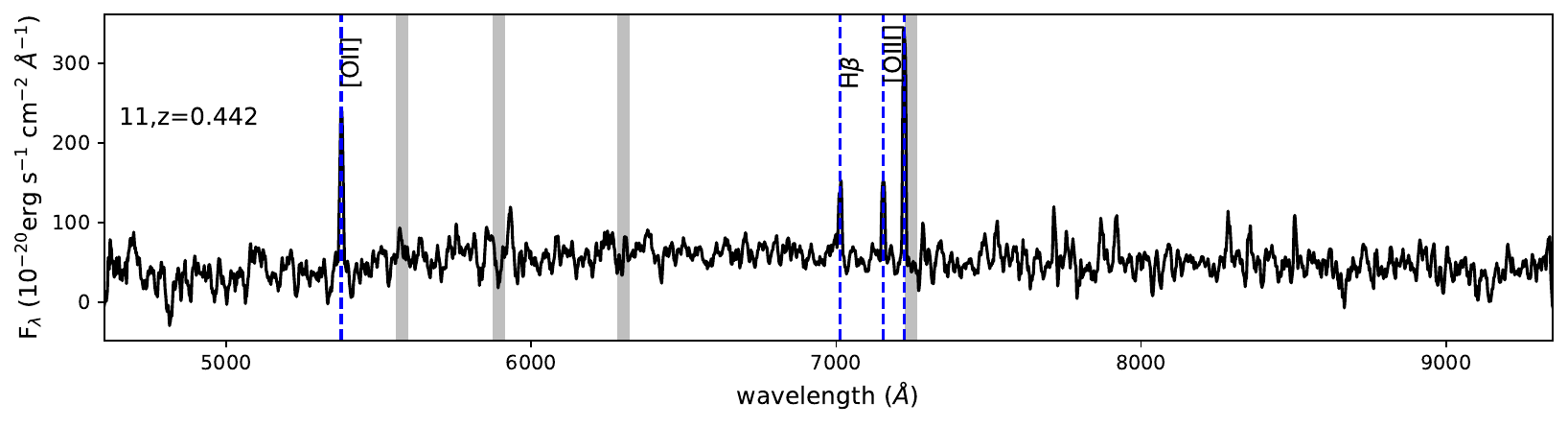}
\includegraphics[width=.85\textwidth]{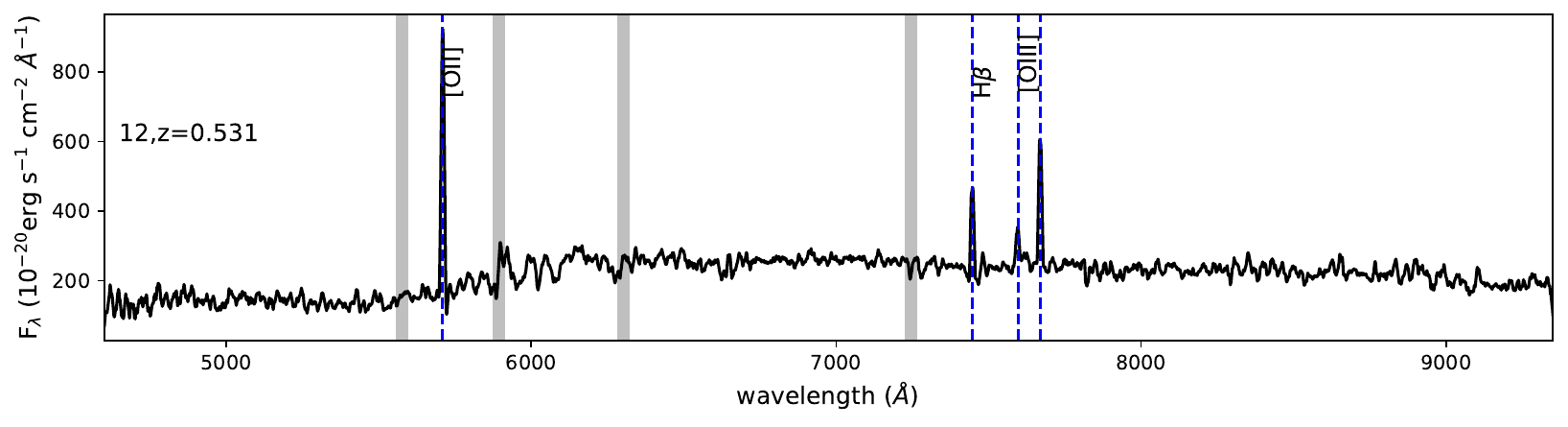}
\includegraphics[width=.85\textwidth]{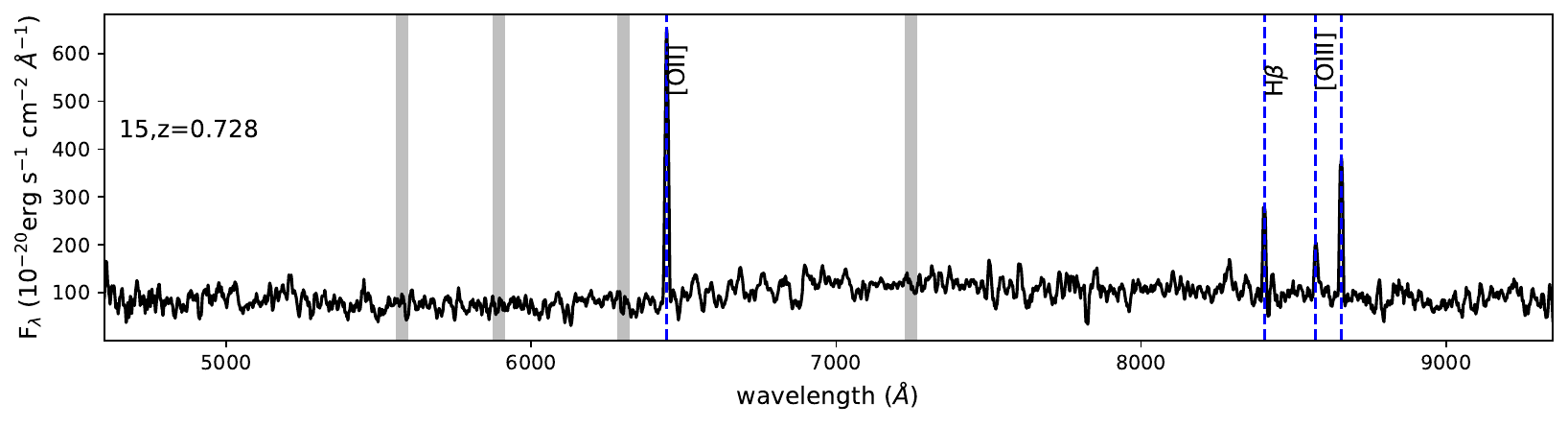}
\caption{Same as Fig.~\ref{apfig:lines_xt1}, but for \XTthree.}
\label{apfig:lines_xt3}
\end{figure}

\begin{figure}[ht!]
\ContinuedFloat
\centering
\includegraphics[width=.85\textwidth]{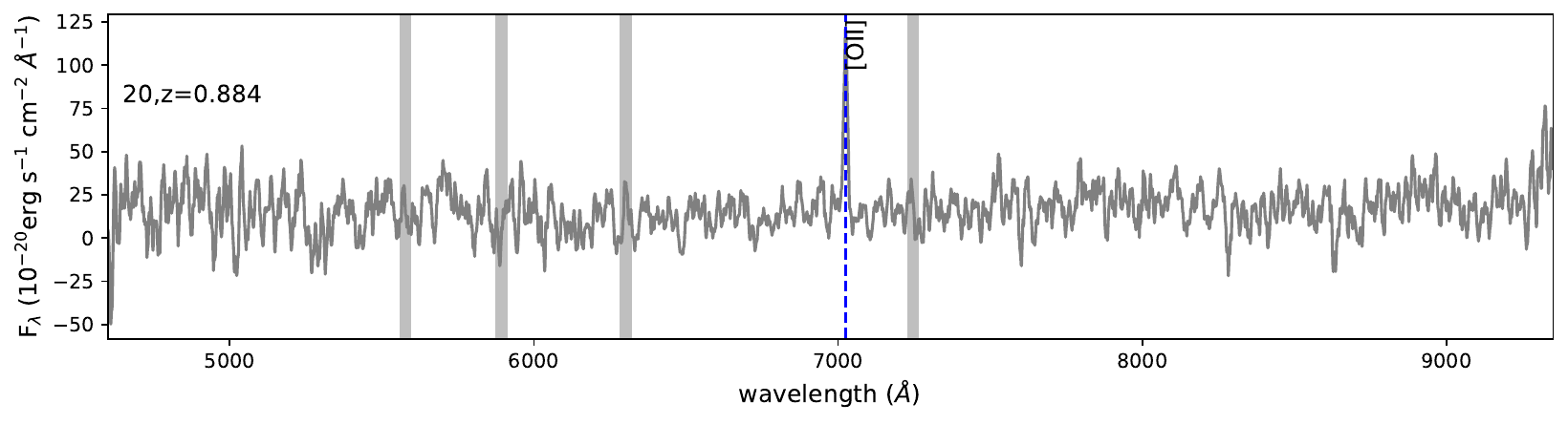}
\includegraphics[width=.85\textwidth]{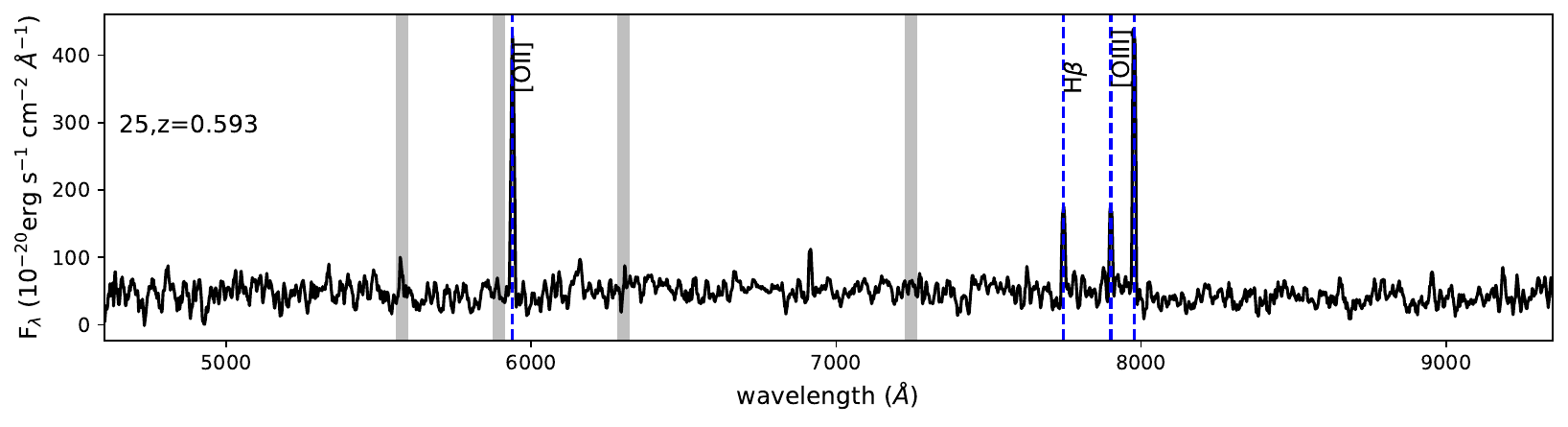}
\includegraphics[width=.85\textwidth]{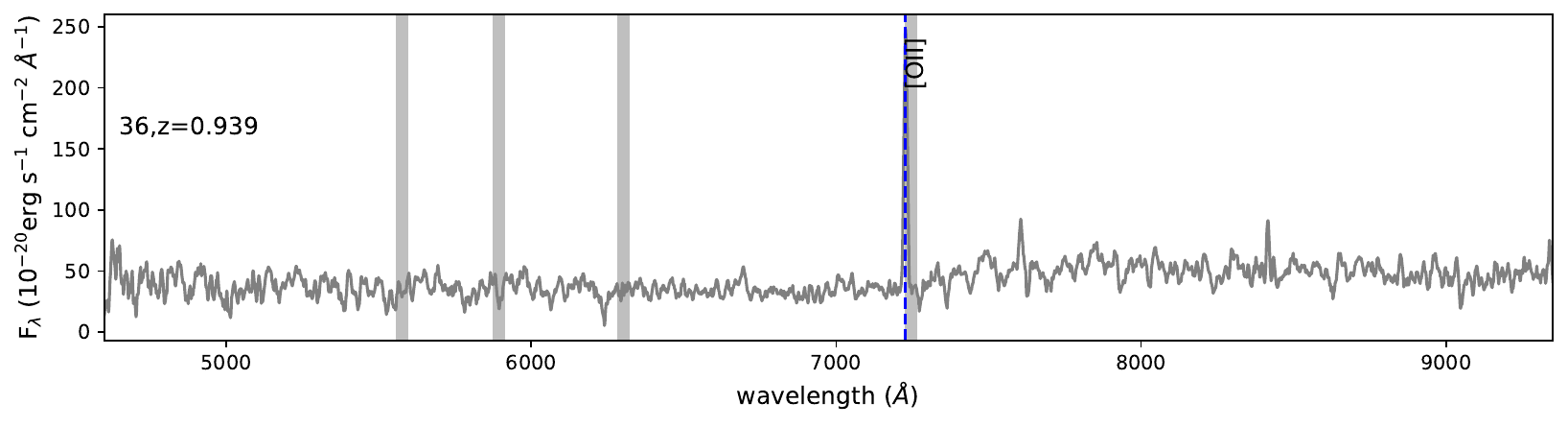}
\includegraphics[width=.85\textwidth]{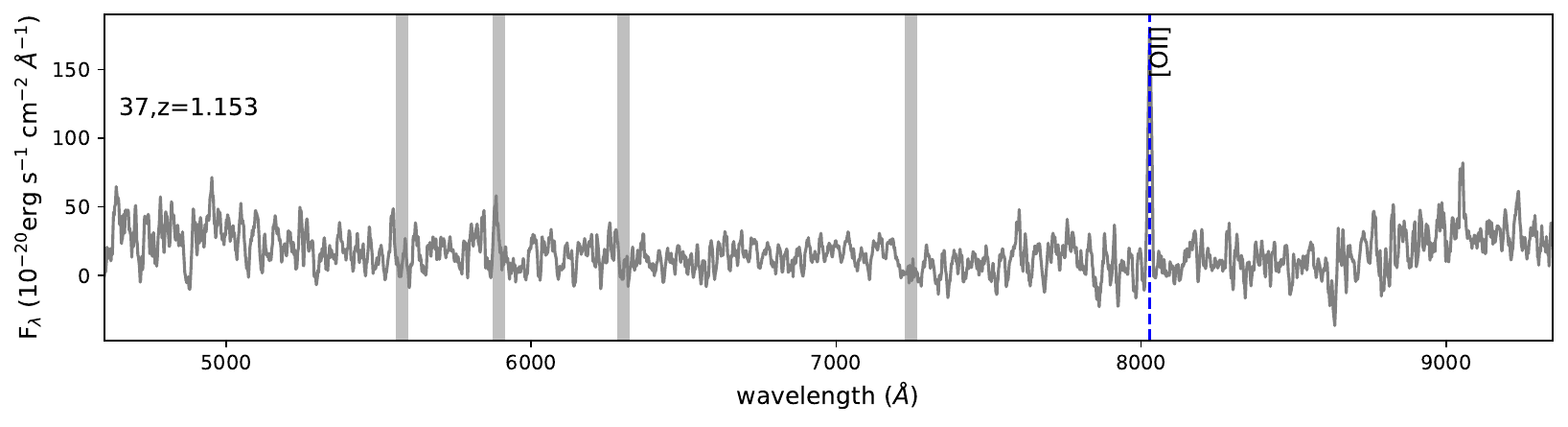}
\includegraphics[width=.85\textwidth]{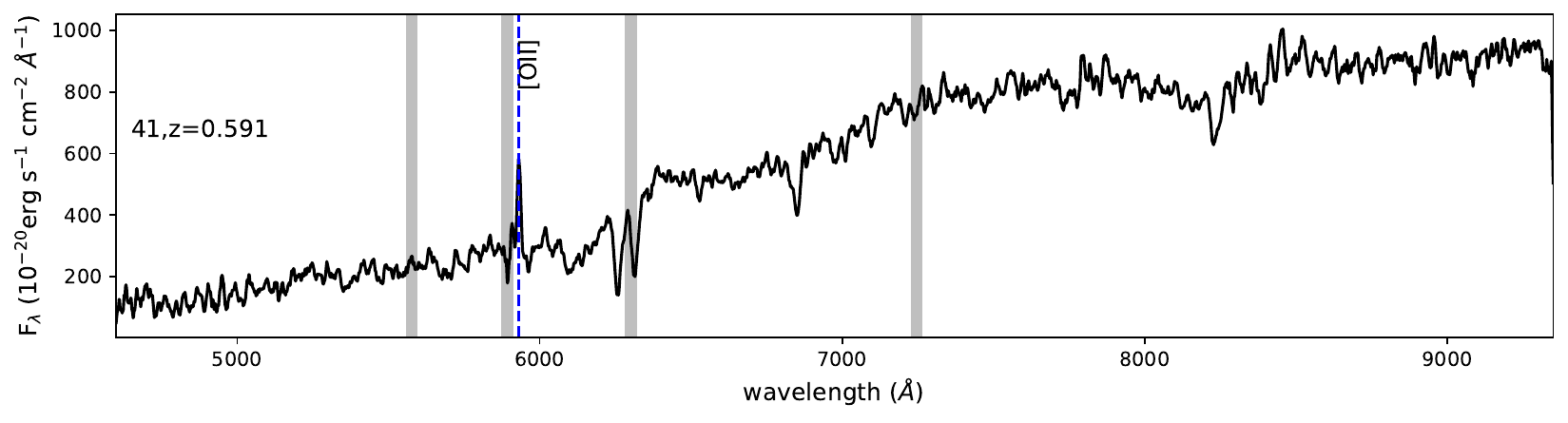}
\caption{Continued}
\end{figure}

\begin{figure}[ht!]
\ContinuedFloat
\centering
\includegraphics[width=.85\textwidth]{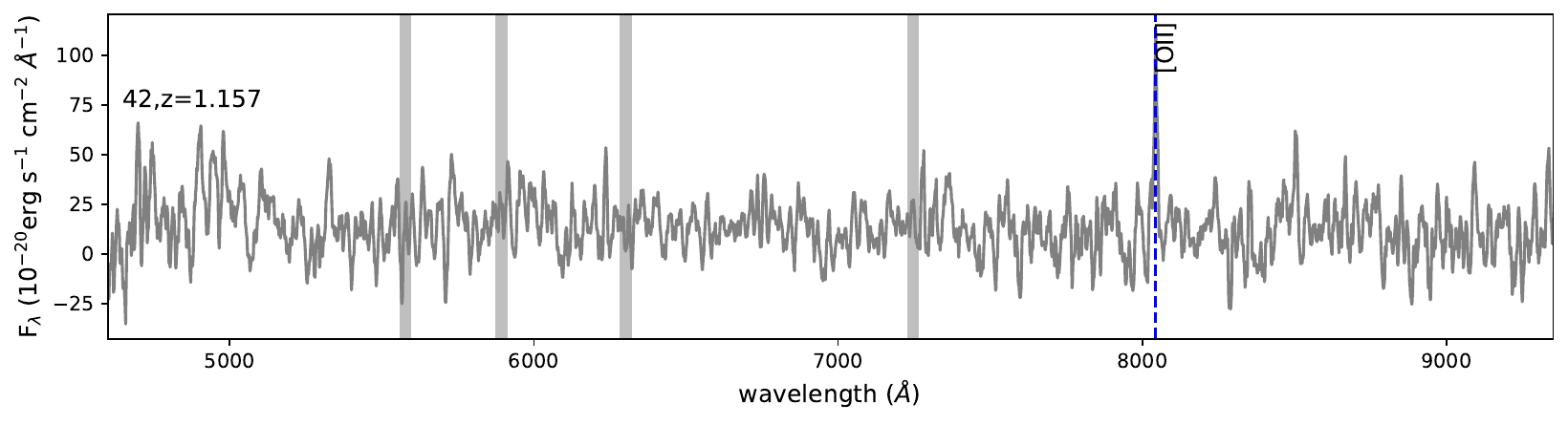}
\includegraphics[width=.85\textwidth]{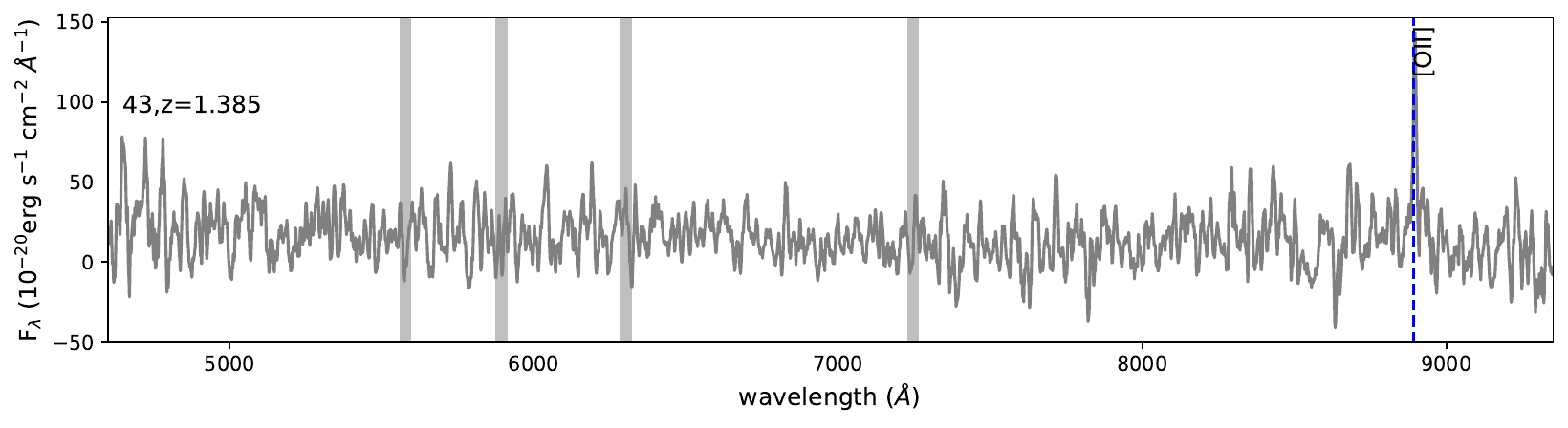}
\includegraphics[width=.85\textwidth]{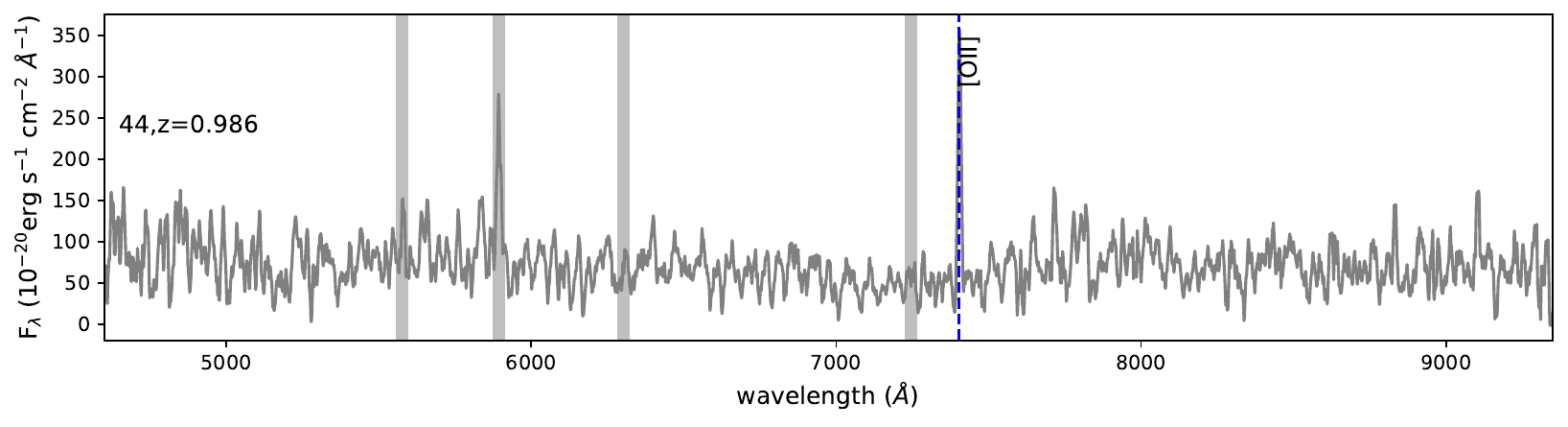}
\includegraphics[width=.85\textwidth]{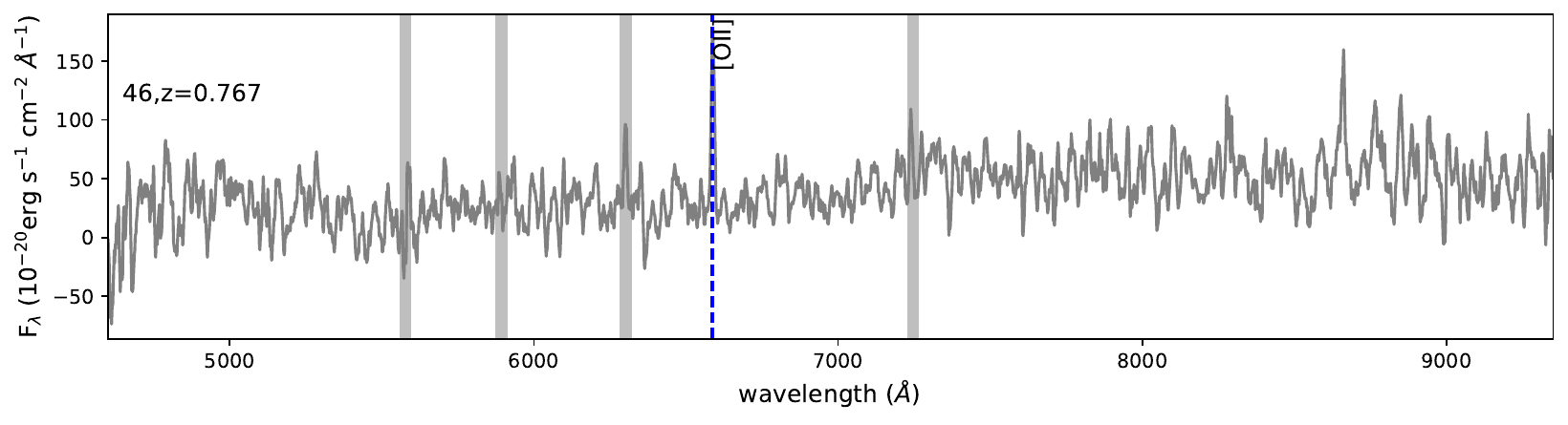}
\includegraphics[width=.85\textwidth]{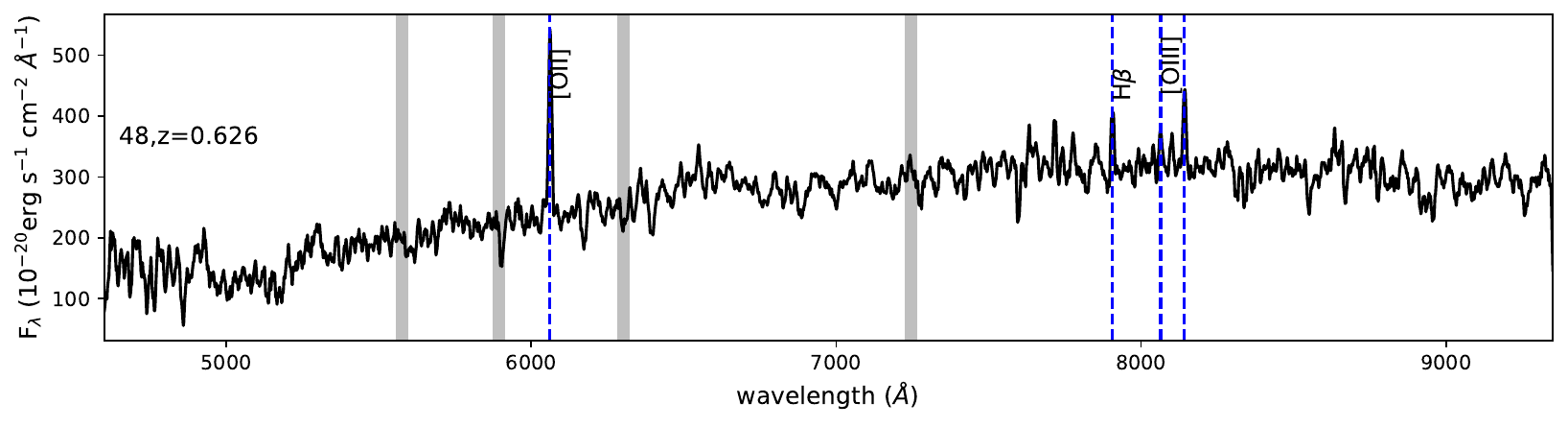}
\caption{Continued. {The line in the spectrum of source 44 coinciding with a grey band indicating a prominent sky line is not at the correct wavelength to be H$\beta$ if the emission line marked with [OII] is actually H$\alpha$. We therefore deem it likely that the emission line is the [OII] doublet.}}
\end{figure}

\begin{figure}[ht!]
\ContinuedFloat
\centering
\includegraphics[width=.85\textwidth]{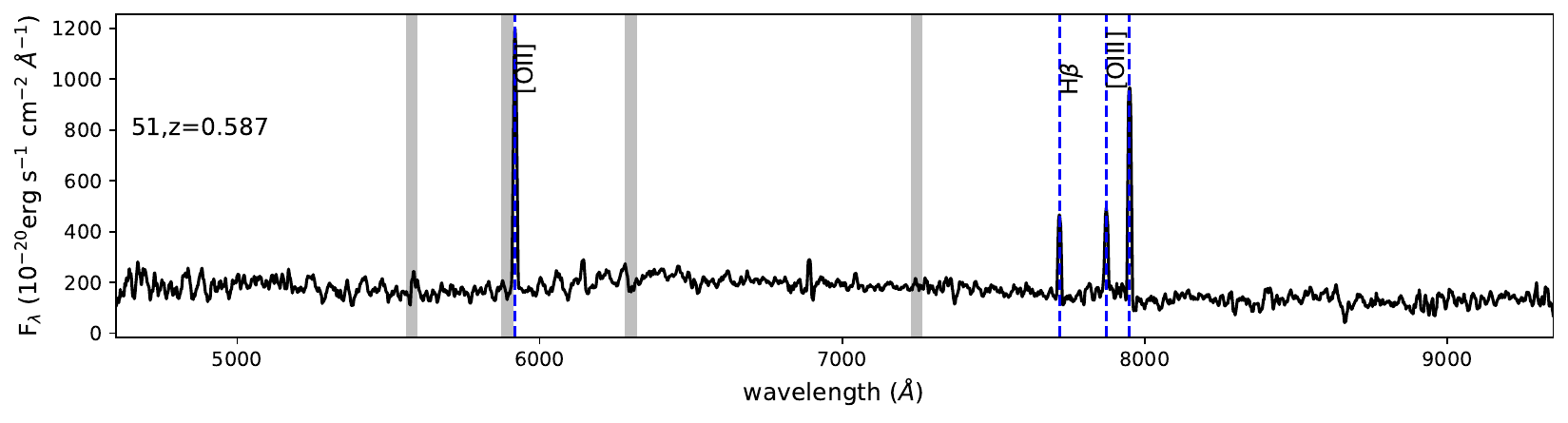}
\includegraphics[width=.85\textwidth]{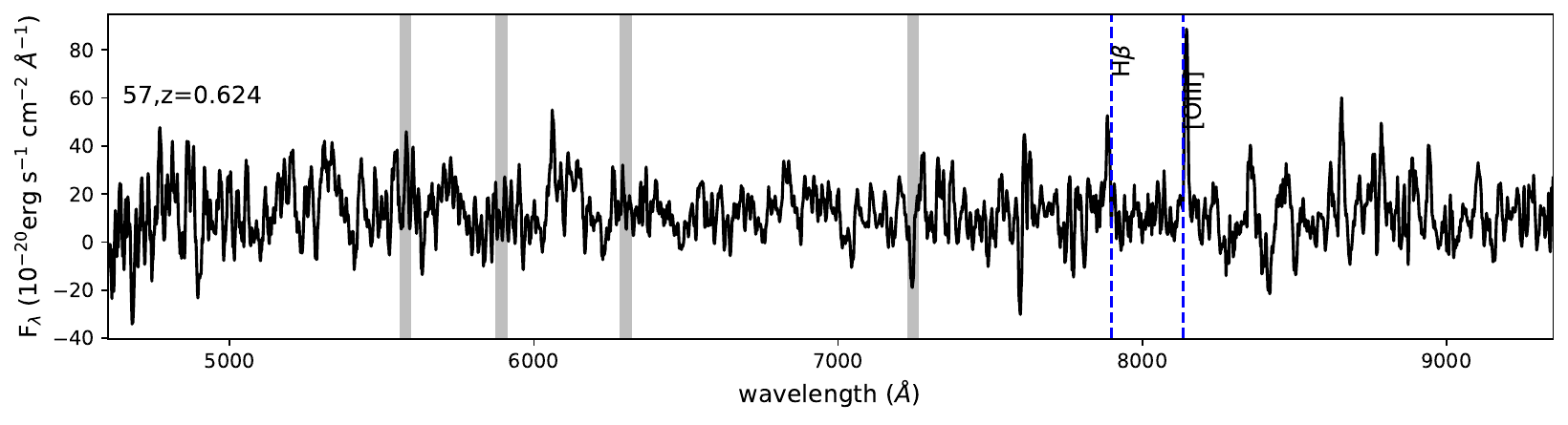}
\includegraphics[width=.85\textwidth]{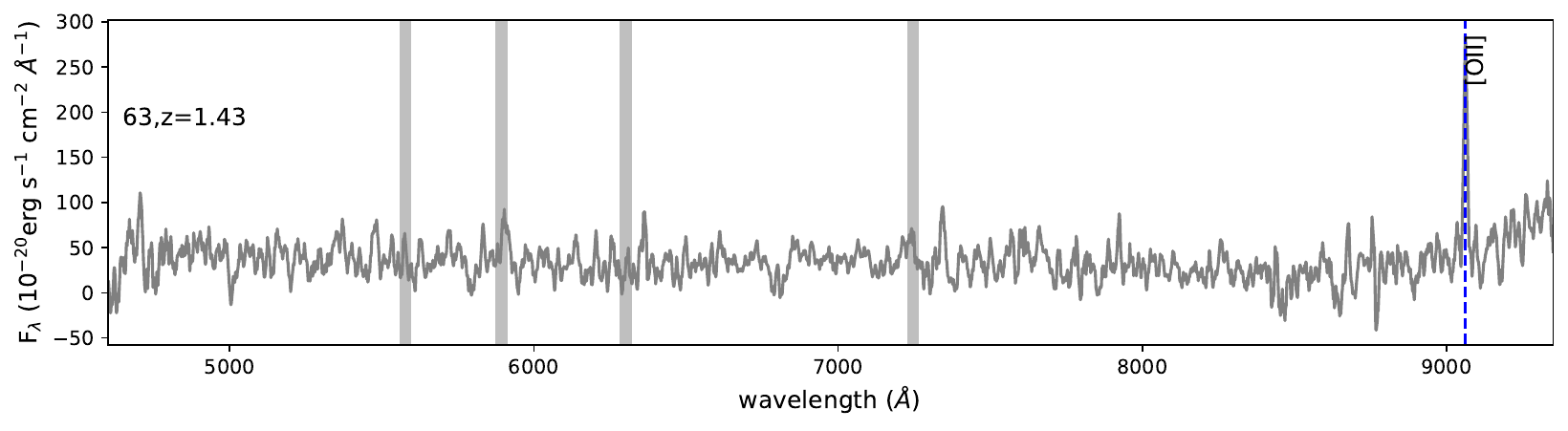}
\caption{Continued}
\end{figure}

\begin{figure}[ht!]
\centering
\includegraphics[width=.85\textwidth]{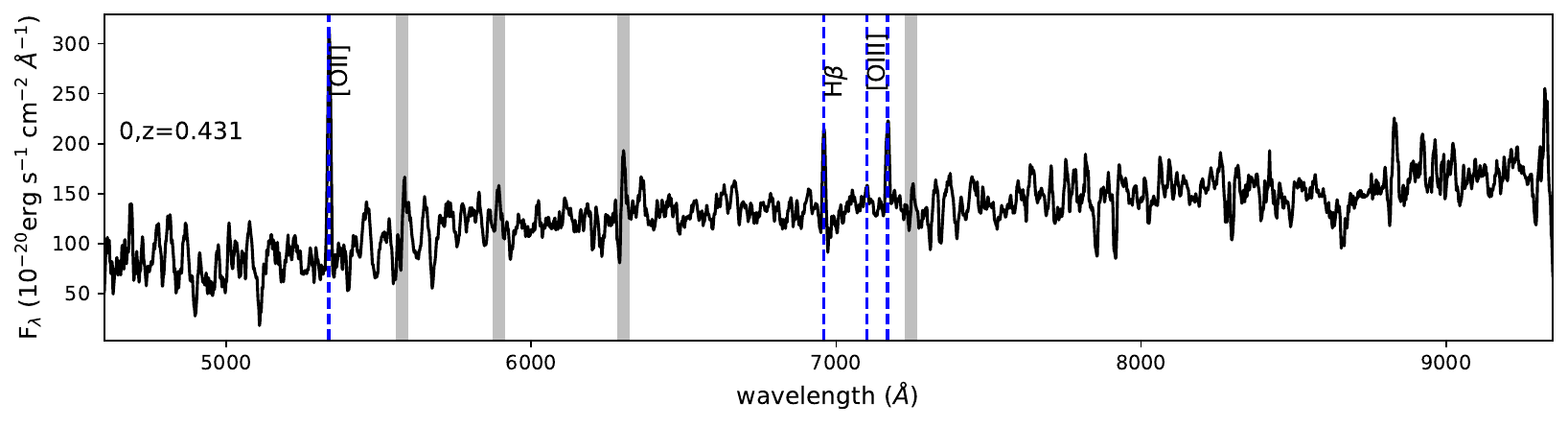}
\includegraphics[width=.85\textwidth]{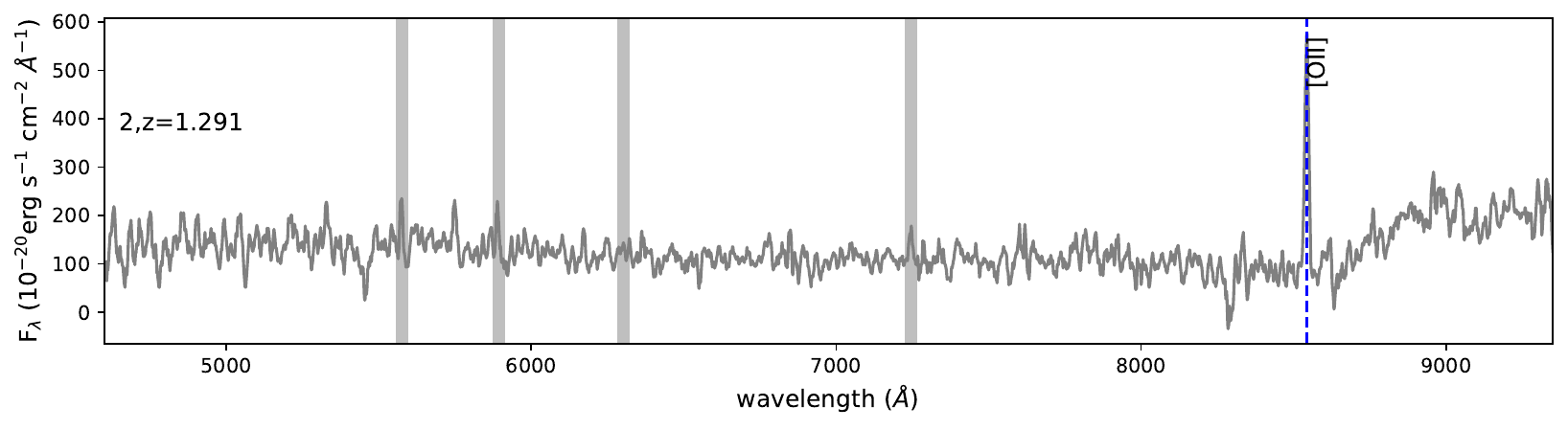}
\includegraphics[width=.85\textwidth]{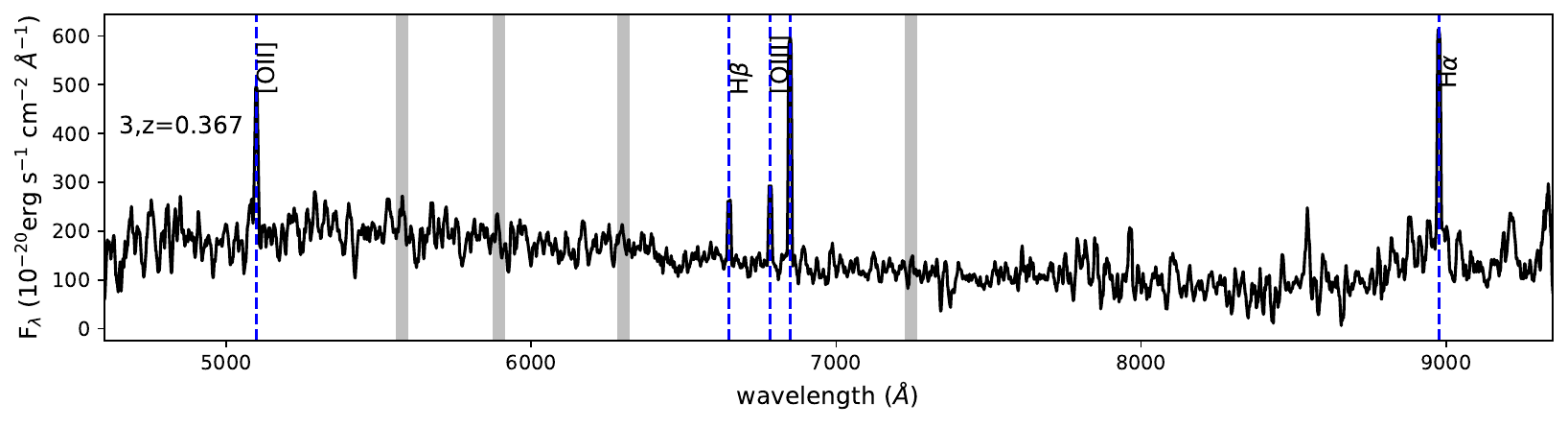}
\includegraphics[width=.85\textwidth]{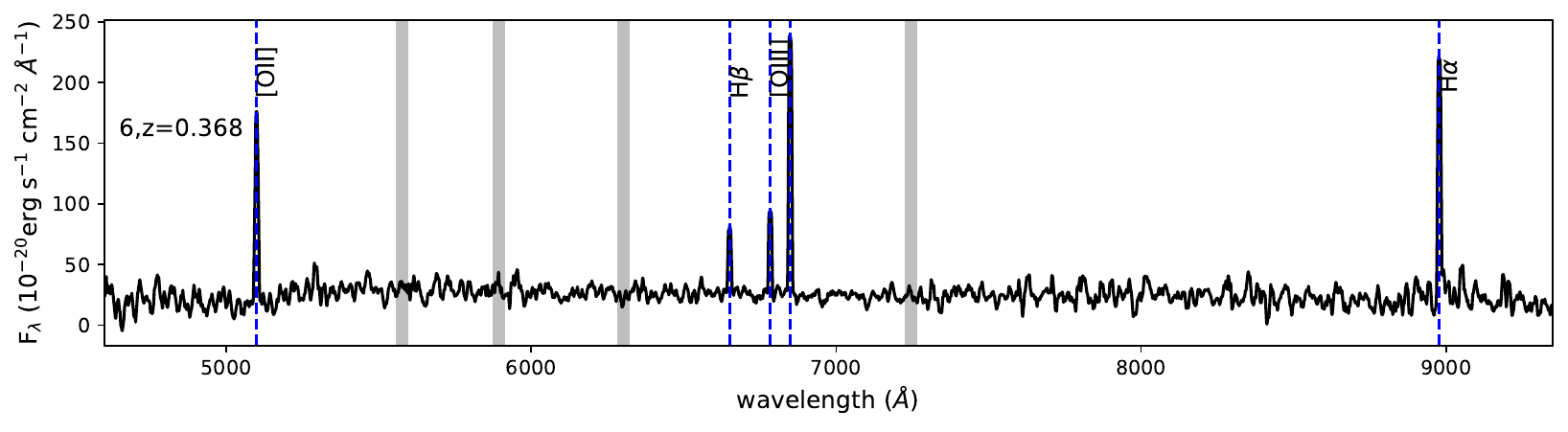}
\includegraphics[width=.85\textwidth]{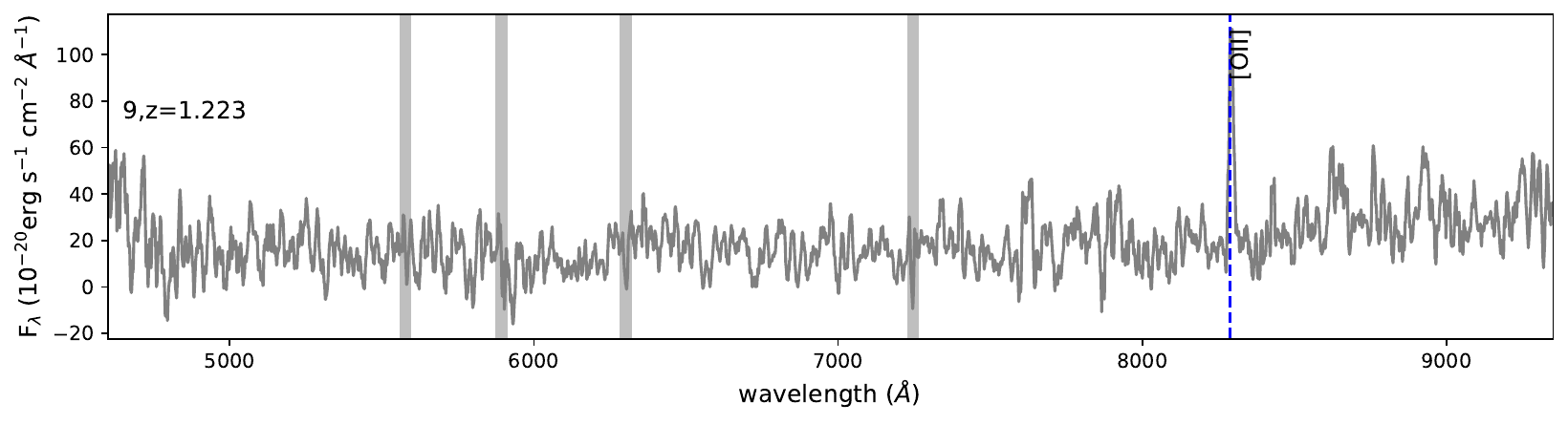}
\caption{Same as Fig.~\ref{apfig:lines_xt1}, but for \XTfour.}
\label{apfig:lines_xt4}
\end{figure}

\begin{figure}[ht!]
\ContinuedFloat
\centering
\includegraphics[width=.85\textwidth]{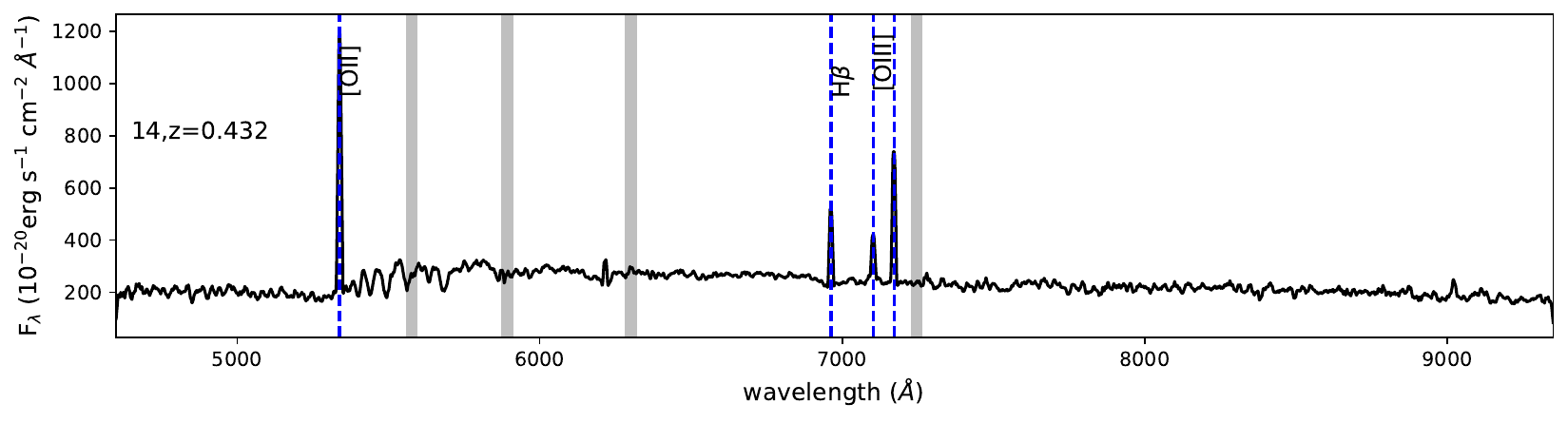}
\includegraphics[width=.85\textwidth]{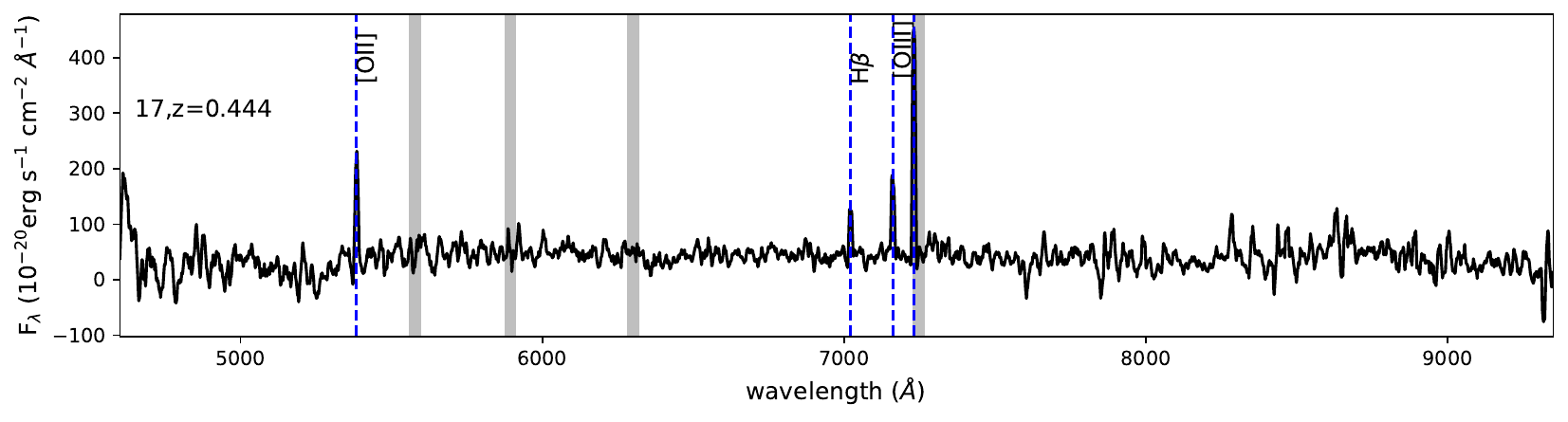}
\includegraphics[width=.85\textwidth]{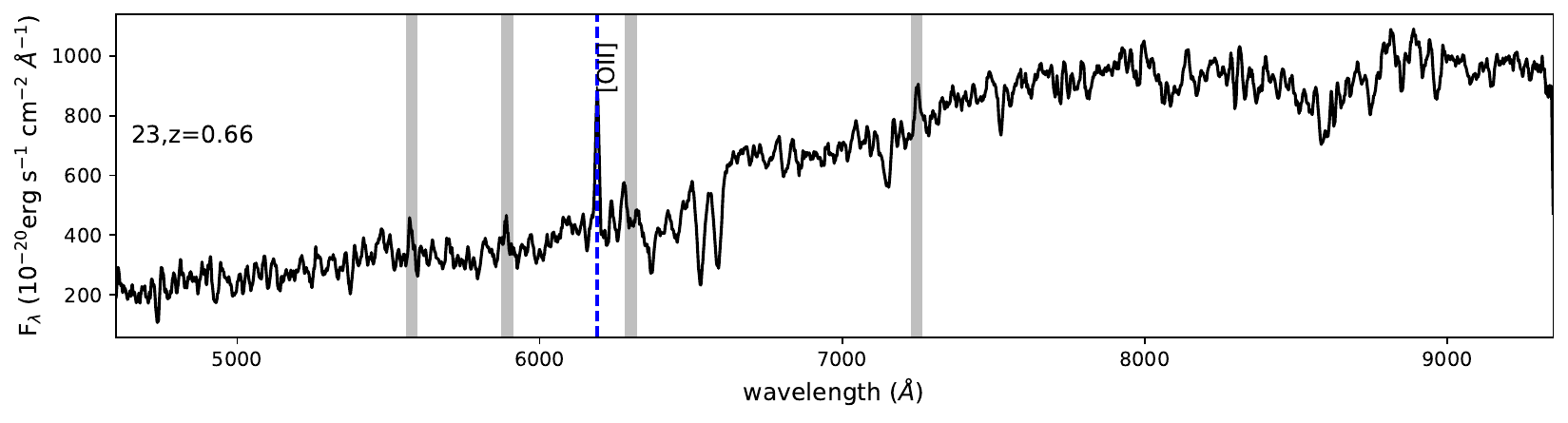}
\includegraphics[width=.85\textwidth]{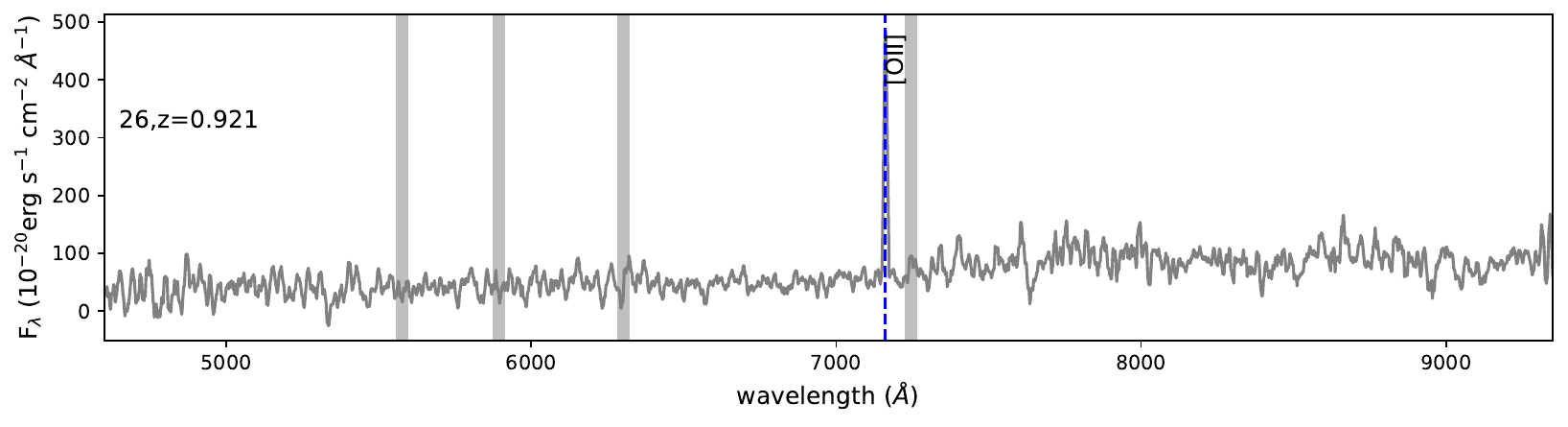}
\includegraphics[width=.85\textwidth]{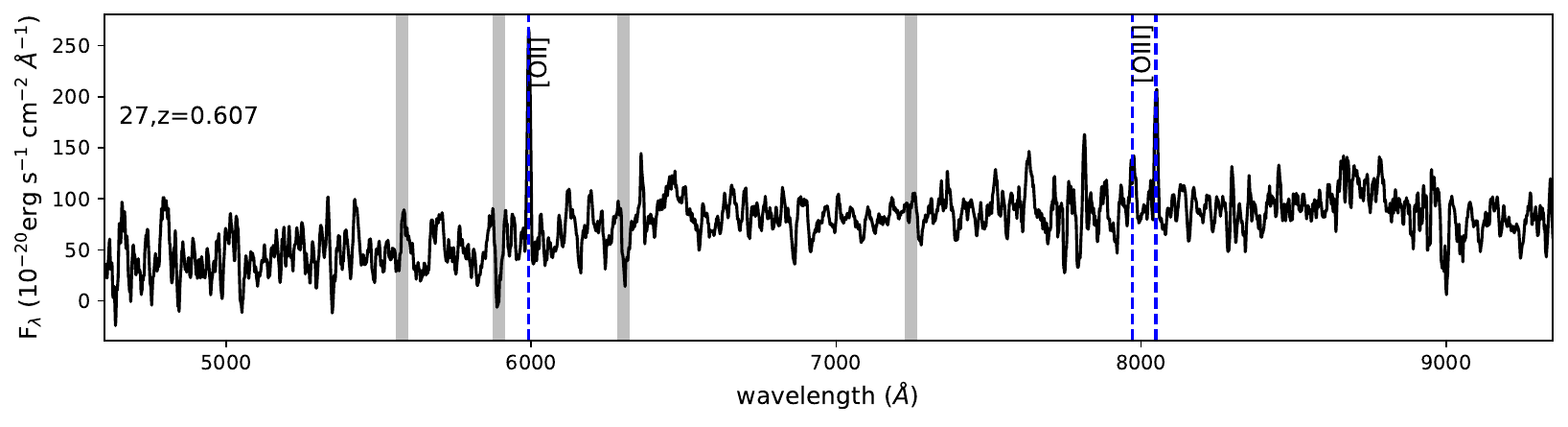}
\caption{Continued}
\end{figure}

\begin{figure}[ht!]
\ContinuedFloat
\centering
\includegraphics[width=.85\textwidth]{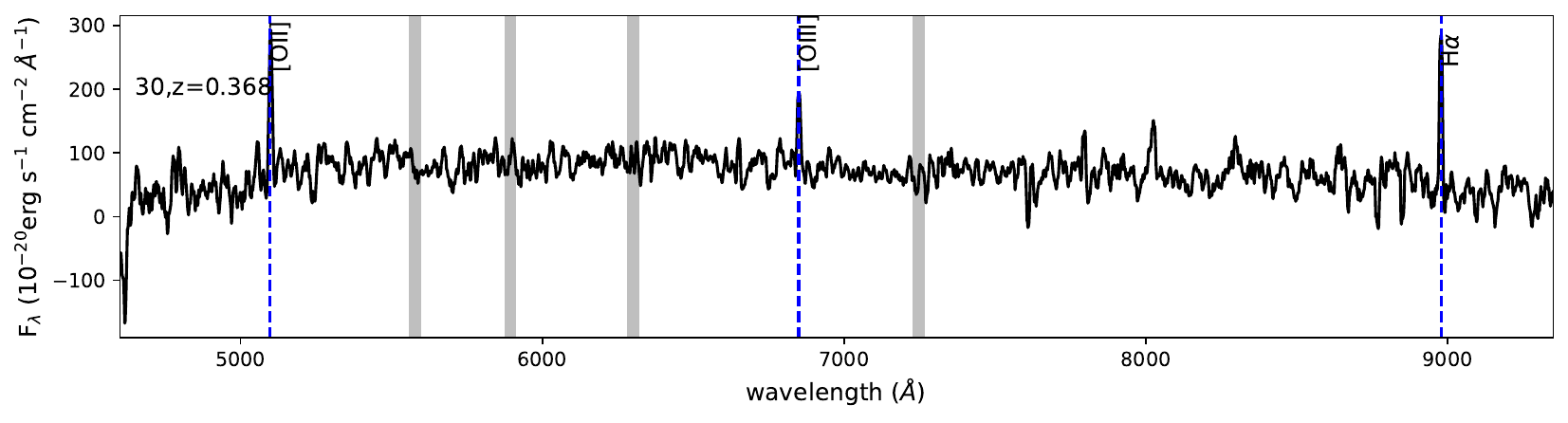}
\includegraphics[width=.85\textwidth]{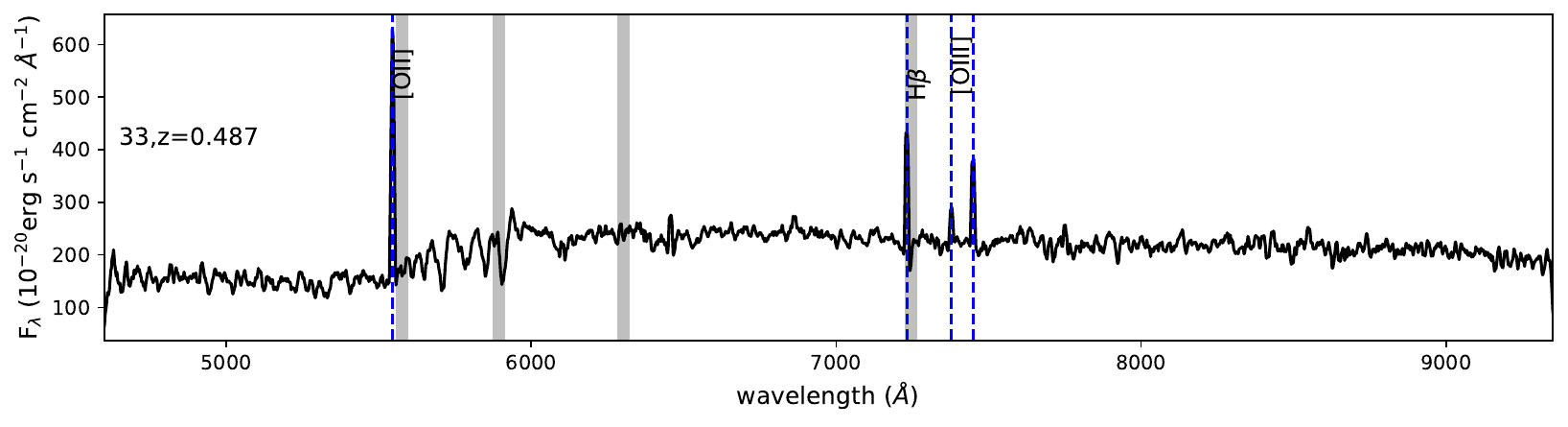}
\caption{Continued}
\end{figure}

\clearpage
\section{{\sc pPXF} results for galaxies}

\begin{figure}[ht!]
\includegraphics[width=0.43\textwidth]{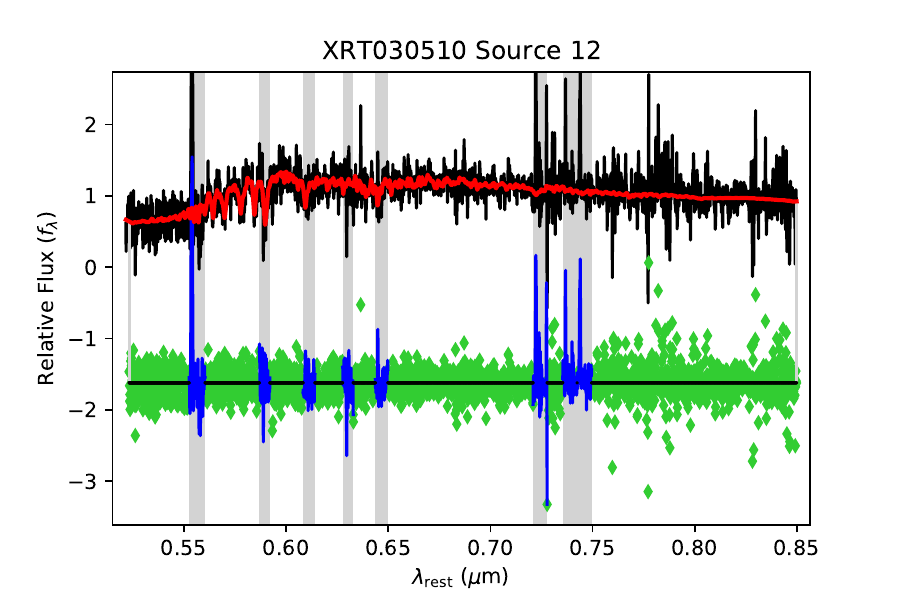}
\includegraphics[width=0.43\textwidth]{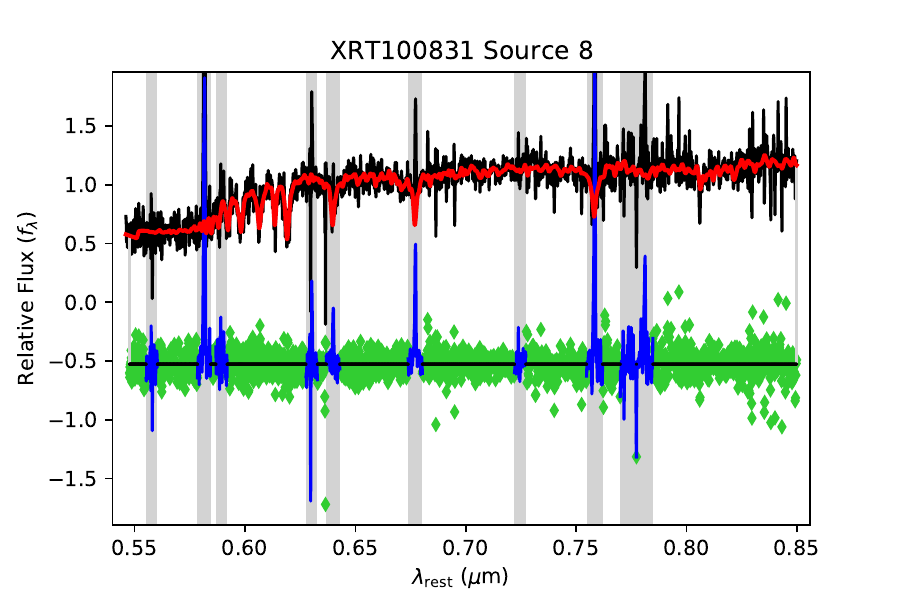}
\includegraphics[width=0.43\textwidth]{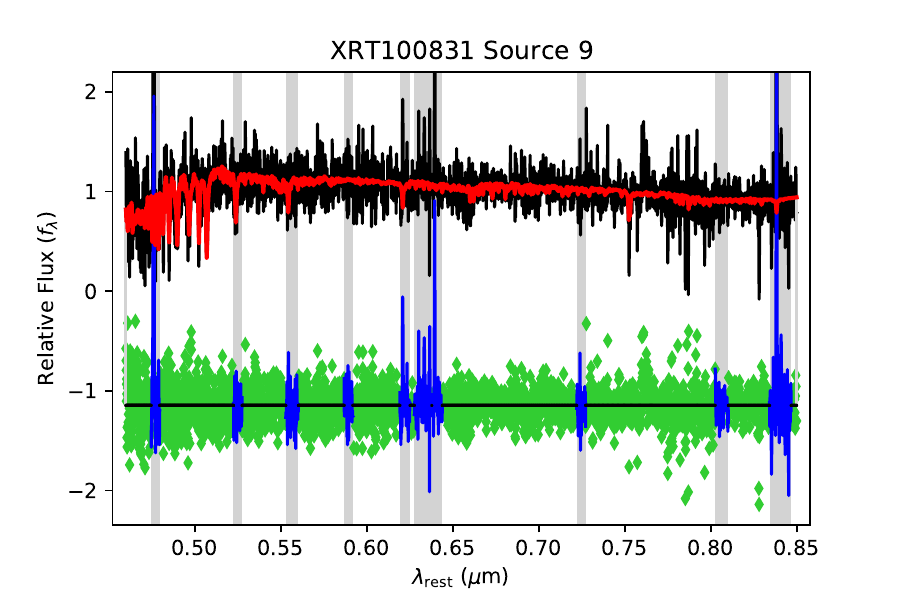}
\includegraphics[width=0.43\textwidth]{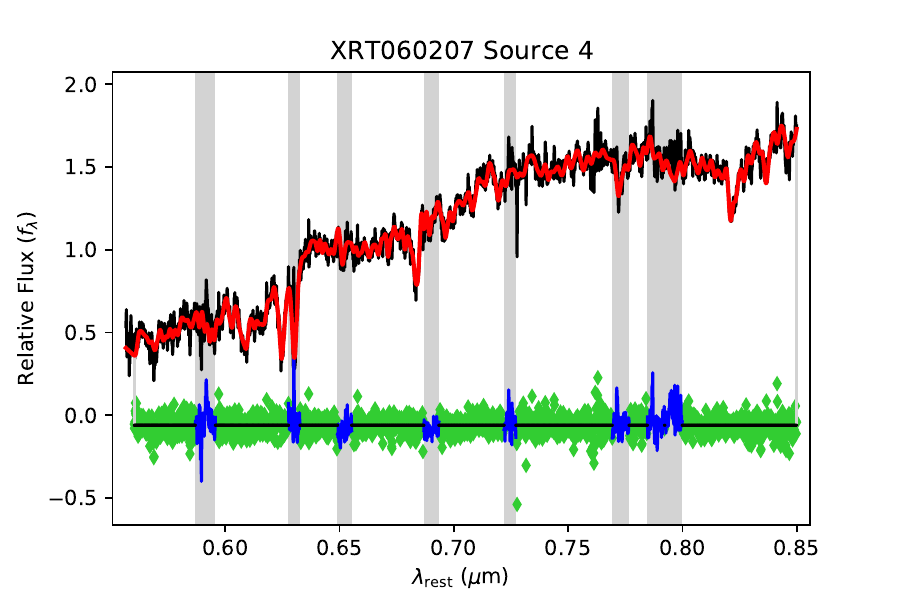}
\includegraphics[width=0.43\textwidth]{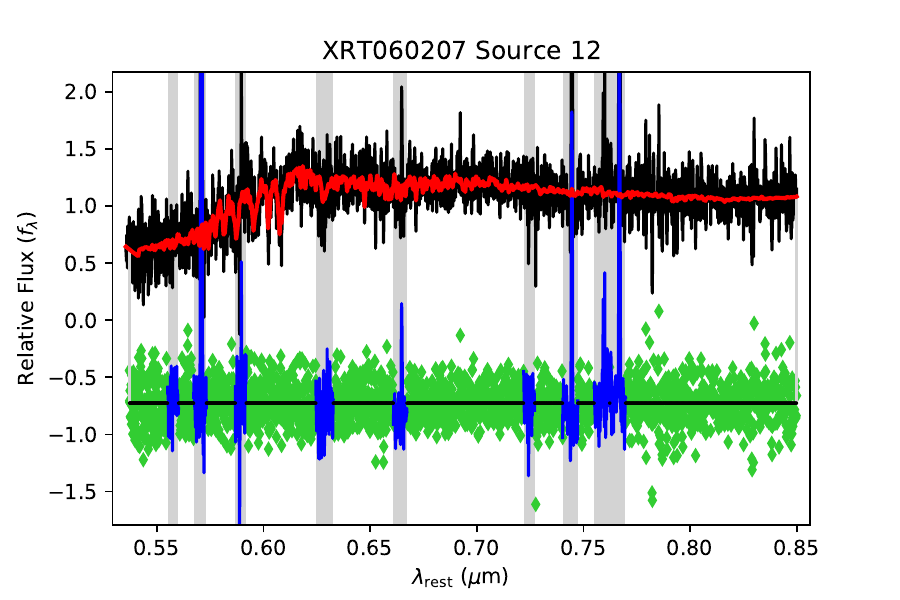}
\includegraphics[width=0.43\textwidth]{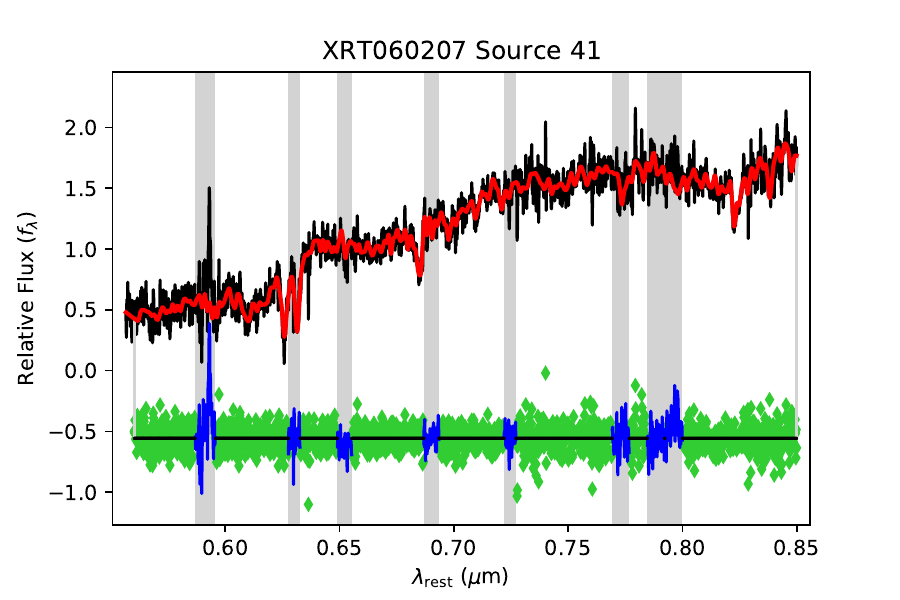}
\includegraphics[width=0.43\textwidth]{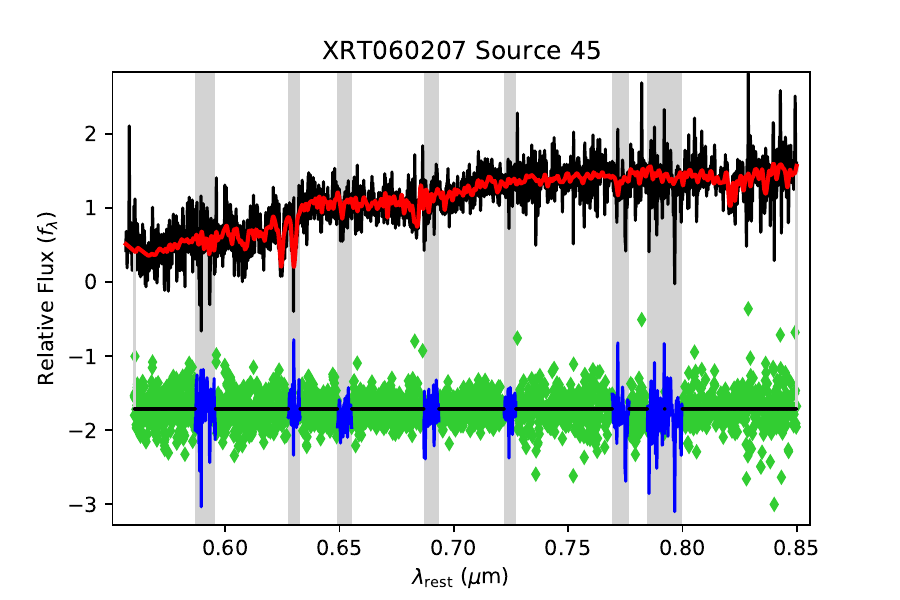}
\hspace*{+2.5cm}\includegraphics[width=0.43\textwidth]{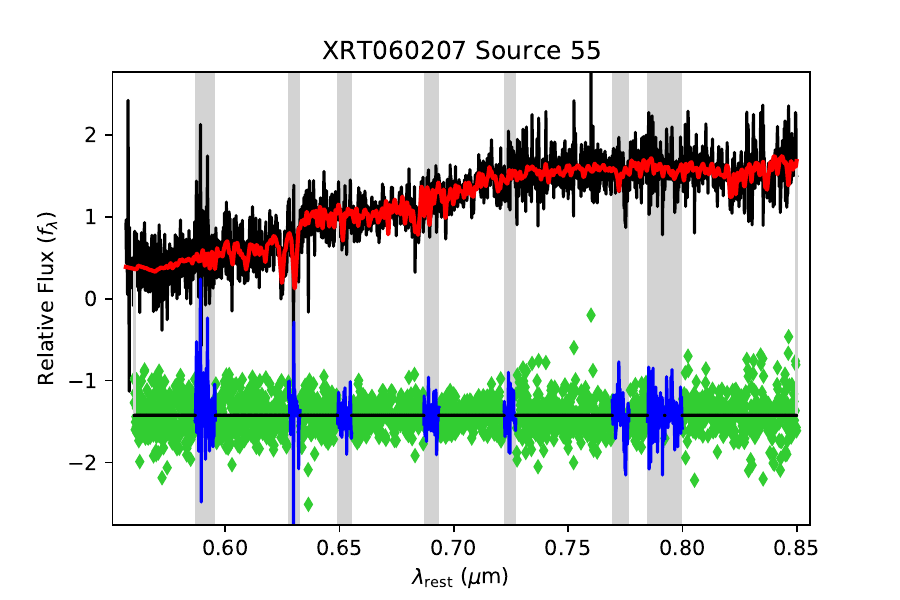}
\caption{The normalised spectra of the sources for which we were able to identify the Ca H+K lines is shown in black. The best fit for the continuum is over-plotted in red and the residuals after subtracting the continuum are plotted in green and red. The grey band correspond to wavelength around host galaxy emission lines or known sky lines and these regions are excluded from the fit.}
\label{apfig:ppxf_galaxies}
\end{figure}

\begin{figure}[ht!]
\ContinuedFloat
\includegraphics[width=0.43\textwidth]{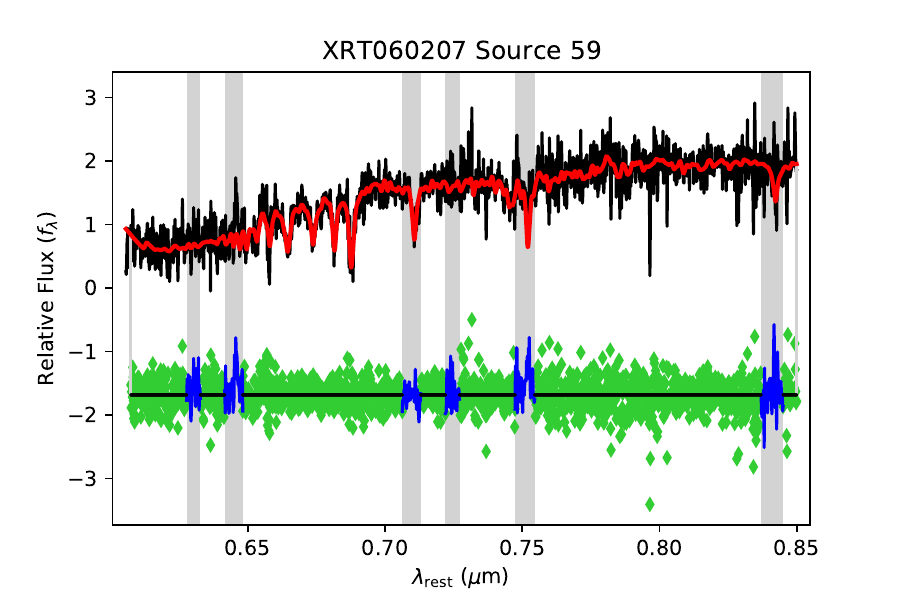}
\includegraphics[width=0.43\textwidth]{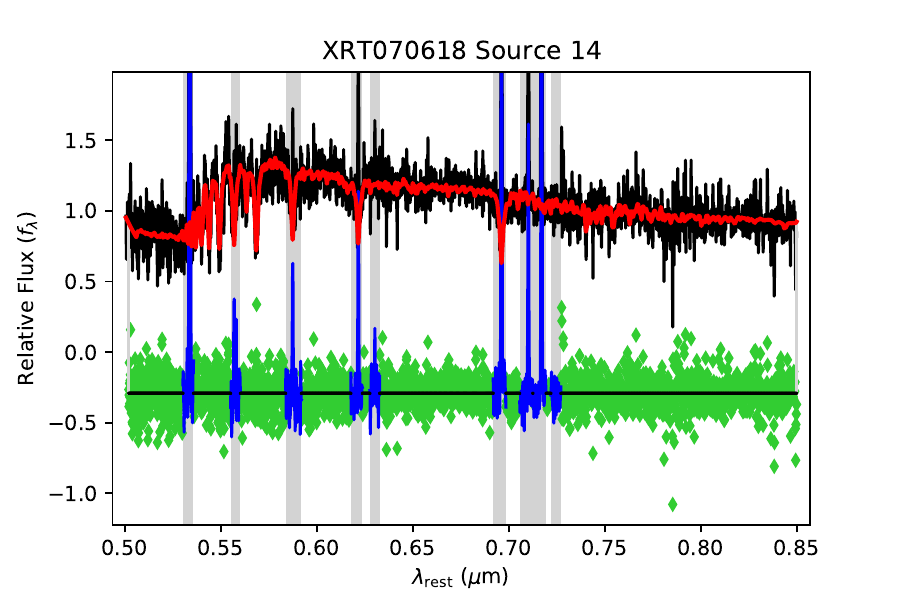}
\includegraphics[width=0.43\textwidth]{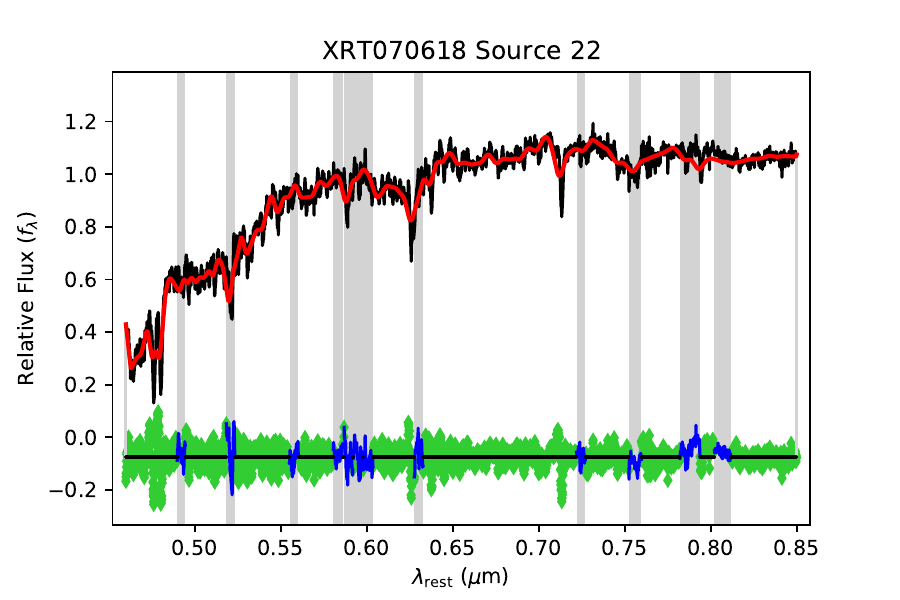}
\includegraphics[width=0.43\textwidth]{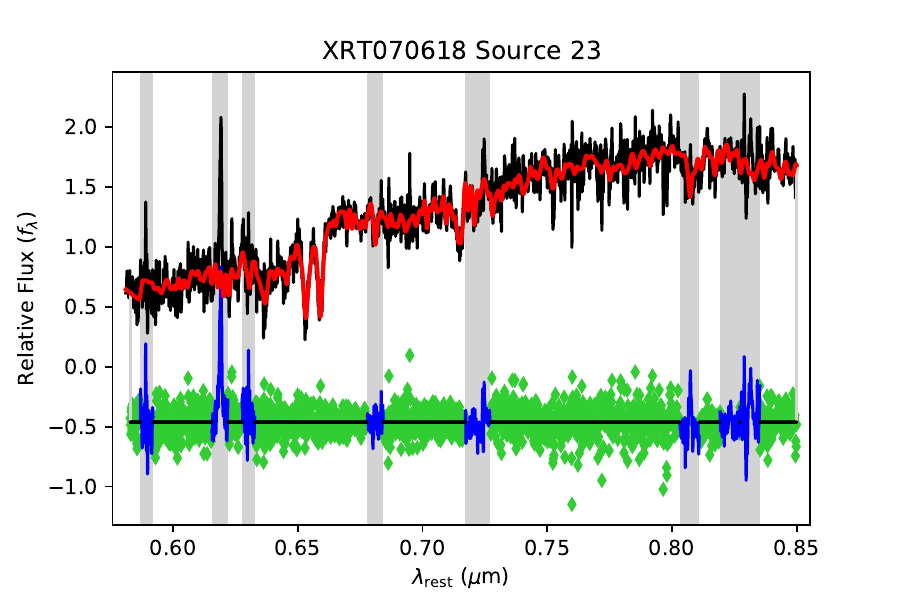}
\includegraphics[width=0.43\textwidth]{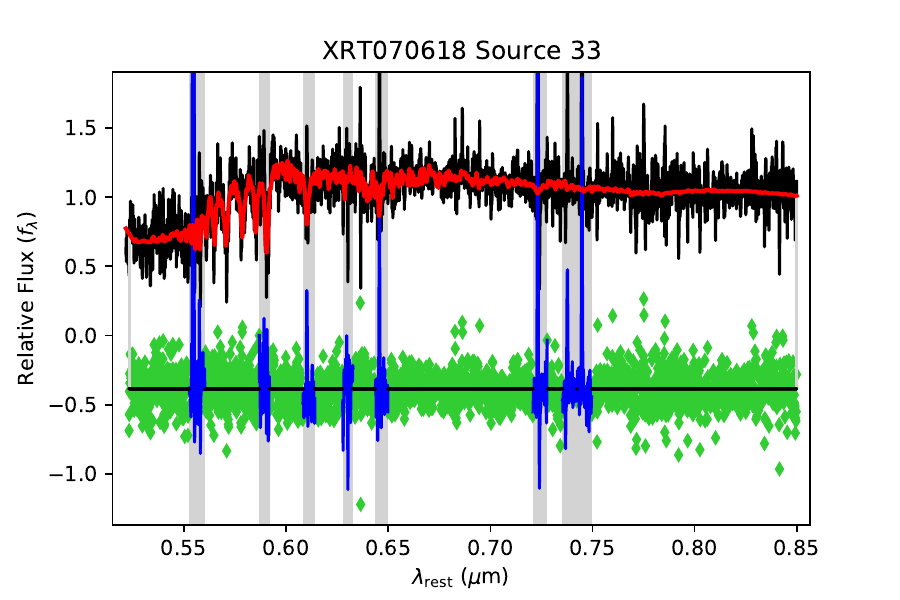}
\caption{Continued}
\end{figure}

\clearpage

\section{{\sc pPXF} results for dwarf stars}

\begin{figure}[ht!]
\includegraphics[width=0.43\textwidth]{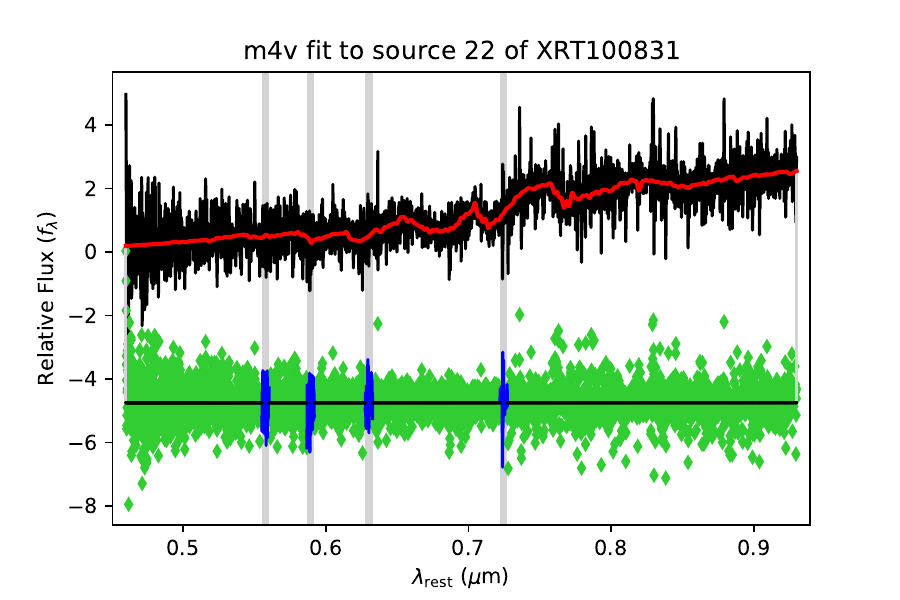}
\includegraphics[width=0.43\textwidth]{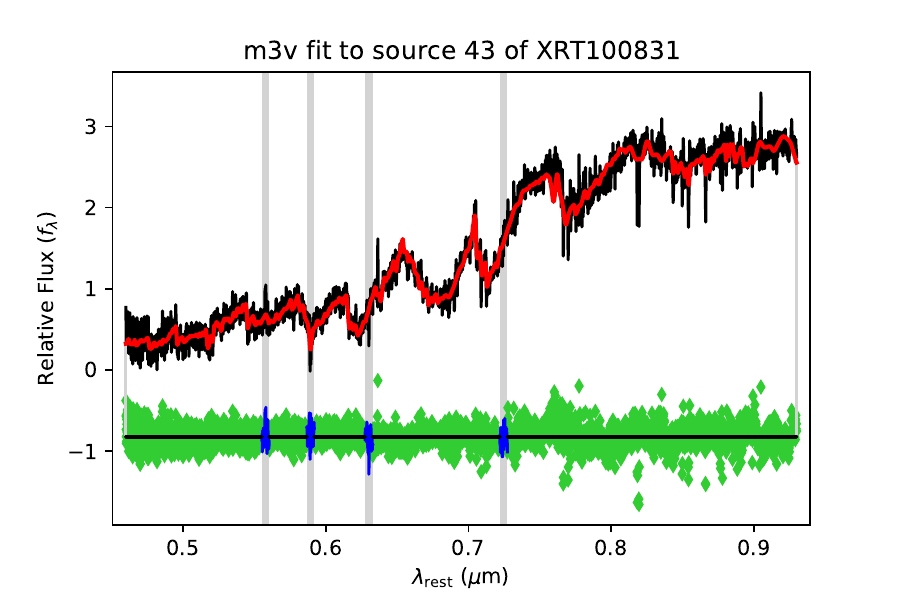}
\includegraphics[width=0.43\textwidth]{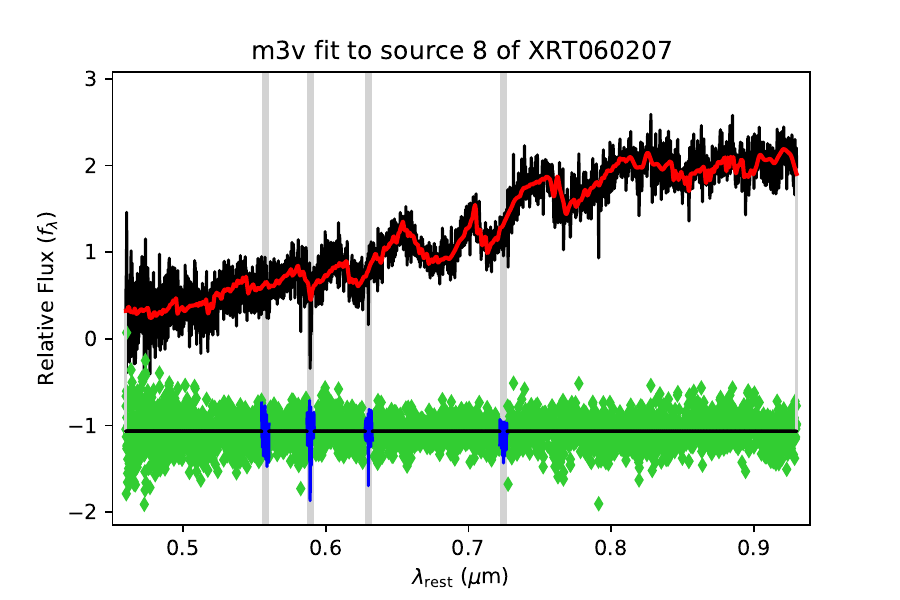}
\includegraphics[width=0.43\textwidth]{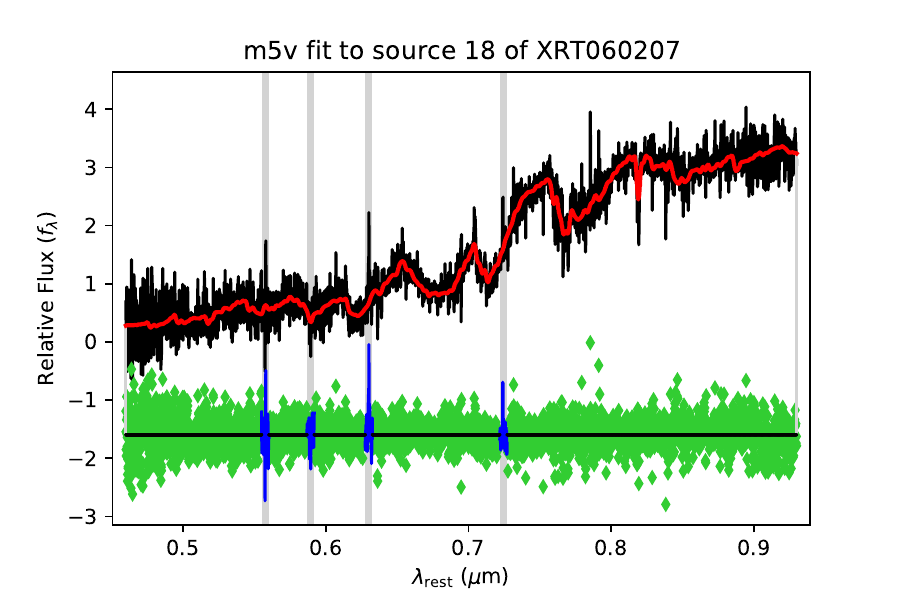}
\includegraphics[width=0.43\textwidth]{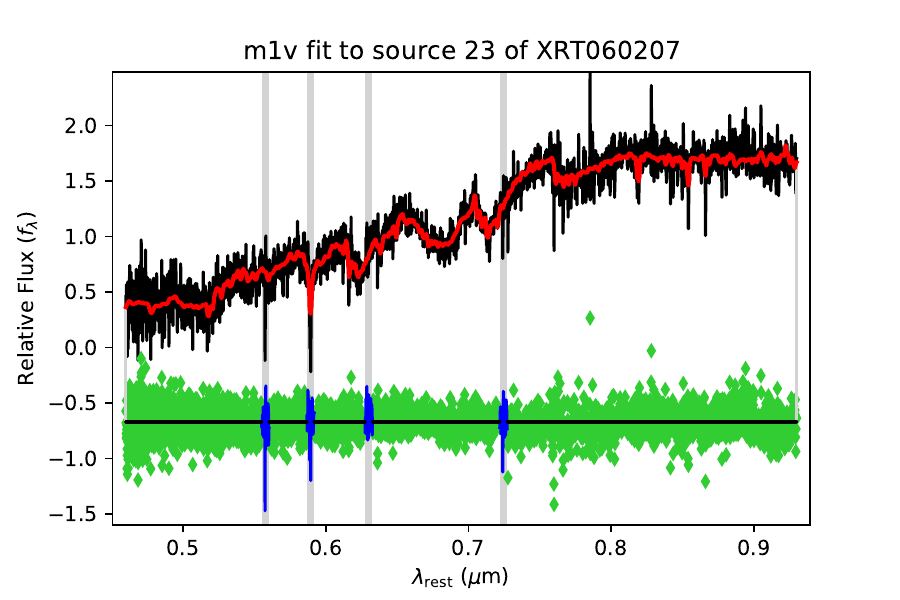}
\includegraphics[width=0.43\textwidth]{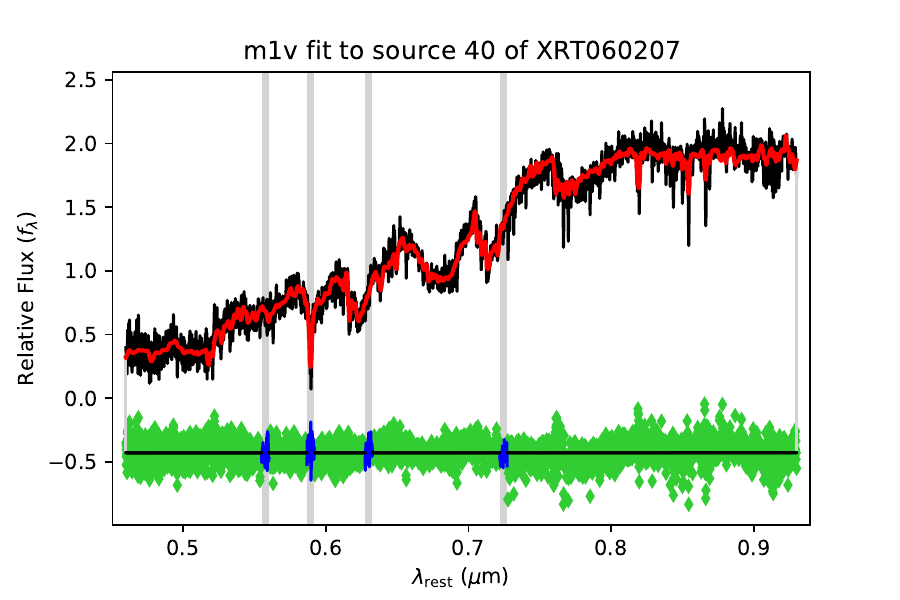}
\includegraphics[width=0.43\textwidth]{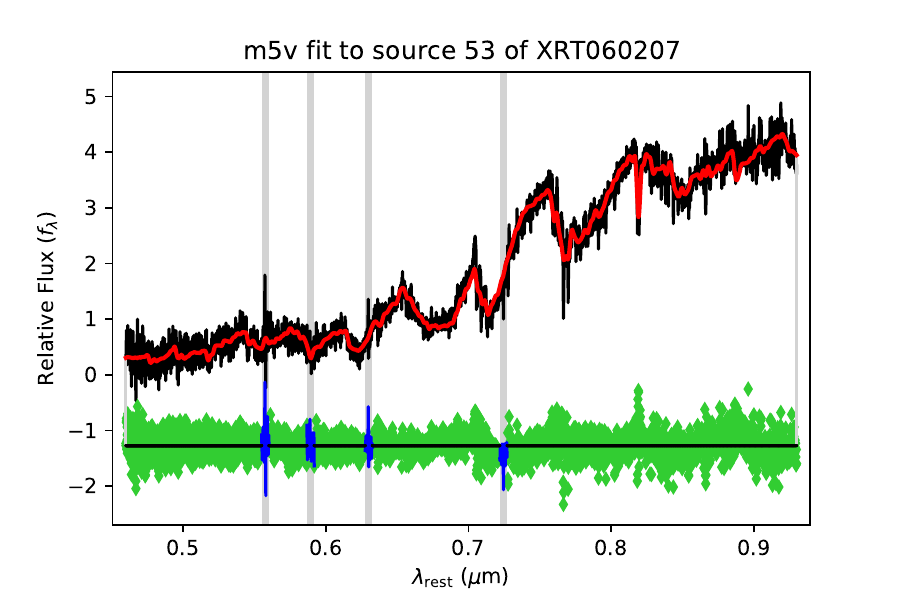}
\hspace*{+1.cm}\includegraphics[width=0.47\textwidth]{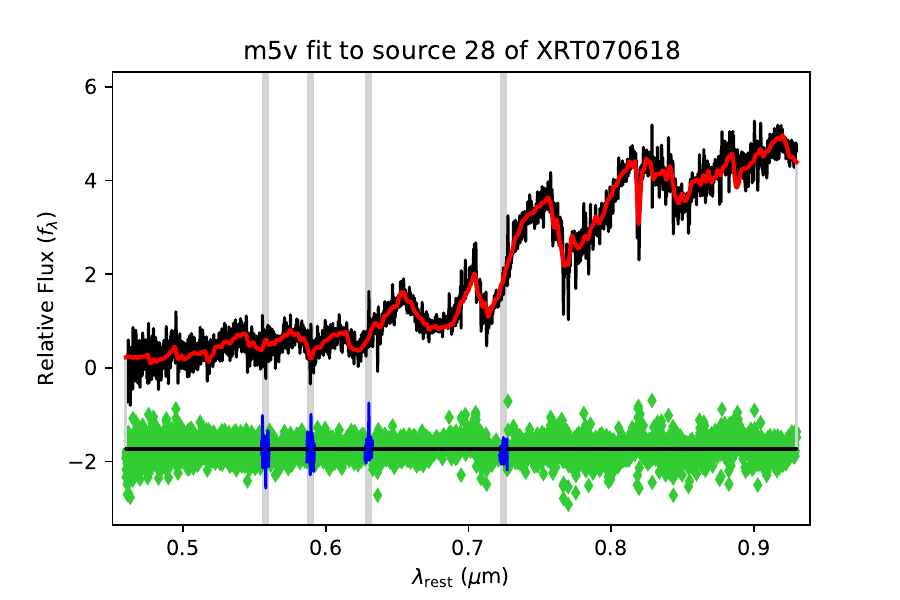}

\caption{The normalised spectra of the sources which we identify as dwarf stars. The best fit for the continuum is over-plotted in red and the residuals after subtracting the continuum are plotted in green and red. The grey band correspond to wavelength around known sky lines and these regions are excluded from the fit. The reduced $\chi^2$ of these fits are listed in Table~\ref{tab:dwarfs}.}
\label{apfig:ppxf_dwarfs}
\end{figure}

\clearpage

\section{Spectra of FXT positions}

\begin{figure}[h!]
\centering
\includegraphics[width=.9\textwidth]{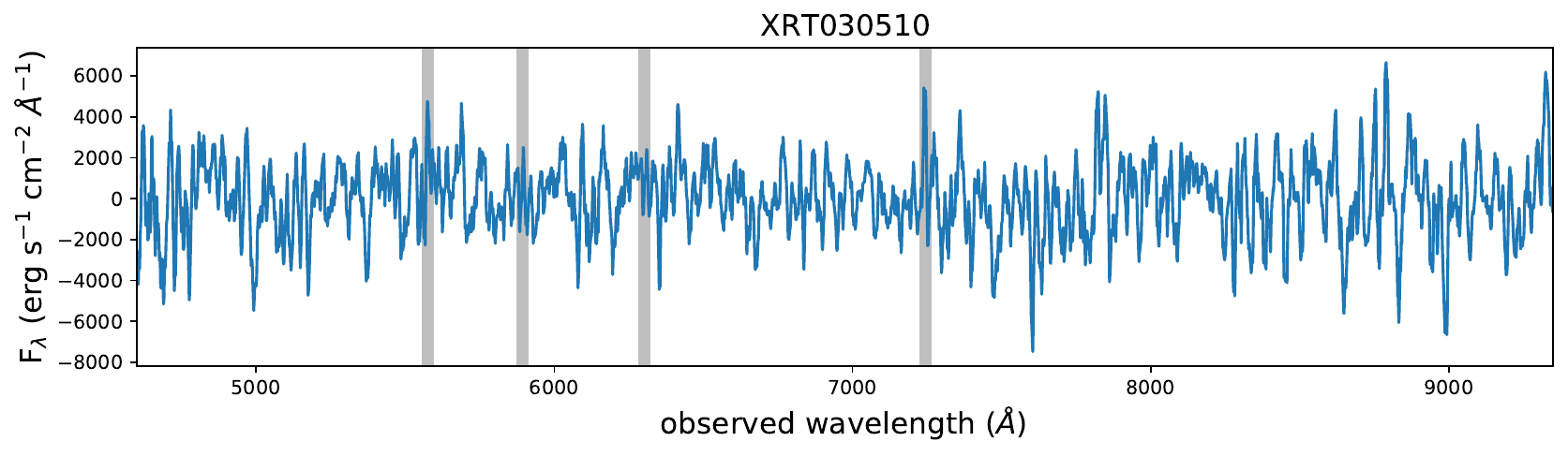}
\includegraphics[width=.9\textwidth]{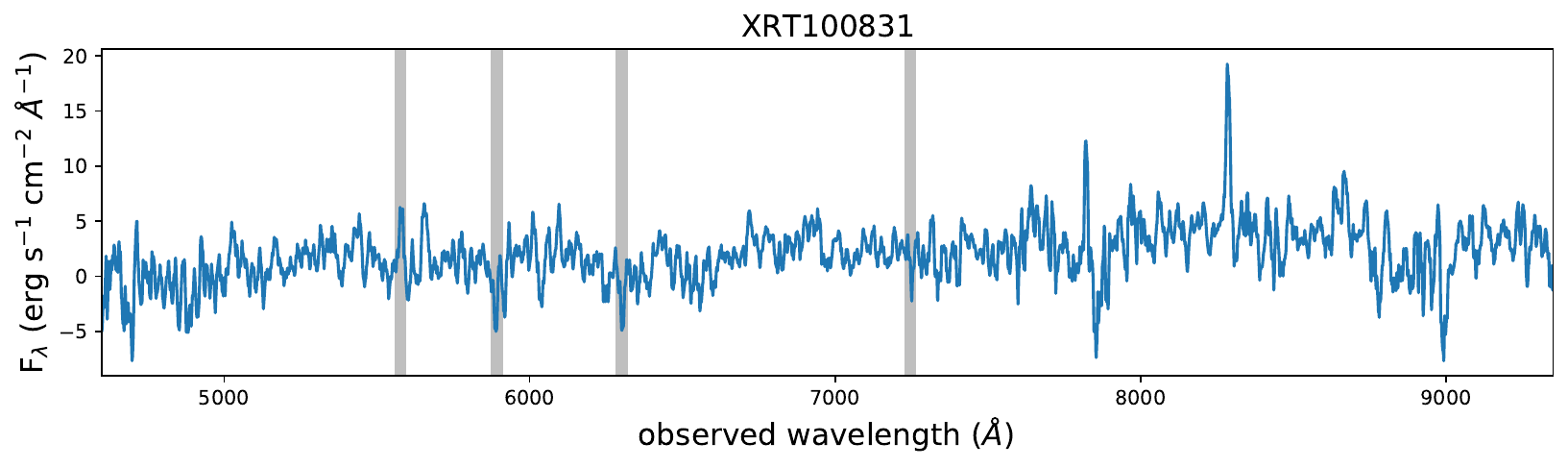}
\includegraphics[width=.9\textwidth]{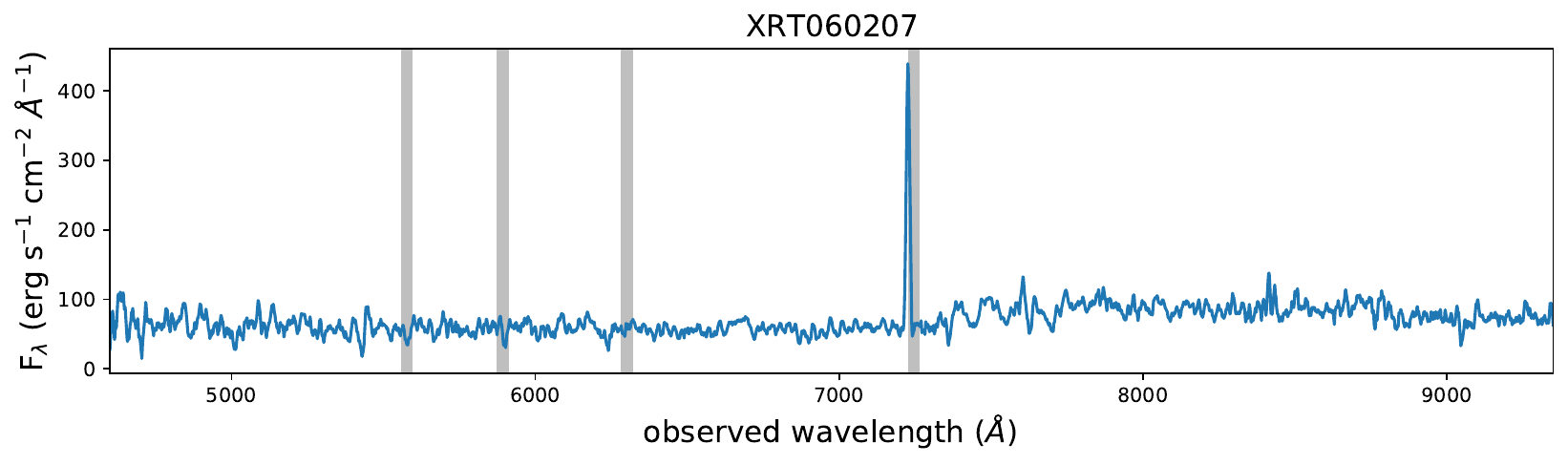}
\includegraphics[width=.9\textwidth]{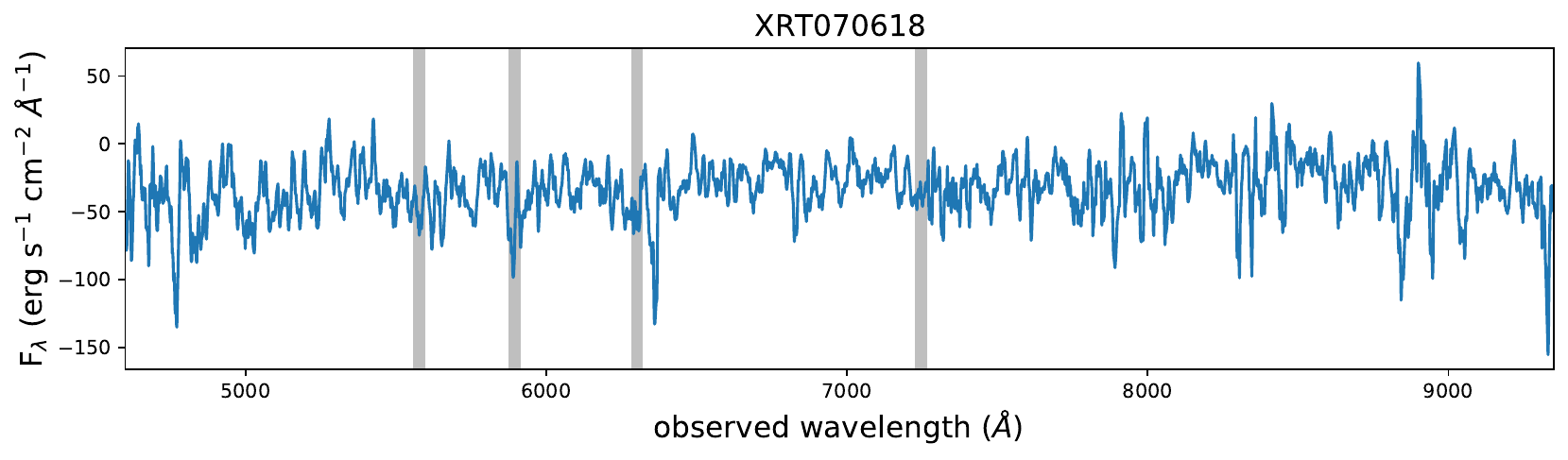}
\caption{The spectra extracted from the MUSE data at the position of the four FXT, smoothed using \texttt{Box1DKernel} with a kernel width of 10~pixels. The area used for the extraction of each of the spectra was equal to the 1$\sigma$ uncertainty on the position of the respective FXT. The vertical grey bands indicate the wavelengths of four prominent sky emission lines. We detect no emission lines in the spectra of \XTone and \XTfour. We detect an emission line in the spectrum of the region of \XTtwo\ and its nature is discussed in Section~\ref{sec:disc_xt2}. The line around 7300\AA\ in the spectrum of the region of \XTthree\ is the same line as in the spectrum of source 36, as this source is within the 1$\sigma$ positional uncertainty region of \XTthree. }
\label{apfig:spectra_fxts}
\end{figure}

\end{appendix}

\end{document}